\documentclass[aps,prd,twocolumn,superscriptaddress,floatfix,showpacs,showkeys,preprintnumbers]{revtex4}
\usepackage[utf8]{inputenc}
\usepackage[T1]{fontenc}
\usepackage{slashed}
\usepackage{times}
\usepackage{amssymb,amsfonts,amsmath,amsthm}
\usepackage{dsfont,bbm}
\usepackage{dcolumn}
\usepackage{epsf}
\usepackage{graphicx}
\usepackage[caption=false]{subfig}
\usepackage{dsfont}
\usepackage{slashed}
\usepackage[active]{srcltx}
\usepackage[usenames]{color}
\newcommand{\up}{{\uparrow}}
\newcommand{\down}{{\downarrow}}
\begin{document}
%\draft
\title{Nucleon Form Factors and Distribution Amplitudes in QCD}
\author{I.V.~Anikin}
\affiliation{Institut f\"ur Theoretische Physik, Universit\"at
   Regensburg,D-93040 Regensburg, Germany}
\affiliation{Bogoliubov Laboratory of Theoretical Physics, JINR, 141980 Dubna, Russia}
\author{V.M.~Braun}
\affiliation{Institut f\"ur Theoretische Physik, Universit\"at
   Regensburg,D-93040 Regensburg, Germany}
\author{N.~Offen}
\affiliation{Institut f\"ur Theoretische Physik, Universit\"at
   Regensburg,D-93040 Regensburg, Germany}
\date{\today}
\begin{abstract}
  \vspace*{0.3cm}
\noindent
We derive light-cone sum rules for the electromagnetic nucleon form factors including the
next-to-leading-order corrections for the contribution of twist-three and twist-four
operators and a consistent treatment of the nucleon mass corrections.
The essence of this approach is that soft Feynman contributions are calculated in terms
of small transverse distance quantities using dispersion relations and duality.
The form factors are thus expressed in terms of nucleon wave functions at small
transverse separations, called distribution amplitudes, without any additional parameters.
The distribution amplitudes, therefore, can be extracted from the comparison with the
experimental data on form factors and compared to the results of lattice QCD simulations.
A selfconsistent picture emerges, with the three valence quarks carrying $40\%:30\%:30\%$
of the proton momentum.
 \end{abstract}
\maketitle
\date{\today}
%%%%%%%%%%%%%%%%%%%%%%%%%%%%%%%%%%%%%%%%%%%%%%%%%%%%%%%%%%%%%%%%%%%%%%%%%%%%%%%%%%%%%
%%%%%%%%%%%%%%%%%%%%%%%%%%%%%%%%%%%%%%%%%%%%%%%%%%%%%%%%%%%%%%%%%%%%%%%%%%%%%%%%%%%%%
\section{Introduction}
\begin{figure}[ht]
\centerline{\includegraphics[width=5cm, clip = true]{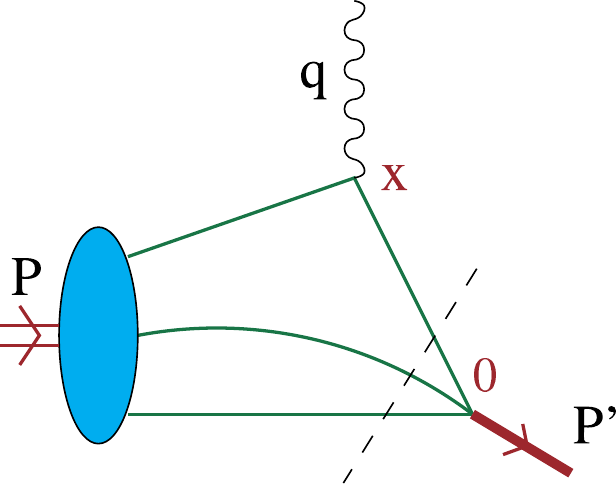}}
%\centerline{\epsfxsize5cm\epsffile{figLCSR}}
\caption{\label{figsum}\small
Schematic structure of the light-cone sum rule for baryon form factors.
}
\end{figure}

\setcounter{equation}{0}
%%%%%%%%%%%%%%%%%%%%%%%%%%%%%%%%%%%%%%%%%%%%%%%%%%%%%%%%%%%%%%%%%%%%%%%%%%%%%%%%%%%%%
%%%%%%%%%%%%%%%%%%%%%%%%%%%%%%%%%%%%%%%%%%%%%%%%%%%%%%%%%%%%%%%%%%%%%%%%%%%%%%%%%%%%%
It is generally accepted that studies of hard exclusive reactions and in particular
hadron form factors at large momentum transfer give access to different aspects of the
internal structure of hadrons as compared to inclusive reactions, so that these two options
are to a large extent complementary to each other.
The QCD factorization approach  to  exclusive processes
\cite{Chernyak:1977as,Radyushkin:1977gp,Lepage:1979zb}
introduces the concept of hadron distribution amplitudes (DAs) which can be thought of as momentum fraction
distributions in configurations with
a fixed number of Fock constituents (quarks, antiquarks and gluons) at small transverse separations.
It is argued that in the formal $Q^2 \to \infty$ limit
form factors can be  written  in a factorized form, as a convolution
of DAs related to  hadrons in the initial and  final state  times a ``short-distance''
coefficient function that is calculable in QCD perturbation theory.
The leading contribution corresponds to DAs with minimal possible number
of constituents --- three for baryons and two for mesons.
Thus, in this framework, measurements of form factors at large momentum transfer $Q$
provide one with the information on valence quark distributions inside hadrons
in rare configurations where they are separated by a small transverse distance of the order of $1/Q$.
This, classical, factorization approach faces conceptional difficulties in the
application to baryons~\cite{Duncan:1979hi,Milshtein:1981cy,Kivel:2010ns} but, probably
more importantly, seems to be failing phenomenologically for realistic momentum transfers
accessible in current or planned experiments. The problem is simply that
each hard gluon exchange is accompanied by the $\alpha_s/\pi$ factor which is a standard
perturbation theory penalty for each extra loop.
If, say,  $\alpha_s/\pi \sim 0.1$, the factorisable contribution to baryon form factors
is suppressed by a factor of 100 compared to the ``soft'' (end-point) contributions which are
suppressed by a power of $1/Q^2$ but do not involve small coefficients.
Hence the collinear factorization regime is approached very slowly.
There is overwhelming evidence from model calculations that ``soft'' contributions
play the dominant role at present energies.
Taking into account soft contributions is challenging because
they involve a nontrivial overlap of nonperturbative wave functions of the initial and the final state
hadrons, and are not factorizable, i.e. cannot be simplified further in terms of simpler inputs.
One possibility is to use transverse-momentum dependent (TMD) light-cone wave functions $\Psi(x,k_\perp)$
in combination with Sudakov suppression of large transverse separations following the
approach suggested initially by Li and Sterman~\cite{Li:1992nu} for the pion form factor.
Another possibility, which we advocate in this
work, is to calculate the soft contributions to the form factors as an expansion in terms of nucleon DAs
of increasing twist using dispersion relations and duality. This technique is known as
light-cone sum rules (LCSRs)~\cite{LCSR}.
It is attractive because in LCSRs  ``soft'' contributions to the form factors are calculated in
terms of the same DAs that enter the pQCD calculation and there is no double counting.
Thus, the LCSRs provide one with the most direct relation of the hadron form factors
and DAs that is available at present, with no other nonperturbative parameters.
The basic object of the LCSR approach to baryon form factors~\cite{Braun:2001tj,Braun:2006hz}
is the correlation function
$$\int\! dx\, e^{-iqx}\langle 0| T \{ \eta (0) j(x) \} | P \rangle $$ in which
$j$  represents the electromagnetic (or weak) probe and $\eta$
is a suitable operator with nucleon quantum numbers.
 The other (in this example, initial state) nucleon  is explicitly represented by its state vector
 $| P\rangle $, see a schematic representation in Fig.~\ref{figsum}.
The LCSR is obtained by comparing (matching) of two different representations for
the correlation function. On the one hand, when both  the momentum transfer  $Q^2$ and
 the momentum $(P')^2 = (P-q)^2$ flowing in the $\eta$ vertex are large and negative, the
 main contribution to the integral comes from the light-cone region $x^2\to 0$ and
 can be studied using the operator product expansion (OPE) of the time-ordered product
$T \{ \eta(0) j(x) \}$.  The  $x^2$-singularity  of a particular contribution
is determined by the twist of the relevant composite operator whose matrix element
$\langle 0|\ldots| P \rangle $ is related to the nucleon DA.
On the other hand, one can represent the answer in form of the dispersion integral
in $(P')^2$ and define the nucleon contribution by the cutoff in the quark-antiquark invariant mass,
the so-called interval of duality $s_0$ (or continuum threshold).
The main role of the interval of duality is that it does not allow large momenta $|k^2| > s_0$ to flow
through the $\eta$-vertex; to the lowest order $O(\alpha_s^0)$ one obtains a purely soft
contribution to the form factor as a sum of terms ordered by twist of the relevant operators and
hence including both the leading- and the higher-twist nucleon DAs.
Note that the contribution of higher-twist DAs is suppressed by powers of
the continuum threshold (or by powers of the Borel parameter after applying the
usual QCD sum rule machinery), but not by powers of $Q^2$, the reason being that
soft contributions are not constrained to small transverse separations.
The LCSR approach is not new and has been used successfully  for the calculations
of pion electromagnetic and also weak $B$-decay form factors.
In both cases this technique has reached a certain degree of maturity; see
Refs.~\cite{Braun:1999uj,Ball:2004ye,Duplancic:2008ix} for several recent state-of-the-art calculations.
The LCSRs for baryon form factors are more complicated and remain to be, comparatively, at exploratory stage.
Following the first formulation of the LCSRs for the electromagnetic form factors
in Ref.~~\cite{Braun:2001tj} there have been several studies aimed at
finding an optimal nucleon interpolation current~\cite{Lenz:2003tq,Braun:2006hz,Aliev:2007qu,Aliev:2008cs}
and extending this technique to other elastic or transition form factors of interest.
LCSRs for the axial nucleon form factor were presented in \cite{Braun:2006hz,Aliev:2007qu,Wang:2006su},
for the scalar form factor in \cite{Wang:2006su} and tensor form factor in \cite{Erkol:2011iw}.
A generalization to the full baryon octet was considered e.g. in ~\cite{Liu:2009mb}.
Application of the same technique to $N\gamma\Delta$ transitions was suggested
in~\cite{Braun:2005be,Aliev:2007qu} and to pion production at threshold in~\cite{Braun:2006td}.
LCRSs for weak baryon decays $\Lambda_b\to p,\Lambda\ell\nu_\ell$ etc. were studied in
\cite{Huang:2004vf,Wang:2009hra,Aliev:2010uy,Khodjamirian:2011jp}, etc.
In order to make the LCSR technique fully quantitative one needs to include the next-to-leading order (NLO)
QCD corrections to the coefficient functions of the DAs, which is the standard accepted in
$B$-decays. Calculation of these corrections for twist-three and twist-four contributions to the
LCSRs for the nucleon electromagnetic form factors $F_1(Q^2)$ and $F_2(Q^2)$ is the goal
and main result of this paper. This task was already partially addressed in Ref.~\cite{PassekKumericki:2008sj};
we will comment on the relation of our calculation to the results of~\cite{PassekKumericki:2008sj} in what
follows. In addition, we are able to organize the higher-twist contributions
related to nucleon mass corrections in a more systematic way, which reduces the corresponding
uncertainties.
The presentation is organized as follows. Sect.~2 is introductory and summarizes the present status
of the LCSR approach. We collect there the necessary
definitions and explain our notation. The general structure of LCSRs is explained and the leading-order
sum rules are given following Ref.~\cite{Braun:2006hz}. We also include new results concerning
the so-called Wandzura-Wilczek contributions to higher-twist DAs.
In Sect.~3 we describe our calculation of the NLO corrections for the contributions of (collinear) twist-three and twist-four operators.
The numerical analysis of the sum rules is presented in Sect.~4 whereas the final Sect.~5 is reserved for a summary
and conclusions.
The paper contains several Appendices. In App.~A we explain a general renormalization scheme for
three-quark operators~\cite{Kraenkl:2011qb} which is used throughout the calculation.
App.~B contains a summary of nucleon DAs and App.~C an update on the light-cone expansion of three-quark currents.
New results there are the twist-four contribution to the three-quark matrix element with generic quark positions
off the light-cone and a new derivation of the twist-five contribution (to LO).
App.~D. contains a summary of special functions that appear in the NLO calculations and their Borel transform.
The final App.~E contains a summary of the NLO coefficient functions to the twist-four accuracy.
%%%%%%%%%%%%%%%%%%%%%%%%%%%%%%%%%%%%%%%%%%%%%%%%%%%%%%%%%%%%%%%%%%%%%%%%%%%%%%%%%%%%%
%%%%%%%%%%%%%%%%%%%%%%%%%%%%%%%%%%%%%%%%%%%%%%%%%%%%%%%%%%%%%%%%%%%%%%%%%%%%%%%%%%%%%
\section{Preliminaries}
\setcounter{equation}{0}
%%%%%%%%%%%%%%%%%%%%%%%%%%%%%%%%%%%%%%%%%%%%%%%%%%%%%%%%%%%%%%%%%%%%%%%%%%%%%%%%%%%%%
%%%%%%%%%%%%%%%%%%%%%%%%%%%%%%%%%%%%%%%%%%%%%%%%%%%%%%%%%%%%%%%%%%%%%%%%%%%%%%%%%%%%%
%%%%%%%%%%%%%%%%%%%%%%%%%%%%%%%%%%%%%%%%%%%%%%%%%%%%%%%%%%%%%%%%%%%%%%%%%%%%%%%%%%%%%
%%%%%%%%%%%%%%%%%%%%%%%%%%%%%%%%%%%%%%%%%%%%%%%%%%%%%%%%%%%%%%%%%%%%%%%%%%%%%%%%%%%%%
\subsection{Distribution amplitudes}\label{ssec:DAs}
%\setcounter{equation}{0}
%%%%%%%%%%%%%%%%%%%%%%%%%%%%%%%%%%%%%%%%%%%%%%%%%%%%%%%%%%%%%%%%%%%%%%%%%%%%%%%%%%%%%
%%%%%%%%%%%%%%%%%%%%%%%%%%%%%%%%%%%%%%%%%%%%%%%%%%%%%%%%%%%%%%%%%%%%%%%%%%%%%%%%%%%%%
The LCSR approach allows one to calculate form factors for the range of momentum transfers accessible
in present day experiments in terms of quark distributions at small transverse separations, dubbed distribution
amplitudes. Conversely, the experimental data on form factors, analyzed in this framework, can be used to determine (constrain)
the DAs which are fundamental nonperturbative functions describing certain aspects of the nucleon structure and are
complementary to usual parton distributions.
The leading-twist-three nucleon (proton) DA
$\varphi_N(x_i,\mu)$ is defined by the matrix element~\cite{Chernyak:1983ej,Braun:2000kw}:
\begin{eqnarray}
\label{varphi-N}
\lefteqn{
\langle 0 |
\epsilon^{ijk}\! \left(u^{\up}_i(a_1 n) C \!\!\not\!{n} u^{\down}_j(a_2 n)\right)
\!\not\!{n} d^{\up}_k(a_3 n)
| P\rangle }
\nonumber\\
&=& - \frac12 f_N\,Pn\! \not\!{n}\, N^\up(P)\! \!\int\! [dx]
\,e^{-i P n \sum x_i a_i}\,
\varphi_N(x_i)\,,
\end{eqnarray}
where $q^{\up(\down)} = (1/2) (1 \pm \gamma_5) q$ are quark fields of given
helicity, $P_\mu$, $P^2=m_N^2$, is the proton momentum, $N(P)$ the usual Dirac spinor in
relativistic normalization, $n_\mu$ an auxiliary light-like vector $n^2=0$ and $C$
the charge-con\-ju\-ga\-tion matrix.
The Wilson lines that ensure gauge invariance are inserted between the quarks;
they are not shown for brevity.
The normalization constant $f_N$ is  defined in such a way that
\begin{equation}
\label{norm}
  \int [dx]\, \varphi_N(x_i) =1\,
\end{equation}
where
\begin{eqnarray}
\label{measure}
&&\int [dx] \,=\, \int_0^1  dx_1 dx_2 dx_3\, \delta\big(\sum x_i-1\big)\,.
\end{eqnarray}
The DA $\varphi_N(x_i,\mu)$ can be viewed, somewhat imprecisely, as the collinear limit of the
light-cone wave function corresponding to the valence three-quark proton state with zero orbital angular momentum~\cite{Lepage:1979zb}
\begin{equation}
 f_N(\mu) \, \varphi_N(x_i,\mu) \sim \int\limits_{|\vec{k}|<\mu} [d^2\vec{k}]\, \Psi_N(x_i,\vec{k}_i)\,,
\end{equation}
where the integration goes over the set of quark transverse momenta $\vec{k}_i$.
Thus, $f_N$ can be interpreted as the nucleon wave function at the origin (in position space).
The DAs are, in general, scheme- and scale-dependent and in the calculation of physical observables this
dependence is cancelled by the corresponding dependence of the coefficient functions.
The DA $\varphi_N(x_i,\mu)$ can be expanded in the set of orthogonal polynomials
$\mathcal{P}_{nk}(x_i)$
defined as eigenfunctions of the corresponding one-loop evolution equation:
\begin{equation}
   \varphi_N(x_i,\mu) = 120 x_1 x_2 x_3 \sum_{n=0}^\infty\sum_{k=0}^n \varphi_{nk}(\mu) \mathcal{P}_{nk}(x_i)
\label{expand-varphi}
\end{equation}
where
\begin{equation}
   \int [dx]\, x_1 x_2 x_3 \mathcal{P}_{nk}(x_i)\mathcal{P}_{n'k'}(x_i) \propto \delta_{nn'}\delta_{kk'}\,
\end{equation}
and to one-loop accuracy
\begin{eqnarray}
  f_N(\mu) &=& f_N(\mu_0) \left(\frac{\alpha_s(\mu)}{\alpha_s(\mu_0)}\right)^{2/(3\beta_0)}\,,
\nonumber\\
  \varphi_{nk}(\mu) &=& \varphi_{nk}(\mu_0)\left(\frac{\alpha_s(\mu)}{\alpha_s(\mu_0)}\right)^{\gamma_{nk}/\beta_0}.
\end{eqnarray}
Here $\beta_0 = 11-\frac23 n_f$ and $\gamma_{nk}$ are the corresponding anomalous dimensions. The double sum in Eq.~(\ref{expand-varphi})
goes over all existing orthogonal polynomials $\mathcal{P}_{nk}(x_i)$, $k=0,\ldots,n$, of degree $n$.
One can show that all eigenfunctions of the evolution equations, $\mathcal{P}_{nk}(x_i)$,
have definite parity under the interchange of the
first and the third argument, i.e. $\mathcal{P}_{nk}(x_3,x_2,x_1)= \pm \mathcal{P}_{nk}(x_1,x_2,x_3)$~\cite{Braun:2008ia}.
The first few polynomials are
\begin{eqnarray}
\mathcal{P}_{00} &= & 1\,,
\nonumber\\
\mathcal{P}_{10} &= &21(x_1-x_3)\,,
\qquad
\mathcal{P}_{11} = 7(x_1-2x_2 + x_3)\,,
\nonumber\\
\mathcal{P}_{20} &= & \frac{63}{10}[3 (x_1-x_3)^2 -3x_2(x_1+x_3)+ 2x_2^2]\,,
\nonumber\\
\mathcal{P}_{21} & = &\frac{63}{2}(x_1-3x_2 + x_3)(x_1-x_3)\,,
\nonumber\\
\mathcal{P}_{22} & = & \frac{9}{5} [x_1^2\!+\!9 x_2(x_1\!+\!x_3)\!-\!12x_1x_3\!-\!6x_2^2\!+\!x_3^2]
\label{lowest-P}
\end{eqnarray}
and the corresponding anomalous dimensions are
\begin{align}
  \gamma_{00} =& 0\,, \qquad\quad\! \gamma_{10} = \frac{20}{9}\,,\qquad \gamma_{11} = \frac83\,,
\notag\\
 \gamma_{20} =& \frac{32}{9}\,,\qquad \gamma_{21} = \frac{40}{9}\,,\qquad \gamma_{22} =\frac{14}{3}\,.
\end{align}
The normalization condition (\ref{norm}) implies that $\varphi_{00}=1$.
In what follows we will refer to the coefficients $\varphi_{nk}(\mu_0)$ with $n=1,2,\ldots $,  as shape parameters.
The set of these coefficients together with the normalization constant $f_N(\mu_0)$ at a reference
scale $\mu_0$ specifies the momentum fraction distribution of valence quarks in the nucleon.
They are nonperturbative quantities that can be related to matrix elements of local gauge-invariant
three-quark operators and calculated, e.g., on the lattice~\cite{Braun:2008ur,lattice2013}.
In the last twenty years there had been mounting evidence that the simple-minded picture of a proton
with the three valence quarks in an S-wave is insufficient, so that for example the proton spin
is definitely not constructed from the quark spins alone and also the electromagnetic Pauli form factor
$F_2(Q^2)$ involves quark orbital angular momenta. As shown in Ref.~\cite{Belitsky:2002kj}, the light-cone
wave functions with $L_z=\pm 1$ are reduced, in the limit of small transverse separation,
to the twist-four nucleon DAs introduced in Ref.~\cite{Braun:2000kw}:
\begin{eqnarray}
\label{twist-4}
\lefteqn{
\langle 0 | \epsilon^{ijk}\!
\left(u^{\up}_i(a_1 n) C\slashed{n} u^{\down}_j(a_2 n)\right)
\!\slashed{p} d^{\up}_k(a_3 n) |P\rangle }
\nonumber\\
&=& -\frac14
\,p n\, \slashed{p}\, N^{\up}(P)\! \!\int\! [dx]
\,e^{-i p n \sum x_i a_i}\,
\nonumber\\
&&\times \left[f_{N}\Phi^{WW}_4(x_i)+\lambda^N_1\Phi_4(x_i)\right],
\nonumber\\
\lefteqn{
\langle 0 | \epsilon^{ijk}\!
\left(u^{\up}_i(a_1 n) C \slashed{n}\gamma_{\perp}\slashed{p} u^{\down}_j(a_2 n)\right)
\gamma^{\perp}\slashed{n} d^{\down}_k(a_3 n) |P\rangle }
\nonumber\\
&=&
-\frac12 m_{N}
\, pn\, \slashed{n}\, N^{\up}(P)\! \!\int\! [dx]
\,e^{-i p  n \sum x_i a_i}\,
\nonumber\\
&&\times \left[f_{N}\Psi^{WW}_4(x_i)-\lambda^N_1\Psi_4(x_i)\right],
%\hspace*{2cm}\phantom{.}
\nonumber\\
\lefteqn{
\langle 0 | \epsilon^{ijk}\!
\left(u^{\up}_i(a_1 n) C\slashed{p}\,\slashed{n} u^{\up}_j(a_2 n)\right)
\!\not\!{n} d^{\up}_k(a_3 n) |P\rangle }
\nonumber\\
&=& \frac{\lambda^N_2}{12}\,m_N\, pn\, \slashed{n} {N}^{\up}(P)\! \!\int\! [dx]
\,e^{-i pn \sum x_i a_i}\,\Xi_4(x_i)\,,
\nonumber \\
\end{eqnarray}
where $\Phi^{WW}_4(x_i)$ and $\Psi^{WW}_4(x_i)$ are the so-called
Wandzura-Wilczek contributions.  They can be expressed in terms of the
leading-twist DA $\varphi_N(x_i)$ as follows~\cite{Braun:2008ia}:
\begin{eqnarray}
\Phi^{WW}_4(x_i) &=& -\sum_{n,k}\frac{240\,\varphi_{nk}}{(n+2)(n+3)}
 \left(n+2-\frac{\partial}{\partial x_3}\right)
\nonumber\\ &&{}\times x_1x_2x_3\mathcal{P}_{nk}(x_1,x_2,x_3)\,,
\nonumber\\
\Psi^{WW}_4(x_i) &=& -\sum_{n,k}\frac{240\,\varphi_{nk}}{(n+2)(n+3)}
 \left(n+2-\frac{\partial}{\partial x_2}\right)
\nonumber\\ &&{}\times x_1x_2x_3\mathcal{P}_{nk}(x_2,x_1,x_3)\,.
\label{WW}
\end{eqnarray}
The two new constants $\lambda_1^N$ and $\lambda_2^N$
are defined in such a way that the integrals of the ``genuine'' twist-4
DAs $\Phi_4$, $\Psi_4$, $\Xi_4$ are normalized to unity, similar to Eq.~(\ref{norm}).
They have the same scale dependence to the one-loop accuracy:
\begin{eqnarray}
  \lambda^N_{1,2}(\mu) &=& \lambda^N_{1,2}(\mu_0) \left(\frac{\alpha_s(\mu)}{\alpha_s(\mu_0)}\right)^{-2/\beta_0}.
\end{eqnarray}
Similar to the leading twist, the twist-4 DAs can be expanded in a set of orthogonal
polynomials that are eigenfunctions of the one-loop evolution equations,
but the difference is that starting from second order one has to take into account mixing with
four-particle (three-quark+gluon) operators. Since at present there is very little information on the
nucleon quark-gluon wave functions (see, however, Ref.~\cite{Braun:2011aw}) in this work we prefer to stay
within a three-quark description, and, for consistency, truncate the expansion of
$\Phi_4$, $\Psi_4$, $\Xi_4$ at the first order. To this accuracy
one obtains~\cite{Braun:2008ia}
\begin{eqnarray}
 \Phi_4(x_i,\mu) &=& 24 x_1 x_2 \Big\{1 + \eta_{10}(\mu) \mathcal{R}_{10}(x_3,x_1,x_2)
\nonumber\\&&{} - \eta_{11}(\mu) \mathcal{R}_{11}(x_3,x_1,x_2)\Big\},
\nonumber\\
 \Psi_4(x_i,\mu) &=& 24 x_1 x_3 \Big\{1 + \eta_{10}(\mu) \mathcal{R}_{10}(x_2,x_3,x_1)
\nonumber\\&&{} + \eta_{11}(\mu) \mathcal{R}_{10}(x_2,x_3,x_1)\Big\},
\nonumber\\
 \Xi_4(x_i,\mu) &=&  24 x_2 x_3 \Big\{1 + \frac{9}{4}\xi_{10}(\mu) \mathcal{R}_{11}(x_1,x_3,x_2)\Big\},
\label{PhiPsi4}
\end{eqnarray}
where
\begin{eqnarray}
 \mathcal{R}_{10}(x_1,x_2,x_3) &=& 4\left(x_1+x_2-\frac32 x_3\right)\,,
\nonumber\\
 \mathcal{R}_{11}(x_1,x_2,x_3) &=& \frac{20}{3}\left(x_1-x_2+\frac12 x_3\right)\,
\end{eqnarray}
and $\eta_{10}(\mu)$, $\eta_{11}(\mu)$, $\xi_{10}(\mu)$ are the  new shape parameters.
The corresponding one-loop anomalous dimensions are~\cite{Braun:2008ia}
\begin{equation}
\gamma_{10}^{(\eta)} =\frac{20}{9}\,,\qquad \gamma_{11}^{(\eta)} = 4\,,\qquad \gamma_{10}^{(\xi)} = \frac{10}{3}\,.
\end{equation}
The three-quark twist-5 distributions are the next in complexity and correspond to taking into account the transverse
momentum dependence (terms $\sim k_\perp^2$) in the collinear limit of the light-cone wave functions with $L_z=0,\pm 1$
and also higher partial waves. They can be written as~\cite{Braun:2000kw}
\begin{eqnarray}
\label{twist-5}
\lefteqn{
\langle 0 | \epsilon^{ijk}\!
\left(u^{\up}_i(a_1 n) C\slashed{p} u^{\down}_j(a_2 n)\right)
\!\slashed{n} d^{\up}_k(a_3 n) |P\rangle }
\nonumber\\
&=& -\frac18 \,m_N^2\, \slashed{n}\, N^{\up}(P)\! \!\int\! [dx] \,e^{-i pn \sum x_i a_i}\,
\nonumber\\
&&\times \left[f_{N}\Phi^{WWW}_5(x_i)+\lambda^N_1\Phi^{WW}_5(x_i) + \Phi_5(x_i)\right],
\nonumber\\
\lefteqn{
\langle 0 | \epsilon^{ijk}\!
\left(u^{\up}_i(a_1 n) C \slashed{p}\gamma_{\perp}\slashed{n} u^{\down}_j(a_2 n)\right)
\gamma^{\perp}\slashed{p} d^{\down}_k(a_3 n) |P\rangle }
\nonumber\\
&=&
-\frac12  m_N \, pn\,\slashed{p}\, N^{\up}(P)\! \!\int\! [dx]
\,e^{-i pn \sum x_i a_i}\,
\nonumber\\
&&\times \left[f_{N}\Psi^{WWW}_5(x_i)-\lambda^N_1\Psi^{WW}_5(x_i)+ \Psi_5(x_i)\right],
\nonumber\\
\lefteqn{
\langle 0 | \epsilon^{ijk}\!
\left(u^{\up}_i(a_1 n) C\slashed{n}\,\slashed{p} u^{\up}_j(a_2 n)\right)
\!\not\!{p} d^{\up}_k(a_3 n) |P\rangle }
\nonumber\\
&=& \frac{1}{12}m_N \, pn\, \slashed{n}\, {N}^{\up}(P)\! \!\int\! [dx]
\,e^{-i P  n \sum x_i a_i}\,
\nonumber\\
&&\times\left[\lambda^N_2\Xi^{WW}_5(x_i)+ \Xi_5(x_i) \right],
\end{eqnarray}
where $\Phi^{WWW}_5(x_i)$ and $\Phi^{WW}_5(x_i)$ (and similar for other DAs) are
the Wandzura-Wilczek-type contributions related to twist-3 and twist-4 operators, respectively.
One can show that
\begin{widetext}
 \begin{eqnarray}
\Phi^{WWW}_5(x_i) &=& \sum_{n,k}\frac{240\,\varphi_{nk}}{(n+2)(n+3)}
\biggl[\left(n+2-\frac{\partial}{\partial x_1}\right)\left(n+1-\frac{\partial}{\partial x_2}\right)-(n+2)^2\biggr]
 x_1x_2x_3\mathcal{P}_{nk}(x_1,x_2,x_3)\,,
\nonumber\\
\Psi^{WWW}_5(x_i) &=& \sum_{n,k}\frac{240\,\varphi_{nk}}{(n+2)(n+3)}
\biggl[\left(n+2-\frac{\partial}{\partial x_3}\right)\left(n+1-\frac{\partial}{\partial x_1}\right)-(n+2)^2\biggr]
x_1x_2x_3\mathcal{P}_{nk}(x_2,x_1,x_3)\,.
\label{WW5a}
\end{eqnarray}
and, for the models in Eq.~(\ref{PhiPsi4}),
 \begin{eqnarray}
\Phi^{WW}_5(x_i) &=&
-24\biggl\{ \frac13 \left(1-\frac{\partial}{\partial x_2}\right) x_2 x_3
 + \frac18 \left(2-\frac{\partial}{\partial x_2}\right) x_2 x_3
\Big[\eta_{10}\mathcal{R}_{10}(x_1,x_3,x_2)+ \eta_{11}\mathcal{R}_{11}(x_1,x_3,x_2) \Big]\biggr\},
\nonumber\\
\Psi^{WW}_5(x_i) &=&
-24\biggl\{ \frac13 \left(1-\frac{\partial}{\partial x_1}\right) x_1 x_2
 + \frac18 \left(2-\frac{\partial}{\partial x_1}\right) x_1 x_2
\Big[\eta_{10}\mathcal{R}_{10}(x_3,x_2,x_1) - \eta_{11}\mathcal{R}_{11}(x_3,x_2,x_1) \Big]\biggr\}.
\nonumber\\
\Xi^{WW}_5(x_i) &=&
\phantom{-}
24\biggl\{ \frac13\Big[ \left(1-\frac{\partial}{\partial x_3}\right) x_1 x_3 - 2 \left(1-\frac{\partial}{\partial x_2}\right) x_1 x_2\Big]
 + \frac{9}{32} \xi_{10} \left(2-\frac{\partial}{\partial x_3}\right) x_1 x_3 \mathcal{R}_{10}(x_2,x_3,x_1)
\nonumber\\&&{}
-
\frac{9}{32} \xi_{10} \left(2-\frac{\partial}{\partial x_2}\right)\Big[\mathcal{R}_{10}(x_3,x_1,x_2)+ \mathcal{R}_{10}(x_3,x_2,x_1)\Big] \biggr\}.
\label{WW5b}
\end{eqnarray}
\end{widetext}
The expressions in Eqs.~(\ref{WW5a}),(\ref{WW5b}) are new results. Their derivation and the generalization of (\ref{WW5b}) to arbitrary DAs
will be presented elsewhere.
The ``genuine'' twist-5 distributions $\Phi_5,\Psi_5,\Xi_5$ are not known apart from that their normalization
integrals and the first moments must vanish from general considerations, e.g.
\begin{equation}
\int [dx]\, \Phi_5(x_i) = \int [dx]\, x_k \Phi_5(x_i) = 0\,,\quad k=1,2,3\,
\end{equation}
and similar for $\Psi_5,\Xi_5$. In our analysis these contributions will be neglected, which is consistent
with neglecting four-particle nucleon DA terms that involve an additional gluon.
In practical calculations it is convenient to work with the expression
for the renormalized three-quark light-ray operator with open Dirac indices in terms
of the DAs. The necessary formulae are collected in App.~\ref{App:NDA}.
%%%%%%%%%%%%%%%%%%%%%%%%%%%%%%%%%%%%%%%%%%%%%%%%%%%%%%%%%%%%%%%%%%%%%%%%%%%%%%%%%%%%%
%%%%%%%%%%%%%%%%%%%%%%%%%%%%%%%%%%%%%%%%%%%%%%%%%%%%%%%%%%%%%%%%%%%%%%%%%%%%%%%%%%%%%
\subsection{LCSRs for nucleon form factors: General structure}\label{ssec:LCSR}
%\setcounter{equation}{0}
%%%%%%%%%%%%%%%%%%%%%%%%%%%%%%%%%%%%%%%%%%%%%%%%%%%%%%%%%%%%%%%%%%%%%%%%%%%%%%%%%%%%%
%%%%%%%%%%%%%%%%%%%%%%%%%%%%%%%%%%%%%%%%%%%%%%%%%%%%%%%%%%%%%%%%%%%%%%%%%%%%%%%%%%%%%
%
%
%
The matrix element of the electromagnetic current
\begin{equation}
\label{em}
 j_{\mu}^{\rm em}(x) = e_u \bar{u}(x) \gamma_{\mu} u(x) + e_d \bar{d}(x) \gamma_{\mu} d(x)
\end{equation}
taken between nucleon states is conventionally written in terms of the
{Dirac} and {Pauli form factors} $F_1(Q^2)$ and $F_2(Q^2)$:
\begin{eqnarray}
\label{F1F2}
\lefteqn{\langle P'| j_{\mu}^{\rm em}(0)|P\rangle =}
\nonumber\\&=&
\bar{N}(P')\left[\gamma_{\mu}F_1(Q^2)-i\frac{\sigma_{\mu\nu}q^{\nu}}{2m_N}F_2(Q^2)\right]N(P),
\end{eqnarray}
where $P_\mu$ is the initial nucleon momentum, $P^2 =m_N^2$, $P'=P-q$, $Q^2 :=-q^2$,
$\sigma_{\mu\nu}=\frac{i}{2}[\gamma_{\mu},\gamma_{\nu}]$ and $N(P)$ is the nucleon spinor.
Experimental data on the  scattering of electrons off nucleons, e.g. $e^- + p \to e^- + p$,
are often presented in terms of the {electric} $G_E(Q^2)$  and {magnetic} $G_M(Q^2)$
Sachs form factors which are related to $F_{1,2}(Q^2)$ as
\begin{eqnarray}
\label{GMGE}
G_M(Q^2) & = & F_1(Q^2)+F_2(Q^2),
\\
G_E(Q^2) & = & F_1(Q^2)-\frac{Q^2}{4m_N^2}F_2(Q^2).
\end{eqnarray}
The LCSR approach allows one to calculate the form factors in terms of the nucleon (proton) DAs
introduced in Sect.~\ref{ssec:DAs}.
To this end we consider the correlation function
\begin{equation}
\label{correlator}
T_{\nu}(P,q) = i\! \int\! d^4 x \, e^{i q x}
\langle 0| T\left[\eta(0) j_{\nu}^{\mathrm{em}}(x)\right] |P\rangle
\end{equation}
where T denotes time-ordering and $\eta(0)$ is the Ioffe interpolating current~\cite{Ioffe:1981kw}
\begin{align}
\label{ioffe}
& \eta(x) = \epsilon^{ijk} \left[u^i(x) C\gamma_\mu u^j(x)\right]\,\gamma_5 \gamma^\mu d^k(x)\,,
\notag\\
& \langle 0| \eta(0)|P\rangle  = \lambda_1 m_N N(P)\,.
\end{align}
We use the standard Bjorken--Drell
convention \cite{BD65} for the metric and the Dirac matrices; in particular,
$\gamma_{5} = i \gamma^{0} \gamma^{1} \gamma^{2} \gamma^{3}$, $C= i\gamma^2\gamma^0$
and the Levi-Civita tensor $\epsilon_{\mu \nu \lambda \sigma}$
is defined as the totally antisymmetric tensor with $\epsilon_{0123} = 1$.
The choice of the nucleon current is discussed at length in Ref.~\cite{Braun:2006hz}.
There is strong evidence that the Ioffe current gives rise to more accurate and reliable sum rules
as compared to other possible choices; for example the QCD sum rule estimates for the corresponding
coupling $\lambda_1$~(see \cite{Gruber:2010bj} for an update and further references)
agree very well with the lattice calculations~\cite{Braun:2008ur}.
The correlation function in Eq.~(\ref{correlator}) contains many different Lorentz structures
that can be separated using light-cone projections.
%Following Refs.~\cite{Braun:2001tj,Braun:2006hz}
We define a light-like vector $n_\mu$ by the condition
\begin{equation}
\label{n}
       q\cdot n =0\,,\qquad n^2 =0
\end{equation}
and introduce the second light-like vector as
\begin{equation}
\label{smallp}
p_\mu  = P_\mu  - \frac{1}{2} \, n_\mu \frac{m_N^2}{P\cdot n}\,,~~~~~ p^2=0\,,
\end{equation}
so that $P \to p$ in the infinite momentum frame $P\cdot n\to \infty$ or
if the nucleon mass can be neglected, $m_N \to 0$.
We also introduce the projector onto the directions orthogonal to $p$ and $n$,
\begin{equation}
       g^\perp_{\mu\nu} = g_{\mu\nu} -\frac{1}{pn}(p_\mu n_\nu+ p_\nu n_\mu)\,
\end{equation}
and will sometimes use a shorthand notation
\begin{equation}
    a_+\equiv a_\mu n^\mu, \qquad  a_-\equiv a_\mu p^\mu\,,
\qquad a_{\bot \mu} \equiv  g^\perp_{\mu\nu} a^{\nu}
\end{equation}
for $\gamma$-matrices and  arbitrary Lorentz vectors $a_\mu$.
The photon momentum can be written as
\begin{eqnarray}
q_{\mu}=q_{\bot \mu}+ n_{\mu}\frac{Pq}{Pn} =q_{\bot \mu}+ n_{\mu}\frac{pq}{pn}\,.
\end{eqnarray}
Last but not least, we define projection operators
\begin{equation}
\Lambda^+ = \frac{\slashed{p}\,\slashed{n}}{2 pn}\,, \qquad
\Lambda^- = \frac{\slashed{n}\,\slashed{p}}{2 pn}
\end{equation}
that pick up the ``plus'' and ``minus'' components of a spinor,
$N^\pm(P) = \Lambda^\pm N(P)$.
Note  the useful relations
\begin{equation}
\label{bwgl}
\slashed{p} N(P) = m_N N^+(P)\,,\qquad \slashed{z} N(P) = \frac{2 pn }{m_N} N^-(P)
\end{equation}
that follow from the Dirac equation
$(\slashed{P}-m_N)\,N(P) =0$.
It is easy to check that $N^+ \sim \sqrt{p^+}$ and $N^- \sim 1/\sqrt{p^+}$ in the infinite momentum frame
$p_+\to \infty$.
Lorentz structures that are most useful for writing the LCSRs are usually those containing
the maximum power of the large momentum $p_+$.
Following Refs.~\cite{Braun:2001tj,Braun:2006hz} we consider in what follows
the ``plus'' spinor projection of the correlation function (\ref{correlator})
involving the ``plus'' component of the electromagnetic current,
which can be parametrized in terms of two invariant functions
\begin{equation}
\label{project4}
  \Lambda_+ T_+ = p_+\left\{ m_N \mathcal{A}(Q^2,P'^2)  +
    \slashed{q}_\perp \mathcal{B}(Q^2,P'^2)\right\}N^+(P)\,,
\end{equation}
where $Q^2=-q^2$ and $P'^2 = (P-q)^2$.
The correlation functions $\mathcal{A}(Q^2,P'^2)$ and $\mathcal{B}(Q^2,P'^2)$
can be calculated in QCD for sufficiently large Euclidean momenta
$Q^2, -P'^2 \gtrsim 1$~GeV$^2$ using OPE (see the next Section). The results can be
presented in the form of a dispersion relation
\begin{eqnarray}
 \mathcal{A}^{\rm QCD}(Q^2,P'^2) &=& \frac1{\pi}\int_0^\infty\frac{ds}{s-P'^2} \text{Im}\, \mathcal{A}^{\rm QCD}(Q^2,s)+\ldots
\nonumber\\
 \mathcal{B}^{\rm QCD}(Q^2,P'^2) &=& \frac1{\pi}\int_0^\infty\frac{ds}{s-P'^2} \text{Im}\, \mathcal{B}^{\rm QCD}(Q^2,s)+\ldots
\nonumber\\
\end{eqnarray}
where the ellipses stand for necessary subtractions.
On the other hand, the same correlation functions can be written in terms of physical spectral densities that
contain a nucleon (proton) pole at $P'^2 \to m_N^2$, the nucleon resonances and the continuum. It is easy to
see that the nucleon contribution is proportional to the electromagnetic form factor, whereas the
contribution of higher mass states can be taken into account using quark-hadron duality:
\begin{eqnarray}
 \mathcal{A}^{\rm phys}(Q^2,P'^2) &=& \frac{2\lambda_1 F_1(Q^2)}{m_N^2-P'^2}
\nonumber\\&&{}+
\frac1{\pi}\int_{s_0}^\infty\frac{ds}{s-P'^2} \text{Im}\, \mathcal{A}^{\rm QCD}(Q^2,s)+\ldots
\nonumber\\
 \mathcal{B}^{\rm phys}(Q^2,P'^2) &=& \frac{\lambda_1 F_2(Q^2)}{m_N^2-P'^2}
\nonumber\\&&{}+
\frac1{\pi}\int_{s_0}^\infty\frac{ds}{s-P'^2} \text{Im}\, \mathcal{B}^{\rm QCD}(Q^2,s)+\ldots
\nonumber\\
\end{eqnarray}
where $s_0\simeq (1.5$~GeV$)^2$ is the interval of duality (also called continuum threshold).
Matching the two above representations and making the Borel transformation that
eliminates subtraction constants
\begin{equation}
  \frac{1}{s-P'^2} \longrightarrow e^{-s/M^2}
\end{equation}
one obtains the sum rules
\begin{align}
 2\lambda_1 F_1(Q^2) &= \frac1{\pi}\int_0^{s_0}ds \, e^{(m_N^2-s)/M^2} \text{Im}\, \mathcal{A}^{\rm QCD}(Q^2,s)\,,
\nonumber\\
  \lambda_1 F_2(Q^2) &= \frac1{\pi}\int_0^{s_0}ds \, e^{(m_N^2-s)/M^2} \text{Im}\, \mathcal{B}^{\rm QCD}(Q^2,s)\,.
\label{eq:LCSRscheme}
\end{align}
The dependence on the Borel parameter $M^2$ is unphysical and has to disappear in the full QCD calculation.
It is in this sense similar to the scale dependence of perturbative QCD calculations at a given order,
and can be used as one of the indicators of the theoretical uncertainty.
The new contribution of this paper is the calculation of the correlation functions
$\mathcal{A}(Q^2,P'^2)$ and $\mathcal{B}(Q^2,P'^2)$ to the NLO accuracy. This calculation is
described in the next Section.
%%%%%%%%%%%%%%%%%%%%%%%%%%%%%%%%%%%%%%%%%%%%%%%%%%%%%%%%%%%%%%%%%%%%%%%%%%%%%%%%%%%%%
%%%%%%%%%%%%%%%%%%%%%%%%%%%%%%%%%%%%%%%%%%%%%%%%%%%%%%%%%%%%%%%%%%%%%%%%%%%%%%%%%%%%%
\subsection{LO LCSRs}\label{sec:LO}
%\setcounter{equation}{0}
%%%%%%%%%%%%%%%%%%%%%%%%%%%%%%%%%%%%%%%%%%%%%%%%%%%%%%%%%%%%%%%%%%%%%%%%%%%%%%%%%%%%%
%%%%%%%%%%%%%%%%%%%%%%%%%%%%%%%%%%%%%%%%%%%%%%%%%%%%%%%%%%%%%%%%%%%%%%%%%%%%%%%%%%%%%
The correlation functions  $\mathcal{A}(Q^2,P'^2)$ and  $\mathcal{B}(Q^2,P'^2)$ can be
written as a sum of contributions of the $u,d$-quarks interacting with the electromagnetic probe,
weighted with the corresponding charges:
\begin{align}
\mathcal{A}= e_d\,\mathcal{A}_d + e_u \mathcal{A}_u\,,
&&
\mathcal{B}= e_d\,\mathcal{B}_d + e_u \mathcal{B}_u\,.
\end{align}
Each of the functions has a perturbative expansion which we write as
\begin{align}
 \mathcal{A} = \mathcal{A}^{\rm LO}  + \frac{\alpha_s(\mu)}{3\pi}\mathcal{A}^{\rm NLO} +\ldots
\label{AB-NLO}
\end{align}
and similar for $\mathcal{B}$; $\mu$ is the renormalization scale.
The leading-order expressions are available from Refs.~\cite{Braun:2001tj,Braun:2006hz}.
For consistency with our NLO calculation we rewrite these results in a somewhat different
form, expanding all kinematic factors in powers of $m_N^2/Q^2$: We keep all corrections
$\mathcal{O}(m_N^2/Q^2)$ but neglect terms $\mathcal{O}(m_N^4/Q^4)$ etc. which
is consistent with taking into account contributions of twist-three, -four, -five
(and, partially, twist-six) in the OPE. It proves to be convenient to write all
expressions in terms of the dimensionless variable~\cite{PassekKumericki:2008sj}
\begin{equation}
   W = 1+P'^2/Q^2 \qquad\text{where} \quad P' = P-q
\end{equation}
so that,  e.g.,
\begin{eqnarray}
(q-xP)^2 &=& Q^2[-1+x W -x\bar x m^2_N/Q^2]\,
\end{eqnarray}
where
\begin{align}
 \bar x = 1-x \,.
\end{align}
We also introduce a set of ``standard'' functions (cf.~App.~\ref{App:D})
\begin{equation}
   g_{k} (x;W) = \frac{1}{[-1+ x W]^k} = \left[\frac{Q^2}{x P'^2 -\bar x Q^2}\right]^k
\label{gk-functions}
\end{equation}
that absorb all momentum dependence. Using the expressions from~ \cite{Braun:2006hz}
one obtains after some algebra:
\begin{widetext}
\begin{eqnarray}
 Q^2 \mathcal{A}^{\mathrm{LO}}_d & = & 2\int [dx_i] \biggl\{
 2 \Big[g_{1}+g_{2}+ x_3\bar x_3 \frac{m_N^2}{Q^2}\Big(g_{2}+2 g_{3}\Big)
\Big](x_3;W) \,{\mathbb{V}}^{(3)}_{2}(x_i)
\nonumber\\&&{}
+x_3 \Big[g_{1} + x_3 \bar x_3 \frac{m_N^2}{Q^2} g_{2}\Big](x_3;W)\mathcal{V}_3(x_i)\biggr\}
+  2 \frac{m_N^2}{Q^2} \int_0^1 dx_3\, x_3^2 g_{2}(x_3;W)\widetilde{\mathcal{V}}_{5}(x_3)\,,
\nonumber\\
 Q^2 \mathcal{A}^{\mathrm{LO}}_u & = & 2 \int [dx_i] \biggl\{
x_2 \Big[g_{1} + x_2 \bar x_2 \frac{m_N^2}{Q^2} g_{2}\Big](x_2;W) \Big( -2 \mathcal{V}_1 + 3\mathcal{V}_3+\mathcal{A}_3 \Big)(x_i)
\nonumber\\&&{}
+ 2\Big[g_{2}+ 2 x_2\bar x_2 \frac{m_N^2}{Q^2} g_{3}\Big](x_2;W) \Big({\mathbb{V}}^{(2)}_{2}+ {\mathbb{A}}^{(2)}_{2}\Big)(x_i)
- 2\Big[g_{1}+ x_2\bar x_2 \frac{m_N^2}{Q^2} g_{2}\Big](x_2;W)  \Big({\mathbb{V}}^{(2)}_{2}- {\mathbb{A}}^{(2)}_{2}\Big)(x_i)
\biggr\}
\nonumber\\&&{}
- 2\frac{m_N^2}{Q^2} \int_0^1 dx_2\, x_2  g_{2}(x_2;W) \biggl[
x_2 \Big(\widehat{\mathcal{V}}_{4} - 2\widehat{\mathcal{V}}_{5}+ \widehat{\mathcal{A}}_{5}\Big)(x_2)
+2 \widehat{\widehat{\mathcal{V}}}_{6}(x_2)
+2 \mathcal{V}_1^{M(u)}(x_2)
\biggr],
\end{eqnarray}
and
\begin{eqnarray}
 Q^2\mathcal{B}^{\mathrm{LO}}_d &=&
-2\int [dx_i]\biggl\{
 \Big[g_{1}+ x_3\bar x_3 \frac{m_N^2}{Q^2} g_{2}\Big](x_3;W) \mathcal{V}_1(x_i)
-  2x_3 \frac{m_N^2}{Q^2} g_{2}(x_3;W){\mathbb{V}}^{(3)}_{2}(x_i)\biggr\}
\nonumber\\&&{}
-2 \frac{m_N^2}{Q^2}\int_0^1 dx_3\,g_{2}(x_3;W) \Big(
x_3 \widetilde{\mathcal{V}}_{5} + \mathcal{V}_1^{M(d)} \Big)(x_3)\,,
\nonumber\\
 Q^2 \mathcal{B}^{\mathrm{LO}}_u &=&
 2 \int [dx_i] \biggl\{
 \Big[g_{1}+ x_2\bar x_2 \frac{m_N^2}{Q^2} g_{2}\Big](x_2;W)\Big(\mathcal{V}_1+\mathcal{A}_1\Big)(x_i)
+ 2 x_2 \frac{m_N^2}{Q^2} g_{2}(x_2;W) \Big({\mathbb{V}}^{(2)}_{2}+{\mathbb{A}}^{(2)}_{2}\Big)(x_i)\biggr\}
\nonumber\\ &&{}
+  2 \frac{m_N^2}{Q^2} \int_0^1 dx_2\,  g_{2}(x_2;W)\biggl[
 x_2 \Big(\widehat{\mathcal{V}}_{4} - 2\widehat{\mathcal{V}}_{5}+\widehat{\mathcal{A}}_{5}\Big)(x_2)
     + \mathcal{V}_1^{M(u)}(x_2)+\mathcal{A}_1^{M(u)}(x_2)
\biggr].
\end{eqnarray}
\end{widetext}
The notations for various DAs are explained in Apps.~\ref{App:NDA}, \ref{App:OPE}.
%%%%%%%%%%%%%%%%%%%%%%%%%%%%%%%%%%%%%%%%%%%%%%%%%%%%%%%%%%%%%%%%%%%%%%%%%%%%%%%%%%%%%
%%%%%%%%%%%%%%%%%%%%%%%%%%%%%%%%%%%%%%%%%%%%%%%%%%%%%%%%%%%%%%%%%%%%%%%%%%%%%%%%%%%%%
\section{NLO LCSRs}\label{sec:NLO}
\begin{figure}[t]
\centerline{\includegraphics[width= 7.5cm, clip = true]{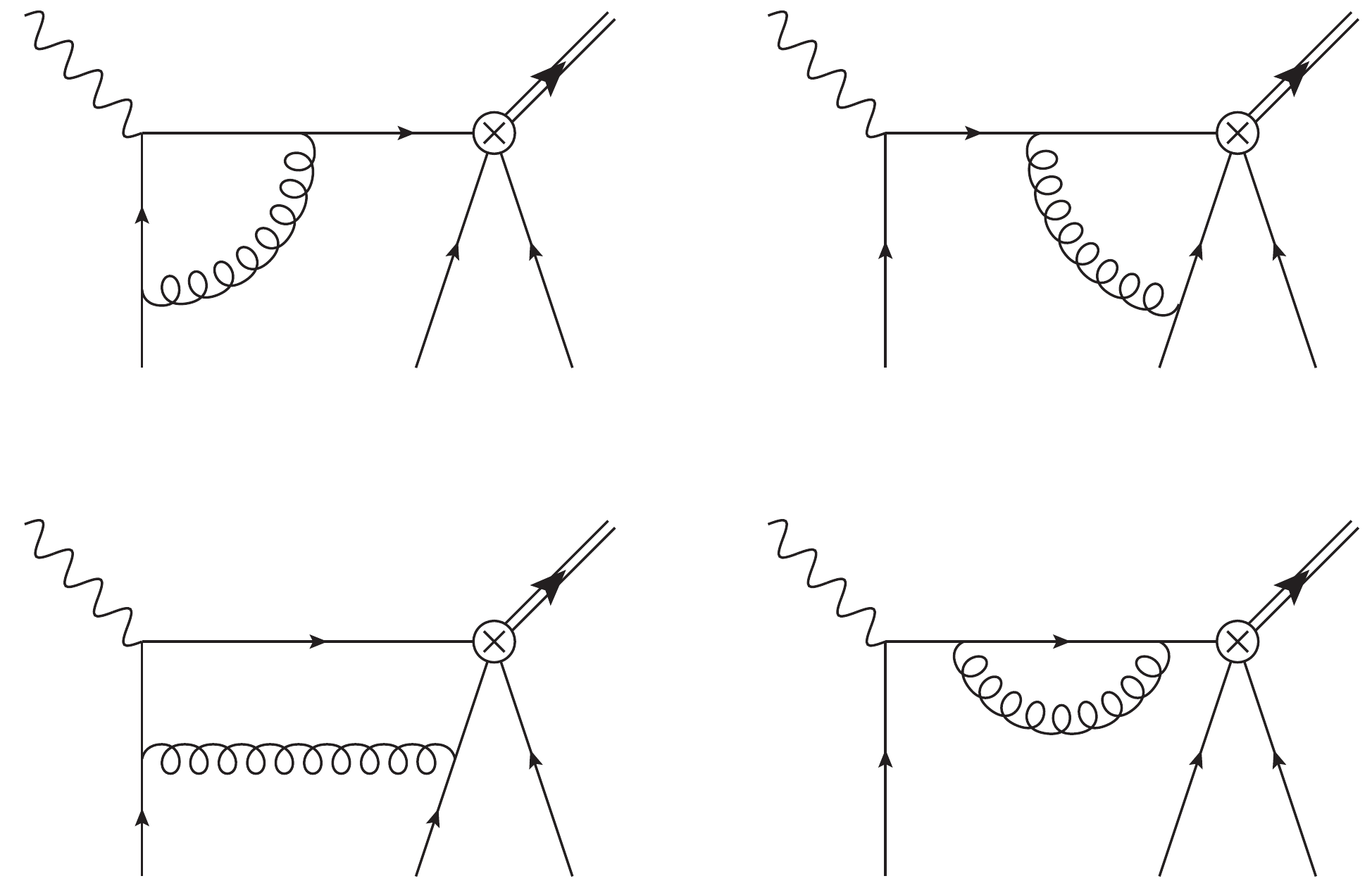}}
%\centerline{\epsfxsize5cm\epsffile{figLCSR}}
\caption{\small
NLO corrections to the light-cone sum rule for baryon form factors.
}
\label{fig:NLO-LCSR}
\end{figure}

%\setcounter{equation}{0}
%%%%%%%%%%%%%%%%%%%%%%%%%%%%%%%%%%%%%%%%%%%%%%%%%%%%%%%%%%%%%%%%%%%%%%%%%%%%%%%%%%%%%
%%%%%%%%%%%%%%%%%%%%%%%%%%%%%%%%%%%%%%%%%%%%%%%%%%%%%%%%%%%%%%%%%%%%%%%%%%%%%%%%%%%%%
The NLO corrections (\ref{AB-NLO}) to the correlation functions $\mathcal{A}(Q^2,P'^2)$ and  $\mathcal{B}(Q^2,P'^2)$
correspond to the Feynman diagrams shown in Fig.~\ref{fig:NLO-LCSR}.
They can be written as a sum of contributions of a given quark flavor $q = u,d$
weighted with the corresponding electromagnetic charges, and
further expanded in contributions of nucleon DAs to the twist-four accuracy as follows:
\begin{eqnarray}
\lefteqn{Q^2{\cal A}^{\rm {NLO}}_q =}
\nonumber\\&=\!&
\!\int[dx_i]\biggl\{ \sum_{k=1,3}\Big[\mathbb{V}_k(x_i) C^{\mathbb{V}_k}_q\!(x_i, W) + \mathbb{A}_k(x_i) C^{\mathbb{A}_k}_q\!(x_i, W)\Big]
\nonumber\\&&{}
+ \sum_{m=1,2,3}\Big[ \mathbb{V}^{(m)}_2(x_i) C^{\mathbb{V}^{(m)}_2}_q\!\!(x_i, W)
\nonumber\\&&{}
+\mathbb{A}^{(m)}_2(x_i) C^{\mathbb{A}^{(m)}_2}_q\!\!(x_i, W)\Big]
\biggr\} + \mathcal{O}(\text{twist-5})
\label{NLO-A}
\end{eqnarray}
and
\begin{eqnarray}
\lefteqn{Q^2{\cal B}^{\rm {NLO}}_q=}
\nonumber\\&=&
\int[dx_i] \biggl[
\mathbb{V}_1(x_i) D^{\mathbb{V}_1}_q(x_i, W) +
\mathbb{A}_1(x_i) D^{\mathbb{A}_1}_q(x_i, W)\biggr]
\nonumber\\&&{} + \mathcal{O}(\text{twist-5}).
\label{NLO-B}
\end{eqnarray}
It turns out that $C^{\mathbb{V}^{(1)}_2}_d\!\!(x_i, W) = C^{\mathbb{A}^{(1)}_2}_d\!\!(x_i, W) =0$. Explicit expressions for the remaining
22 nontrivial coefficient functions are collected in App.~\ref{App:NLO-functions}.
The leading-twist NLO corrections to the  $\mathcal{B}$-function,  $D^{\mathbb{V}_1}_q$ and $D^{\mathbb{A}_1}_q$, were previously calculated
in Ref.~\cite{PassekKumericki:2008sj} in a different, ``naive'' dimensional regularization scheme. The other functions
have been calculated for the first time. Note that the twist-4 NLO contributions are only present in  $\mathcal{A}(Q^2,P'^2)$;
the corresponding corrections to $\mathcal{B}(Q^2,P'^2)$ are effectively collinear-twist-five and are beyond the accuracy of this paper.
Each coefficient function has a generic form
\begin{eqnarray}
  C^{\mathbb{F}}_q &=& C_0(x_i,W) \ln\frac{Q^2}{\mu^2} + C_1(x_i,W)
\end{eqnarray}
where $\mu$ is the factorization scale. Here
\begin{eqnarray}
 C_0&=& c_{10}(x_i,W)\ln (1-xW) + c_{00}(x_i,W)\,,
\nonumber\\
C_1 &=&  c_{21}(x_i,W)\ln^2 (1-xW) + c_{11}(x_i,W)\ln (1-xW)
\nonumber\\&&{} + c_{01}(x_i,W)\,,
\label{eq:Sudakov}
\end{eqnarray}
where $x$ is one of the quark momentum fractions (or their combination), and
the functions $c_{nk}$ can further be expanded in powers of $1/W$ or $1/(1-x W)$ but do not contain
logarithms.
This structure is expected and similar to what has been found in  previous studies of the LCSRs for mesons, e.g.~\cite{Braun:1999uj}.
The factorization scale dependence cancels to leading order by the scale dependence of nucleon DAs and the Ioffe coupling constant.
This cancellation was verified for twist-three contributions in Ref.~\cite{PassekKumericki:2008sj}.
The Sudakov-type logarithms in Eq.~(\ref{eq:Sudakov}) after integration over the momentum fractions and the subtraction of the continuum
produce terms $\sim \ln Q^2/s_0$. Such contributions can, in principle, be resummed to all orders (cf. Ref.~\cite{Li:1992nu,Kivel:2010ns}) but
the effect of the resummation in the medium momentum transfer region $Q^2 \leq 10-20$~GeV$^2$ is likely to be marginal.
In the remaining part of this section we discuss two important technical aspects of our calculation.
%%%%%%%%%%%%%%%%%%%%%%%%%%%%%%%%%%%%%%%%%%%%%%%%%%%%%%%%%%%%%%%%%%%%%%%%%%%%%%%%%%%%%
%%%%%%%%%%%%%%%%%%%%%%%%%%%%%%%%%%%%%%%%%%%%%%%%%%%%%%%%%%%%%%%%%%%%%%%%%%%%%%%%%%%%%
\subsection{Renormalization scheme}\label{sec:RS}
%\setcounter{equation}{0}
%%%%%%%%%%%%%%%%%%%%%%%%%%%%%%%%%%%%%%%%%%%%%%%%%%%%%%%%%%%%%%%%%%%%%%%%%%%%%%%%%%%%%
%%%%%%%%%%%%%%%%%%%%%%%%%%%%%%%%%%%%%%%%%%%%%%%%%%%%%%%%%%%%%%%%%%%%%%%%%%%%%%%%%%%%%
It is well known that for generic composite operators the celebrated $\overline{MS}$ prescription
does not fix a renormalization scheme completely because of the existence of evanescent operators
in non-integer $d$ dimensions which do not have four-dimensional analogues. Such operators cannot
be neglected because they mix with physical operators under renormalization. A common
approach~\cite{Dugan:1990df} is to get rid of this mixing by a suitable finite renormalization.
The choice of evanescent operators and hence a precise renormalization condition is not
unique~\cite{Herrlich:1994kh} and has to be specified in detail.
The necessity of extra finite renormalization was overlooked in Ref.~\cite{PassekKumericki:2008sj}.
The scheme~\cite{Dugan:1990df} was suggested originally for treatment of the four-fermion operators that
appear in the effective weak Hamiltonian, but it can be used for three-quark operators as well.
We find, however, that an alternative scheme suggested by Krankl and Manashov
\cite{Kraenkl:2011qb} (KM--scheme in what follows) is more convenient for our purposes.
The KM--scheme is described in Appendix A. Its advantage is the guaranteed vanishing
of evanescent operators in $d=4$ dimensions so that one can work with physical (four-dimensional)
operators only. As a consequence, the renormalization procedure preserves Fierz identities between
renormalized operators. These attractive features come at the cost of a certain complication
of the algebraic structure of the anomalous dimensions, which do not pose a problem of principle,
however. The self-consistency of the KM--scheme has been checked to the three-loop accuracy in
Ref.~\cite{Gracey:2012gx}.
The basic idea is to consider operator renormalization with free spinor indices.
For a generic  three-quark operator
\begin{align}
\mathcal{Q}_{\alpha\beta\gamma}=\epsilon^{ijk} q^{i,a}_\alpha q^{j,b}_\beta q^{k,c}_\gamma\,,
\end{align}
the renormalized operator,  $[\mathcal{Q}]_{\alpha\beta\gamma}$, is defined as
\begin{eqnarray}
{}[\mathcal{Q}]_{\alpha\beta\gamma}&=&
\mathcal{Z}_{\alpha\beta\gamma}^{\alpha'\beta'\gamma'} Z_q^{-3}
\mathcal{Q}^{\rm bare}_{\alpha'\beta'\gamma'}\,,
\end{eqnarray}
where $Z_q$ is the quark field renormalization constant and
$\mathcal{Z}_{\alpha\beta\gamma}^{\alpha'\beta'\gamma'}$ corresponds to the subtraction
of the divergent part of the corresponding vertex function. It has the structure
\begin{align}
 \mathcal{Z}_{\alpha\beta\gamma}^{\alpha'\beta'\gamma'} = 1 +
  \sum_{lmn} g_{lmn}(\epsilon)(\Gamma_{lmn})^{\alpha'\beta'\gamma'}_{\alpha\beta\gamma}\,,
\end{align}
where $g_{lmn}(\epsilon)$ are given by a series in $1/\epsilon$,
\begin{align}
 g_{lmn}(\epsilon) = \sum_{p=1}^\infty \epsilon^{-p} a_{lmn}^{(p)}(\alpha_s)\,, \qquad d=4-2\epsilon\,,
\end{align}
and the gamma-matrix structures $(\Gamma_{lmn})^{\alpha'\beta'\gamma'}_{\alpha\beta\gamma}$ are
defined as
\begin{align}
 (\Gamma_{lmn})^{\alpha'\beta'\gamma'}_{\alpha\beta\gamma} = \gamma^{(l)}_{\alpha\alpha'} \otimes \gamma^{(m)}_{\beta\beta'}
\otimes \gamma^{(n)}_{\gamma\gamma'}\,,
\label{Gamma1}
\end{align}
where
\begin{align}
  \gamma^{(n)}_{\mu_1,\mu_2,\ldots,\mu_n} = \gamma_{[\mu_1}\gamma_{\mu_2}\ldots\gamma_{\mu_n]}
\end{align}
are the antisymmetrized (over Lorentz indices) products of gamma-matrices, cf.~\cite{Dugan:1990df}.
For example the renormalized Ioffe current (\ref{ioffe}) is defined as
\begin{eqnarray}
{}[\eta]_\gamma &=& \mathcal{P}^{(\eta)}_{\alpha\beta,\gamma\gamma'} [\epsilon^{ijk} u^{i}_{\alpha} u^{j}_{\beta} d^{k}_{\gamma'}]\,.
\end{eqnarray}
where
\begin{eqnarray}
  \mathcal{P}^{(\eta)}_{\alpha\beta,\gamma\gamma'} &=& (C\gamma^\mu)_{\alpha\beta} (\gamma_\mu)_{\gamma\gamma'}
\end{eqnarray}
is the projector that is applied to the \emph{renormalized} three-quark operator, i.e. in four dimensions.
Similarly, renormalized nucleon DAs are defined as matrix elements of the renormalized light-ray
operators
 \begin{eqnarray}
\label{defdisamp}
\lefteqn{
4 \langle 0| [\epsilon^{ijk} u_\alpha^i(a_1 n) u_\beta^j(a_2 n) d_\gamma^k(a_3 n)]
|P\rangle =}
\nonumber \\[1mm]
&&\hspace*{2.7cm} =~ V_1  \big(\slashed{p}C \big)_{\alpha \beta} \big(\gamma_5 N^+\big)_\gamma + \ldots
\end{eqnarray}
where the ellipses stand for the other existing Dirac structures, cf.~Eq.~(\ref{defdisamp}).
Here, again, the strings of $\gamma$--matrices on the r.h.s. are in four dimensions so that
the relations between different DAs that are a consequence of Fierz identities are fulfilled
identically (for renormalized DAs).
The coefficient functions of light-ray operators are calculated as finite parts of the amplitudes on free quark states
\begin{widetext}
\begin{eqnarray}
\label{green3}
(\mathcal{M}_{\nu})^{\alpha\beta\gamma}_{\alpha'\beta'\gamma'}(q,p_1,p_2,p_3)= i\! \int\! d^4 x \, e^{i q x}
\langle 0| T\left[\epsilon^{ijk} u^{i}_{\alpha}(0) u^{j}_{\beta}(0) d^{k}_{\gamma}(0) j_{\nu}^{\mathrm{em}}(x)\right] | u^{i'}_{\alpha'}(p_1) u^{j'}_{\beta'}(p_2) d^{k'}_{\gamma'}(p_3) \rangle
\end{eqnarray}
\end{widetext}
applying the same decomposition of gamma-matrix structures (\ref{Gamma1}) and using appropriate projection operators (in four dimensions)
to separate different contributions. In contrast, in Ref.~\cite{PassekKumericki:2008sj} the subtraction has been applied
to the correlation functions after multiplication with projection operators. This procedure is valid, but it has to be complemented
by additional finite renormalization in order to get rid of contributions of evanescent operators~\cite{Dugan:1990df}.
It is easy to convince oneself that these subtleties do not affect terms in $\ln Q^2/\mu^2$ and also the leading Sudakov double-logarithms.
To this accuracy our results agree with \cite{PassekKumericki:2008sj}; the Sudakov single-logarithms and constant terms are, however,
somewhat different.
%%%%%%%%%%%%%%%%%%%%%%%%%%%%%%%%%%%%%%%%%%%%%%%%%%%%%%%%%%%%%%%%%%%%%%%%%%%%%%%%%%%%%
%%%%%%%%%%%%%%%%%%%%%%%%%%%%%%%%%%%%%%%%%%%%%%%%%%%%%%%%%%%%%%%%%%%%%%%%%%%%%%%%%%%%%
\subsection{Twist-four contributions}\label{sec:T4}
%\setcounter{equation}{0}
%%%%%%%%%%%%%%%%%%%%%%%%%%%%%%%%%%%%%%%%%%%%%%%%%%%%%%%%%%%%%%%%%%%%%%%%%%%%%%%%%%%%%
%%%%%%%%%%%%%%%%%%%%%%%%%%%%%%%%%%%%%%%%%%%%%%%%%%%%%%%%%%%%%%%%%%%%%%%%%%%%%%%%%%%%%
Contributions of the leading-twist DA $\varphi_N(x_i) = V_1(x_i)-A_1(x_1)$ correspond to
contributions of local (geometric) twist-three operators in the OPE of the product
$T(\eta(0)j_\mu(x)$ for $x^2\to 0$:
\begin{align}
  \big(D^{k_1}_+ u_+)(0)  \big(D^{k_2}_+ u_+)(0)  \big(D^{k_3}_+ d_+)(0) \,.
\label{eq:local-t3}
\end{align}
Here $D_+\equiv n^\mu D_\mu$ and $q_+\equiv \Lambda_+ q$ are the ``plus'' components of the
covariant derivative and the quark field, respectively. The color structure is not shown
for brevity. Equivalently, the leading-twist contributions can be attributed to the
single light-ray operator
\begin{align}
  u_+(a_1 n) u_+(a_2 n) d_+(a_3n)\,,
\label{eq:lightray-t3}
\end{align}
where $a_i$ are (real) numbers and the gauge links are implied.
Expansion of the light-ray operator (\ref{eq:lightray-t3}) at short distances $a_i\to 0$
generates a formal Taylor series in local twist-three operators.
Either way, the corresponding coefficient functions can be calculated from the amplitude (\ref{green3})
with on-shell quarks with collinear momenta $p_i = x_i p$, $p^2=0$  or, in position space,
with the three quark fields on a light ray $y_i = a_i n$.
Going over to the next-to-leading twist the situation becomes more complicated.
A twist-four operator can be constructed in two different ways: either changing
the ``plus'' projection of one of the quark fields to the ``minus'', or adding
a transverse derivative, e.g.
\begin{align}
   &\big(D^{k_1}_+ u_-)(0)  \big(D^{k_2}_+ u_+)(0)  \big(D^{k_3}_+ d_+)(0)\,,
\notag\\
   &\big(D^{k_1}_+ D_\perp u_+)(0)  \big(D^{k_2}_+ u_+)(0)  \big(D^{k_3}_+ d_+)(0)
\end{align}
(and similar operators with the minus projection or transverse derivative on the $d$-quark).
Contributions of the first type correspond to the nonlocal light-ray operators
$u_-(a_1 n) u_+(a_2 n) d_+(a_3n)$ and $u_+(a_1 n) u_+(a_2 n) d_-(a_3n)$.
The corresponding coefficient functions can be calculated in the same way as the leading-twist-three contributions, considering the matrix elements over
free quarks with collinear momenta and taking a different spinor projection at the end.
The contributions of operators involving a transverse derivative are more complicated and can be
obtained from the light-cone expansion of the nonlocal three-quark operator
\begin{align}
     u_+(y_1) u_+(y_2) d_+(y_3)\,, \qquad y_i = a_i n + b_{i,\perp}
\end{align}
where $b_\perp \to 0$ is an auxiliary transverse vector. The twist-four contribution
(one transverse derivative) corresponds to picking up terms of first order, $\mathcal{O}(b_\perp)$, in the
light-cone expansion. Note that $y_i^2 = b^2_{i,\perp}$ can be neglected to this accuracy, so that the quarks can still be
considered as being on the light-cone (but not on the same light-ray).
This means that the twist-four coefficient functions (of the second type) can be
calculated by considering the matrix elements with quark momenta $p_i = x_i p + p_{i,\perp}$ and expanding
to the first order in $p_{i\perp}$ along the collinear direction $p_{i,\perp}\to 0$. In this calculation the
quark virtualities can be neglected $p_i^2 = - p^2_{i,\perp } \to 0$.
As an example, consider the contribution of the twist-four DA $\mathbb{V}_2^{(2)}(x_i)$ defined in Eq.~(\ref{mathbbVA}),
which we can rewrite as
\begin{eqnarray}
\lefteqn{4 \langle 0|[ \epsilon^{ijk} u_\alpha^i(y_1) u_\beta^j(y_2) d_\gamma^k(y_3)] |P\rangle = }
\nonumber\\&=&
\mathcal{P}^{\mathbb{V}_2^{(2)}}_{\rho;\alpha\beta\gamma}  y_2^\rho \int [dx_i]\, \mathbb{V}^{(2)}_2(x_i)\, e^{-iP\sum x_i y_i} + \ldots
\nonumber\\&=&
i \mathcal{P}^{\mathbb{V}_2^{(2)}}_{\rho;\alpha\beta\gamma} \int [dx_i]\,\mathbb{V}^{(2)}_2(x_i) \frac{\partial}{\partial p_2^\rho} e^{-i\sum p_i y_i}\Big|_{p_k = x_k p} + \ldots
\label{eq:mathbbV2}
\end{eqnarray}
where
\begin{align}
 \mathcal{P}^{\mathbb{V}_2^{(2)}}_{\nu;\alpha\beta\gamma} = (\slashed{P}C)_{\alpha\beta}(\gamma_\nu \gamma_5 N(P))_\gamma\,.
\end{align}
Note that the exponential factor $e^{-iP\sum x_i y_i}$ in the second line in Eq.~(\ref{eq:mathbbV2}) can be written as
$e^{-i(Pn)\sum x_i a_i} = e^{-i(pn)\sum x_i a_i}$ so that the quark momenta $p_i \equiv x_i p$
are collinear and the dependence on the transverse separation is contained entirely in the prefactor $y_2^\rho = a_2 n^\rho + b_{2,\perp}^\rho$.
In the last line in  Eq.~(\ref{eq:mathbbV2}) the quark momenta can be set to the same collinear values only \emph{after} taking the derivative.
The corresponding contribution to the correlation function (\ref{correlator}) can be written as
\begin{eqnarray}
\Lambda_+ n^\nu T^{\mathbb{V}^{(2)}_2}_\nu(P,q) &=&  \frac{i}{4} \int [dx_i]\,\mathbb{V}^{(2)}_2(x_i) (\Lambda_+)_\delta
\mathcal{P}_{\alpha\beta;\delta\gamma}^{(\eta)}
\nonumber\\&&{}\hspace*{-1.4cm}
\times \mathcal{P}^{\mathbb{V}_2^{(2)}}_{\rho;\alpha\beta\gamma} \frac{\partial}{\partial p_2^\rho} [n^\nu\mathcal{M}_\nu]^{\alpha\beta\gamma}_{\alpha'\beta'\gamma'}(q,p_i)
\Big|_{p_k = x_k p}
\end{eqnarray}
where $[\mathcal{M}_\nu]$ is the renormalized amplitude calculated on free quarks (\ref{green3}).
It is given by the sum of Feynman diagrams shown in Fig.~\ref{fig:NLO-LCSR} (to the NLO accuracy).
The derivative over the second quark momentum can be written as a sum of contributions corresponding to the longitudinal and transverse components
\begin{align}
  \mathcal{P}^{\mathbb{V}_2^{(2)}}_{\rho;\alpha\beta\gamma}\frac{\partial}{\partial p_2^\rho} =
 \frac{n^\rho}{pn} \mathcal{P}^{\mathbb{V}_2^{(2)}}_{\rho;\alpha\beta\gamma} \frac{d}{dx_2}
+\mathcal{P}^{\mathbb{V}_2^{(2)}}_{\perp;\alpha\beta\gamma} \frac{\partial}{\partial p_2^\perp}
+ \mathcal{O}(\text{twist-5}).
\end{align}
The first contribution involves the amplitude calculated on collinear quarks; the derivative $d/dx_2$ can be dispensed off using integration
by parts. The derivative over the quark transverse momentum in the second contribution is applied to each propagator on the second quark line.
Thanks to the Ward identity
\begin{align}
  \frac{\partial}{\partial p^\perp} \frac{\slashed{p} + \slashed{\ell}}{(p+\ell)^2\!+i\epsilon}
 = - \frac{\slashed{p} + \slashed{\ell}}{(p+\ell)^2\!+i\epsilon} \gamma^\perp \frac{\slashed{p} + \slashed{\ell}}{(p+\ell)^2+\!i\epsilon}
\end{align}
a derivative is equivalent to the insertion of $\gamma^\perp$-matrix in the quark line. Thus one ends up with the sum of
Feynman diagrams with collinear quarks and extra $\gamma^\perp$-insertions along the quark line.
The calculation in this work was done using computer algebra. To this end two codes have been written using FORM and FeynCalc, respectively, and
produced identical results. The results are summarized in App.~\ref{App:NLO-functions} and are also available as a MATHEMATICA package that can be
requested from the authors.
    %%%%%%%%%%
    %%%%%%%%%% Begin Table 1
    %%%%%%%%%%
\begin{table*}[t]
\renewcommand{\arraystretch}{1.2}
\begin{center}
\begin{tabular}{@{}l|l|l|l|l|l|l|l|l|l|l@{}} \hline
Model &  Method    & $f_N/\lambda_1 $ & $\varphi_{10}$ & $\varphi_{11}$ & $\varphi_{20}$ & $\varphi_{21}$ & $\varphi_{22}$ & $\eta_{10}$   & $\eta_{11}$    & Reference \\ \hline
ABO1  & LCSR (NLO) & $-0.17$         & $0.05$        & $0.05$        & $0.075(15)$   & $-0.027(38)$ & $0.17(15)$    & $-0.039(5)$ & $0.140(16)$   & this work \\ \hline
ABO2  & LCSR (NLO) & $-0.17$         & $0.05$        & $0.05$        & $0.038(15)$   & $-0.018(37)$ & $-0.13(13)$   & $-0.027(5)$ & $0.092(15)$   & this work \\ \hline
BLW   & LCSR (LO)  & $-0.17$         & $0.0534$      & $0.0664$      & -             & -            & -             & $0.05$      & $0.0325$      & \cite{Braun:2006hz}    \\ \hline
BK    & pQCD       & -               & $0.0357$      & $0.0357$      & -             & -            & -             & -           & -             & \cite{Bolz:1996sw}     \\ \hline
COZ   & QCDSR (LO) & -               & $0.163$       & $0.194$       & $0.41$        & $0.06$       & $-0.163$      & -           & -             & \cite{Chernyak:1987nv} \\ \hline
KS    & QCDSR (LO) & -               & $0.144$       & $0.169$       & $0.56$        & $-0.01$      & $-0.163$      & -           & -             & \cite{King:1986wi}     \\ \hline
      & QCDSR (NLO)& $-0.15$         & -             & -             & -             & -            & -             & -           & -             & \cite{Gruber:2010bj}   \\ \hline
LAT09 & LATTICE    & $-0.083(6)$     & $0.043(15)$   & $0.041(14)$   & $ 0.038(100)$ & $-0.14(15)$  & $-0.47(33)$   & -           & -             & \cite{Braun:2008ur}    \\ \hline
LAT13 & LATTICE    & $-0.075(5)$     & $0.038(3)$    & $0.039(6)$    & $-0.050(80)$  & $-0.19(12)$  & $-0.19(14)$   & -           & -             & \cite{lattice2013}     \\ \hline
\end{tabular}
\end{center}
\caption[]{\sf Parameters of the nucleon distribution amplitudes at the scale $\mu^2=2$~GeV$^2$.
 For the lattice results \cite{lattice2013} only statistical errors are shown.
 }
\label{tab:shape}
\renewcommand{\arraystretch}{1.0}
\end{table*}
    %%%%%%%%%%
    %%%%%%%%%% End Table 1
    %%%%%%%%%%
%%%%%%%%%%%%%%%%%%%%%%%%%%%%%%%%%%%%%%%%%%%%%%%%%%%%%%%%%%%%%%%%%%%%%%%%%%%%%%%%%%%%%
%%%%%%%%%%%%%%%%%%%%%%%%%%%%%%%%%%%%%%%%%%%%%%%%%%%%%%%%%%%%%%%%%%%%%%%%%%%%%%%%%%%%%
\section{Results}\label{sec:Results}
%\setcounter{equation}{0}
%%%%%%%%%%%%%%%%%%%%%%%%%%%%%%%%%%%%%%%%%%%%%%%%%%%%%%%%%%%%%%%%%%%%%%%%%%%%%%%%%%%%%
%%%%%%%%%%%%%%%%%%%%%%%%%%%%%%%%%%%%%%%%%%%%%%%%%%%%%%%%%%%%%%%%%%%%%%%%%%%%%%%%%%%%%
%%%%%%%%%%%%%%%%%%%%%%%%%%%%%%%%%%%%%%%%%%%%%%%%%%%%%%%%%%%%%%%%%%%%%%%%%%%%%%%%%%%%%
%%%%%%%%%%%%%%%%%%%%%%%%%%%%%%%%%%%%%%%%%%%%%%%%%%%%%%%%%%%%%%%%%%%%%%%%%%%%%%%%%%%%%
\subsection{Discussion of parameters}\label{sec:ResultsPar}
%\setcounter{equation}{0}
%%%%%%%%%%%%%%%%%%%%%%%%%%%%%%%%%%%%%%%%%%%%%%%%%%%%%%%%%%%%%%%%%%%%%%%%%%%%%%%%%%%%%
%%%%%%%%%%%%%%%%%%%%%%%%%%%%%%%%%%%%%%%%%%%%%%%%%%%%%%%%%%%%%%%%%%%%%%%%%%%%%%%%%%%%%
Main nonperturbative input in the LCSR calculation of form factors is provided by
normalization constants and shape parameters of nucleon DAs. The existing information,
together with our final choices explained below, is summarized in Table~\ref{tab:shape}.
The nucleon coupling to the (Ioffe) interpolation current (\ref{ioffe}), $\lambda_1$,  simultaneously
determines the normalization of twist-four DAs and cancels out between the l.h.s. and the r.h.s.
so that the sum rule effectively only involves the ratio of twist-three and twist-four couplings,
$f_N/\lambda_1$, which is given in the Table. All entries in Table~\ref{tab:shape} except for the
Bolz-Kroll model~\cite{Bolz:1996sw} are rescaled to $\mu^2=2$~GeV$^2$ using one-loop anomalous
dimensions collected in Sec.~\ref{ssec:DAs}.
The other parameters that enter LCSRs are the interval of duality (continuum threshold) $s_0$, Borel parameter $M^2$ and
factorization scale $\mu^2$. In this work we use the standard value $s_0=2.25$~GeV$^2$ that is accepted
in most studies. Variations of $s_0$ with respect to this value can be studied, but have to be accompanied
by the corresponding variations of the effective nucleon coupling to the Ioffe current.
This is usually done using the ratio method, in which $\lambda_1$ on the l.h.s of the LCSR
is substituted by the corresponding QCD sum rule with the same interval of duality.
The experience of such calculations is that the sensitivity of the sum rules to the precise value of $s_0$
is greatly reduced and is not significant as compared to other sources of uncertainty.
The Borel parameter $M^2$ corresponds, loosely speaking, to the inverse imaginary time (squared)
at which matching of the QCD calculation is done to the expansion in hadronic states.
One usually tries to take $M^2$ as small as possible in order to reduce sensitivity to the contributions
of higher-mass states, which is the main irreducible uncertainty of the sum rule method.
In two-point sum rules that are used to determine the nucleon mass and the coupling~\cite{Ioffe:1981kw}
the default values are in the range $M^2=1.0-1.5$~GeV$^2$. The light-cone sum rules are somewhat different
in that the expansion parameter in the QCD calculation is $1/(\langle x \rangle M^2)$ rather than $1/M^2$
in two-point sum rules, where $\langle x \rangle$ is a typical quark momentum fraction~\cite{Ali:1993vd}.
Thus one has to go over to somewhat higher $M^2$ values in order to ensure the same suppression of
(uncalculated) contributions of very high twist. In this work we take $M^2=1.5$~GeV$^2$ and  $M^2=2$~GeV$^2$
as two acceptable choices.
Finally, natural values of the factorization scale $\mu^2$ are determined by the virtuality of the quark
interacting with the hard probe
\begin{align}
 \mu^2 \sim (1-x) Q^2 - x P'^2.
\end{align}
In the sum rules $-P'^2 \to M^2 $ and the integration over the quark momentum fraction
is restricted to the end-point region  $x > x_0 = Q^2/(s_0+Q^2)$. Thus
\begin{align}
 \mu^2 \leq (1-x_0) Q^2 + x_0 M^2 \leq \frac{2s_0 Q^2}{s_0+Q^2} < 2 s_0\,,
\end{align}
where we assumed (for simplicity)  that $M^2 \simeq s_0 < Q^2 $.
Thus for $Q^2 \sim 1-10 $~GeV$^2$ the natural scale is $\mu^2 \sim 1-3$~GeV$^2$ and is not rising with $Q^2$
(or rising very slowly). In our calculations we take $\mu^2 = 2$~GeV$^2$ as the default value.
The renormalization scale is taken to be equal to the factorization scale. We use a two-loop expression
for the QCD coupling with $\Lambda^{(4)}_{QCD}=326$~MeV resulting in the value $\alpha_s(2~\text{GeV}^2) = 0.374$.
%%%%%%%%%%%%%%%%%%%%%%%%%%%%%%%%%%%%%%%%%%%%%%%%%%%%%%%%%%%%%%%%%%%%%%%%%%%%%%%%%%%%%
%%%%%%%%%%%%%%%%%%%%%%%%%%%%%%%%%%%%%%%%%%%%%%%%%%%%%%%%%%%%%%%%%%%%%%%%%%%%%%%%%%%%%
\subsection{Results for the form factors}\label{sec:ResultsFF}
%\setcounter{equation}{0}
%%%%%%%%%%%%%%%%%%%%%%%%%%%%%%%%%%%%%%%%%%%%%%%%%%%%%%%%%%%%%%%%%%%%%%%%%%%%%%%%%%%%%
%%%%%%%%%%%%%%%%%%%%%%%%%%%%%%%%%%%%%%%%%%%%%%%%%%%%%%%%%%%%%%%%%%%%%%%%%%%%%%%%%%%%%
%
\begin{figure*}[t]
   \begin{center}
%\begin{picture}(210,140)(0,0)
\includegraphics[width= 7.5cm, clip = true]{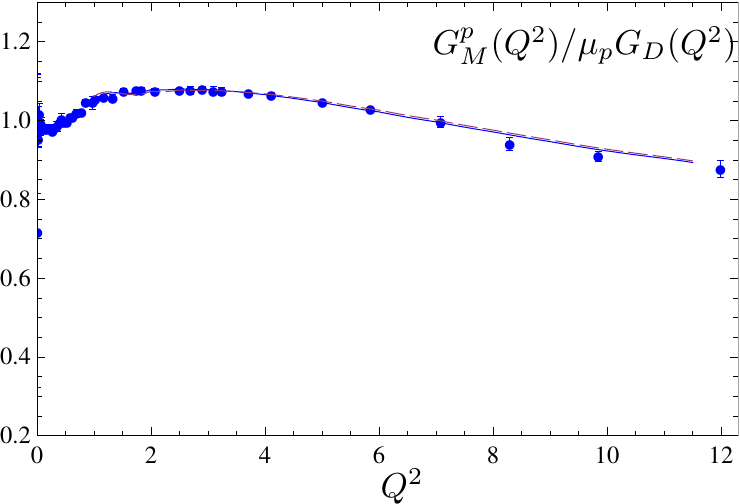}
%\put(-5,0){\epsfxsize7.5cm\epsffile{GMpGDp-borel1p5-scale2-pfit.eps}}
%\put(105,-8){$Q^2$}
%\put(120,120){$G_M^p(Q^2)/\mu_pG_D(Q^2)$}
%\end{picture}
\qquad
%\begin{picture}(210,140)(0,0)
%\put(-5,0){\epsfxsize7.5cm\epsffile{GMnGDn-borel1p5-scale2-pfit.eps}}
%\put(105,-8){$Q^2$}
%\put(120,120){$G_M^n(Q^2)/\mu_nG_D(Q^2)$}
%\end{picture}
\includegraphics[width= 7.5cm, clip = true]{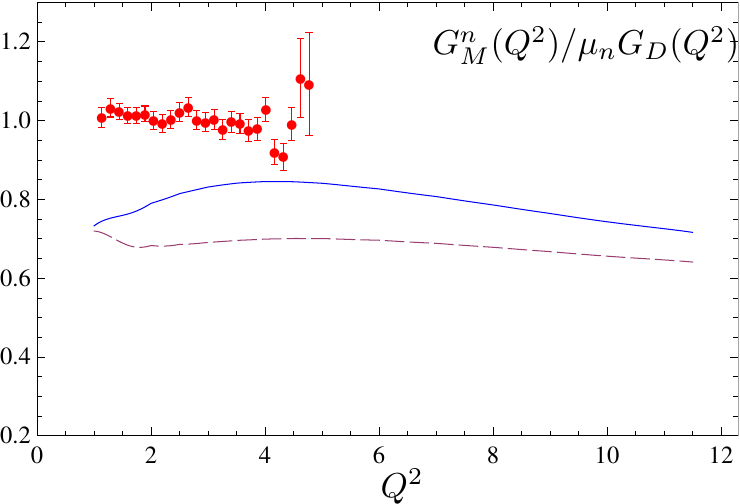}
\\[5mm]
%\begin{picture}(210,140)(0,0)
%\put(-5,0){\epsfxsize7.5cm\epsffile{GEpGMp-borel1p5-scale2-pfit.eps}}
%\put(105,-8){$Q^2$}
%\put(120,120){$\mu_pG_E^p(Q^2)/G_M^p(Q^2)$}
%\end{picture}
\includegraphics[width= 7.5cm, clip = true]{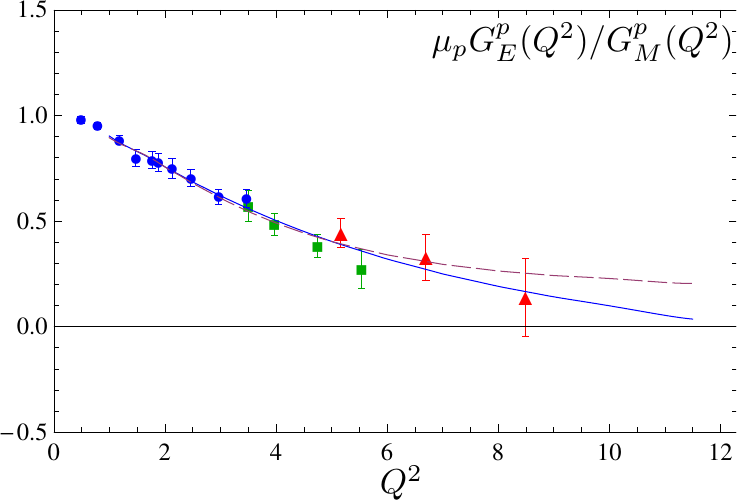}
\hspace*{0.5cm}
%\begin{picture}(210,140)(0,0)
%\put(-5,0){\epsfxsize7.5cm\epsffile{GEnGMn-borel1p5-scale2-pfit.eps}}
%\put(105,-8){$Q^2$}
%\put(120,120){$\mu_n G_E^n(Q^2)/G_M^n(Q^2)$}
%\end{picture}
\includegraphics[width= 7.5cm, clip = true]{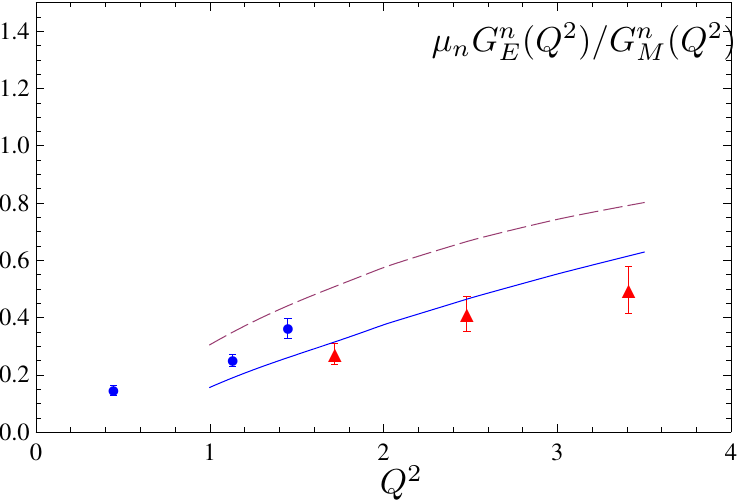}
   \end{center}
\caption{Nucleon electromagnetic form factors from LCSRs compared to the experimental data
\cite{Arrington:2007ux,Lachniet:2008qf,Gayou:2001qd,Punjabi:2005wq,Puckett:2010ac,Plaster:2005cx,Riordan:2010id}.
Parameters of the nucleon DAs correspond to the sets ABO1 and ABO2 in Table~\ref{tab:shape} for the solid and dashed
curves, respectively. Borel parameter $M^2=1.5$~GeV$^2$ for ABO1 and $M^2=2$~GeV$^2$ for ABO2.
}
\label{fig:ABO1}
\end{figure*}
\begin{figure}[ht]
   \begin{center}
\includegraphics[width= 7.5cm, clip = true]{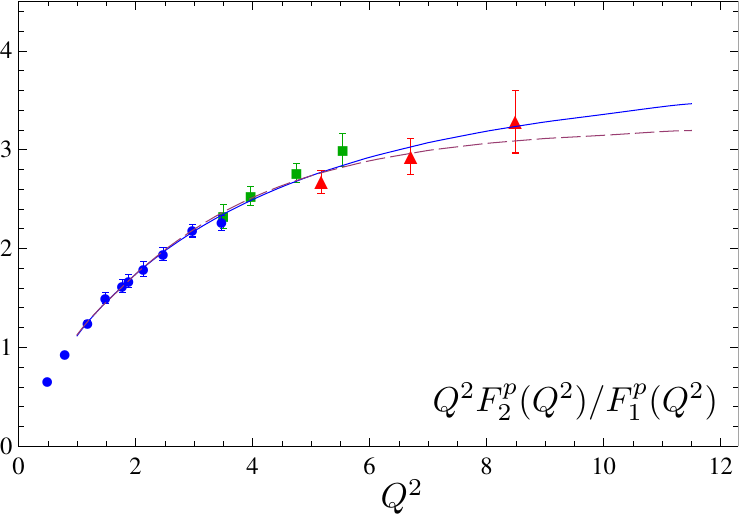}
   \end{center}
\caption{The ratio of Pauli and Dirac electromagnetic proton form factors from LCSRs compared to the experimental data
\cite{Gayou:2001qd,Punjabi:2005wq,Puckett:2010ac}.
Parameters of the nucleon DAs correspond to the sets ABO1 and ABO2 in Table~\ref{tab:shape} for the solid and dashed
curves, respectively. Borel parameter $M^2=1.5$~GeV$^2$ for ABO1 and $M^2=2$~GeV$^2$ for ABO2.}
\label{fig:ABO1-F2F1}
\end{figure}
\begin{figure*}[ht]
   \begin{center}
\includegraphics[width= 7.5cm, clip = true]{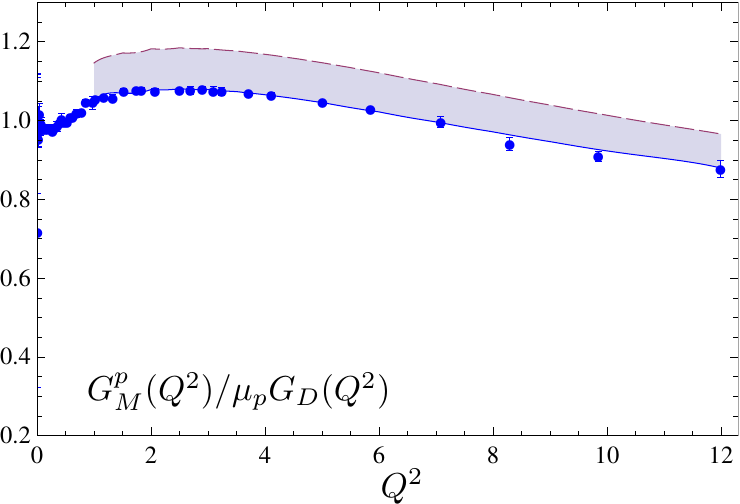}
\qquad
\includegraphics[width= 7.5cm, clip = true]{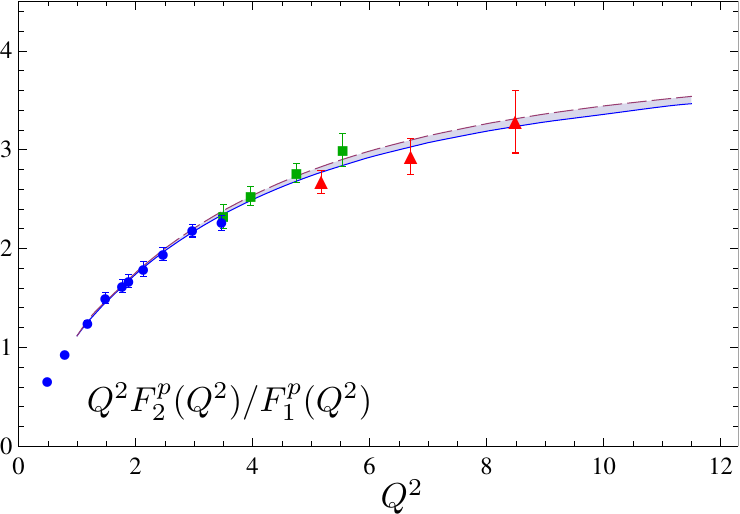}
   \end{center}
\caption{Borel parameter dependence of the magnetic proton form factor (left panel) and the $F_2^p(Q^2)/F_1^p(Q^2)$-ratio
for the given parameter set, ABO1, of nucleon DAs. The shaded areas correspond to variation of the form factors
in the range between $M^2=1.5$~GeV$^2$ (default value for the fit) and $M^2=2$~GeV$^2$.
}
\label{fig:ABO1-Borel}
\end{figure*}

As it is seen from Table~\ref{tab:shape}, at present there exist quantitative estimates
for the ratio of the couplings $f_N/\lambda_1$ and the first-order shape parameters
$\varphi_{10}$, $\varphi_{11}$ of the leading twist DA. The other parameters, in contrast,
are very weakly constrained. From the comparison with the experimental data it turns
out that larger values of $f_N/\lambda_1$ are preferred so that we fix $f_N/\lambda_1=-0.17$
and also take $\varphi_{10}=\varphi_{11} = 0.05$ in agreement with lattice calculations and
the previous LO LCSR studies~\cite{Braun:2006hz}. We then make a fit to the experimental
data on the magnetic proton form factor $G_M^p(Q^2)$ and the electric-to-magnetic form factor ratio
$G_E^p/G_M^p$ in the interval $1 < Q^2 < 8.5$~GeV$^2$ with all other entries as free parameters.
Since the data on the magnetic form factor are much more accurate than for the ratio $G_E^p/G_M^p$
we have increased the corresponding error bars by 50\% in order to give an equal weighting
to both data sets in our fit. We do not include the uncertainty in the Borel parameter
in the error estimates, but do separate fits for $M^2=1.5$~GeV$^2$ and  $M^2=2$~GeV$^2$
that are referred in what follows as ABO1 and ABO2, respectively. The resulting values of shape parameters
are collected in Table~\ref{tab:shape} and the corresponding form factors (solid curves for the set ABO1 and dashed for ABO2) are
shown in Fig.~\ref{fig:ABO1} for the proton (left two panels) and the neutron (right two panels).
For the magnetic form factors we plot the ratios to the dipole formula
\begin{align}
 G_D(Q^2) = 1/(1+ a Q^2)^2, \qquad a = 1/0.71~\text{GeV}^2
\end{align}
and use in all plots the proton and neutron magnetic moments for normalization,
$\mu_p= 2.793$, $\mu_n = -1.913$.
The ratio $Q^2 F^p_2(Q^2)/F^p_1(Q^2)$ of Pauli and Dirac form factors in the proton is
shown in Fig.~\ref{fig:ABO1-F2F1}.
The quality of the two fits of the proton data is roughly similar, whereas the description of neutron form factors
(that are not fitted) is slightly worse for ABO2 compared to ABO1. In both fits the neutron magnetic form factor
comes out to be 20-30\% below the data. This feature is rather robust. In contrast,
the description of the neutron electric-to-magnetic form factor ratio $G_E^n/G_M^n$
can easily be improved by choosing somewhat larger values of the first-order shape parameters
$\varphi_{10},\varphi_{11} \sim 0.06-0.07$, cf. Table~\ref{tab:shape}.
\begin{figure*}[ht]
   \begin{center}
\includegraphics[width= 7.5cm, clip = true]{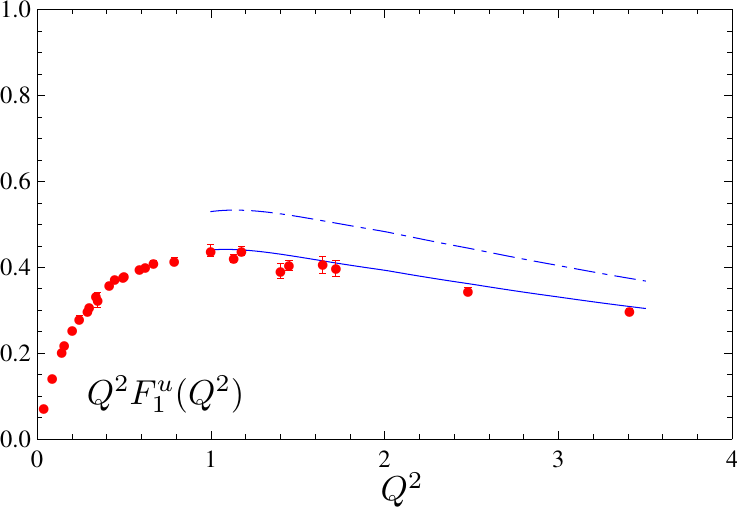}
\qquad
\includegraphics[width= 7.5cm, clip = true]{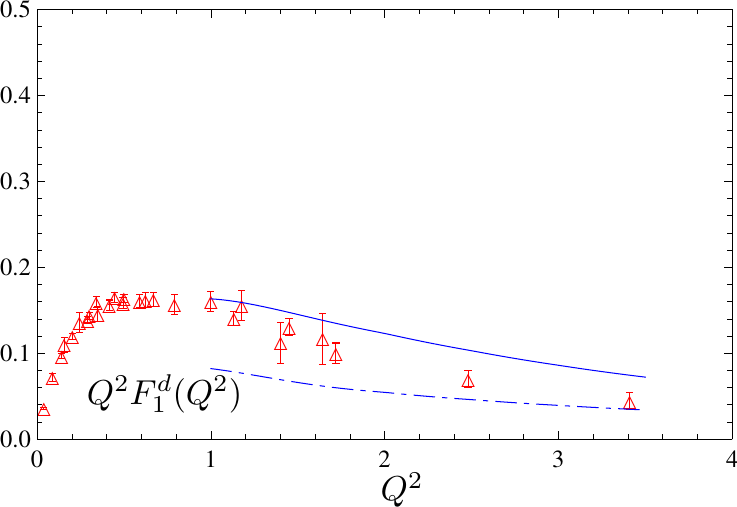}
\\[5mm]
\includegraphics[width= 7.5cm, clip = true]{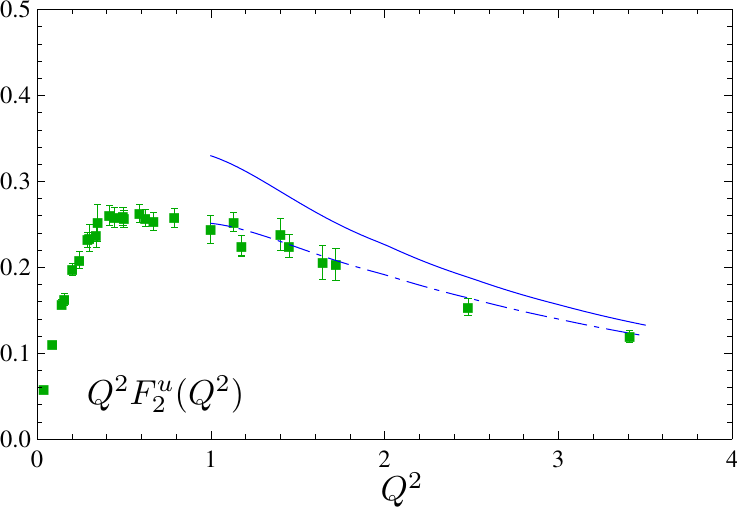}
\hspace*{0.5cm}
\includegraphics[width= 7.5cm, clip = true]{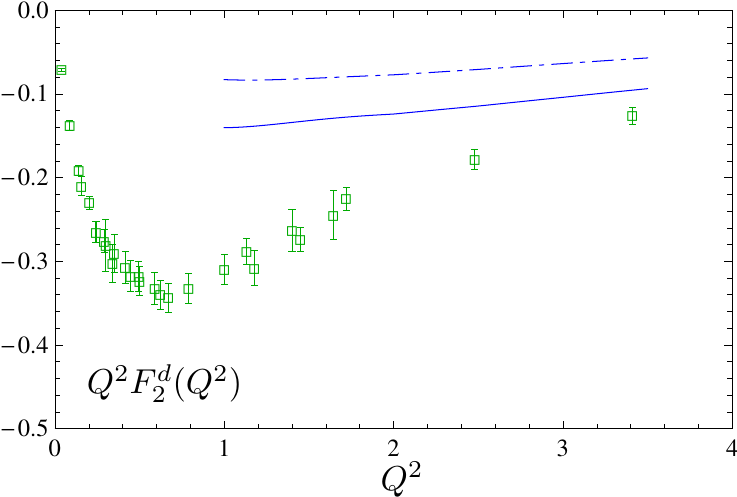}
   \end{center}
\caption{Contributions of different quark flavors to the proton electromagnetic form factors
 compared to the compilation of experimental data in Ref.~\cite{Diehl:2013xca}.
 The corresponding leading-order results are shown by the dash-dotted curves for comparison.
Parameters of the nucleon DAs correspond to the set ABO1 in Table~\ref{tab:shape}.
}
\label{fig:ABO1-DiehlKroll}
\end{figure*}
The underlying reason for this difficulty becomes more clear from the  results on the
contributions of different quark flavors to the proton form factors $F^p_1$ and $F^p_2$.
The LCSR calculation (ABO1) is compared to the compilation of the experimental data by Diehl and Kroll~\cite{Diehl:2013xca}
in Fig.~\ref{fig:ABO1-DiehlKroll}. One sees that the $u$- and $d$-quark contributions to $F_1(Q^2)$ are described rather well,
whereas there are considerable deviations in the Pauli form factor in the smaller $Q^2$ region. This feature is not unexpected and
is due to the structure of the twist expansion in these two cases. Recall that Dirac and Pauli form factors are extracted from the
correlation functions $\mathcal{A}(Q^2,P'^2)$ and $\mathcal{B}(Q^2,P'^2)$, respectively, cf. Eq.~(\ref{eq:LCSRscheme}).
%Although formally $\mathcal{A}$ and $\mathcal{B}$ are of the same collinear twist,
The light-cone expansion of
the latter is much more involved so that we are able to calculate less terms. Hence the sum rules are less accurate.
E.g. the calculation of the radiative corrections to the contributions of the next-to-leading twist nucleon DAs for the
$\mathcal{B}$-function requires taking into account second order corrections in the expansion over quark transverse momenta which is beyond
the scope of this paper. One should expect that the
$\mathcal{B}$-function at smaller values of $Q^2$ also receives large contributions of very high twist e.g. due to
factorizable five-quark DAs (e.g. quark condensate times a leading-twist DA). This question requires a separate study.
On the same plot the results of the corresponding leading-order calculations are shown by dash-dotted curves for comparison.
One sees that the NLO corrections are of the order of 20\% for $u$-quark contributions and much larger for $d$-quarks.
Hence the $d$-quark contributions are more affected by QCD corrections and generically less precise.
This pattern is probably due to the specific spin-flavor structure of the Ioffe current that is used in our calculations.
By virtue of isospin symmetry $d$-quark contributions to the proton form factors are equal to the $u$-quark contributions
for the neutron but are weighted in the latter case with a larger electric charge $e_d\to e_u$. This reweighting
is the simple reason behind a worse description of neutron form factors as compared to the proton ones.
We remind that the two sets of shape parameters of DAs in Table~\ref{tab:shape} are obtained from the fits of the proton
form factor using different values of the Borel parameter,  $M^2=1.5$~GeV$^2$ for ABO1 and $M^2=2$~GeV$^2$ for ABO2.
The difference in the fitted values in ABO1 and ABO2 sets is, therefore, a measure of the Borel parameter dependence
that is an intrinsic uncertainty of the sum rule method. Another, more direct possibility to quantify the Borel parameter
dependence is to compare the LCSR predictions for $M^2=1.5$~GeV$^2$ and $M^2=2$~GeV$^2$ for a given DA parameter set.
It turns out that increasing  the Borel parameter from $1.5$ to $2$~GeV$^2$ leads to an increase of all
form factors by the same amount $\sim 10\%$ (for all $Q^2$) so that the form factor ratios are affected only weakly.
This is illustrated in Fig.~\ref{fig:ABO1-Borel} where the proton magnetic form factor and the ratio $F_2^p/F_1^p$
%$F_2^p(Q^2)/F_1^p(Q^2)$
are shown on the left and the right panel, respectively. The effect on the neutron form factors is very similar.
The last remark concerns factorization scale dependence. Our calculations are done for the default value
$\mu^2 = 2$~GeV$^2$.  Varying $\mu^2$ in the interval $1-4$~GeV$^2$ and taking into account one-loop anomalous dimensions
the form factors $F_1^u, F_1^d, F_2^u, F_2^d$ change by $\pm 1\%, 2\%, 8\%, 8\%$ at $Q^2=1$~GeV$^2$ and
$\pm 10\%, 1\%, 14\%, 14\%$ at $Q^2=10$~GeV$^2$, respectively. Note that the uncertainty gets larger with increasing
$Q^2$, which is consistent with the expected dominant role of hard scattering corrections at asymptotically large
momentum transfers. Such corrections enter the LCSRs for the nucleon form factors starting at the next-to-next-to-leading
order (NNLO), cf.~\cite{Braun:1999uj}.
%%%%%%%%%%%%%%%%%%%%%%%%%%%%%%%%%%%%%%%%%%%%%%%%%%%%%%%%%%%%%%%%%%%%%%%%%%%%%%%%%%%%%
%%%%%%%%%%%%%%%%%%%%%%%%%%%%%%%%%%%%%%%%%%%%%%%%%%%%%%%%%%%%%%%%%%%%%%%%%%%%%%%%%%%%%
\subsection{Results for the  nucleon DAs}\label{sec:resultsDA}
\begin{figure*}[ht]
\centerline{\includegraphics[width=15cm, clip = true]{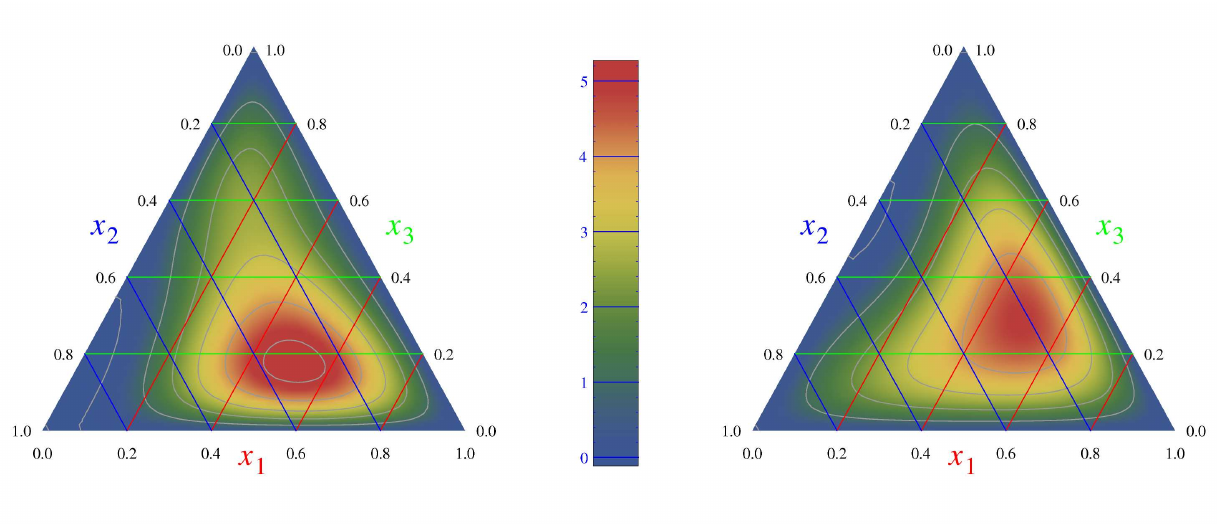}}
\caption{Leading twist distribution amplitude of the proton $\varphi(x_i)$
for the parameter sets ABO1 (left) and ABO2 (right) in Table~\ref{tab:shape}.
Central values are used for the second order parameters.
}
\label{fig:nucleonDA}
\end{figure*}

%\setcounter{equation}{0}
%%%%%%%%%%%%%%%%%%%%%%%%%%%%%%%%%%%%%%%%%%%%%%%%%%%%%%%%%%%%%%%%%%%%%%%%%%%%%%%%%%%%%
%%%%%%%%%%%%%%%%%%%%%%%%%%%%%%%%%%%%%%%%%%%%%%%%%%%%%%%%%%%%%%%%%%%%%%%%%%%%%%%%%%%%%
The DAs corresponding to our parameter sets ABO1 and ABO2 are shown in barycentric coordinates in
Fig.~\ref{fig:nucleonDA}.
The main physical conclusion from our study is that the existing experimental data on the
nucleon form factors are consistent with the nucleon wave function at small transverse distances,
the nucleon DA, that deviates  somewhat from its asymptotic form, although the difference seems to be
much less dramatic as compared to ``old'' QCD sum rule predictions~\cite{Chernyak:1987nv,King:1986wi}.
In particular the shape parameters of the first order, $\varphi_{10}$ and $\varphi_{11}$ are rather well
constrained by lattice calculations and appear to be, roughly, factor three below the QCD sum rule estimates.
The values accepted in our models, $\varphi_{10}=\varphi_{11} =0.05$, correspond to 40\% of the proton momentum carried  by the $u$-quark with the
same helicity, $\langle x_1 \rangle =0.4$, and the other two quarks carrying equal momentum fractions $\langle x_2 \rangle = \langle x_3 \rangle = 0.3$.
To this approximation the nucleon DA is symmetric under the interchange of the valence quarks with compensating helicities
\begin{align}
  \varphi_N(x_1,x_2,x_3) \simeq  \varphi_N(x_1,x_3,x_2)\,.
\label{eq:23symmetry}
\end{align}
This symmetry was conjectured originally in the diquark picture, cf.~\cite{Bolz:1996sw}.
Note, however, that the symmetry (\ref{eq:23symmetry}) cannot be exact since $\varphi_{10}$ and $\varphi_{11}$
have different anomalous dimensions.
Our fits of the proton form factor data, in particular $G^p_E/G^p_M$, indicate a small but nonvanishing
second order coefficient
\begin{align}
 \varphi_{20} = 0.06(3)\,,
\end{align}
an order of magnitude smaller than QCD sum rule estimates~\cite{Chernyak:1987nv,King:1986wi} and also smaller
than the accuracy of the present lattice data~\cite{Braun:2008ur,lattice2013}.
The remaining two second order coefficients, $\varphi_{21}$ and $\varphi_{22}$, are comparable with zero within the
error bars, see Table~\ref{tab:shape}.
The ``diquark symmetry'' (\ref{eq:23symmetry}) translates to the following relation between the $\varphi_{2k}$:
\begin{align}
 \varphi_{20}-5\phi_{21}+2\phi_{22} = 0\,.
\end{align}
It is satisfied approximately for the set of parameters ABO2 and violated by $\sim 2$~standard deviations for the set ABO1
so that we do not have a definite conclusion. For illustration we show the DAs corresponding to the central values of our parameter sets ABO1 and ABO2
in barycentric coordinates in Fig.~\ref{fig:nucleonDA}. Although ABO1 leads to a somewhat better overall description of the form factors
as compared to ABO2,  the difference is not significant in view of the intrinsic uncertainties of the method. Thus the
difference of the two pictures in Fig.~\ref{fig:nucleonDA} should be regarded as the uncertainty of our calculation.
%
%%%%%%%%%%%%%%%%%%%%%%%%%%%%%%%%%%%%%%%%%%%%%%%%%%%%%%%%%%%%%%%%%%%%%%%%%%%%%%%%%%%%%
%%%%%%%%%%%%%%%%%%%%%%%%%%%%%%%%%%%%%%%%%%%%%%%%%%%%%%%%%%%%%%%%%%%%%%%%%%%%%%%%%%%%%
\section{Summary and conclusions}\label{sec:summary}
%\setcounter{equation}{0}
%%%%%%%%%%%%%%%%%%%%%%%%%%%%%%%%%%%%%%%%%%%%%%%%%%%%%%%%%%%%%%%%%%%%%%%%%%%%%%%%%%%%%
%%%%%%%%%%%%%%%%%%%%%%%%%%%%%%%%%%%%%%%%%%%%%%%%%%%%%%%%%%%%%%%%%%%%%%%%%%%%%%%%%%%%%
We have given the state-of-the-art analysis of nucleon electromagnetic
form factors in the LCSR approach. As explained in the Introduction, the main challenge
in the QCD description of form factors is the calculation of soft overlap contributions
corresponding to the so-called Feynman mechanism to transfer the large momentum.
The LCSR approach is attractive because soft contributions are calculated in
terms of the same DAs that enter the pQCD calculation of hard rescattering contributions,
and there is no double counting. Thus, the LCSRs provide one with the most direct relation of
the hadron form factors and DAs that is available at present, at the cost of slight model dependence
of the nucleon separation from the higher-mass background.
Our calculation incorporates the following new elements as compared to previous studies in the same framework~\cite{Braun:2001tj,Braun:2006hz}:
\begin{itemize}
\item{} Next-to-leading order QCD corrections to the contributions of twist-three and twist-four DAs, Sec.~\ref{sec:NLO} and App.~\ref{App:NLO-functions}.
\item{} Exact account of ``kinematic'' contributions to the nucleon DAs of twist-four and twist-five
        induced by lower geometric twist operators (Wandzura-Wilczek terms), Eqs.~(\ref{WW}), (\ref{WW5a}), (\ref{WW5b}).
\item{} Light-cone expansion to the twist-four accuracy of the three-quark matrix elements with generic quark positions,
        Eqs.~(\ref{mathbbVA}), (\ref{mathbbV2}), (\ref{mathbbA2}).
\item{} A new calculation of twist-five off-light cone contributions, Eqs.~(\ref{eq:y20}), (\ref{eq:y20}).
\item{} A more general model for the leading-twist DA, including contributions of second-order polynomials.
\end{itemize}
The numerical analysis of the LCSRs is presented in Sec.~\ref{sec:Results}. The main message
is that electromagnetic form factors can be described to the expected 10-20\% accuracy using
the nucleon DA with comparatively small corrections to its asymptotic form.
We believe that a combination of LCSRs and lattice calculations of moments of DA allows one to
obtain quantitative information on the structure of the nucleon at small interquark separations.
In particular the valence quark average momentum fractions can be determined to a few percent accuracy.
The present study can be extended in several directions, in particular updating the existing LO LCSR calculations of the electroproduction
of negative parity resonances~\cite{Braun:2009jy} and threshold pion electroproduction~\cite{Braun:2006td}.
This is needed, on the one hand, in view of the existing CLAS data \cite{Aznauryan:2009mx} \cite{Park:2012yf} \cite{Khetarpal:2012vs}
and the experimental program for the 12 GeV upgrade at Jefferson laboratory~\cite{Aznauryan:2012ba}. On the other hand, a global fit to the
nucleon DAs from different hard reactions would be extremely interesting and increase our confidence in the emerging picture.
Several technical aspects of the LCSRs deserve further study, e.g. contributions of factorizable multiquark nucleon DAs to $F_2(Q^2)$.
Also the magnetic transition form factor for the electroexcitation of Delta-resonance~\cite{Braun:2005be}
needs to be reexamined.
%%%%%%%%%%%%%%%%%%%%%%%%%%%%%%%%%%%%%%%%%%%%%%%%%%%%%%%%%%%%%%%%%%%%%%%%%%%%%%%%%%%%
\section*{Acknowledgements}
We are grateful to A.~N.~Manashov for the possibility to use his results on the
Wandzura-Wilczek contributions in  Eqs.~(\ref{WW5a}) and  (\ref{WW5b}) prior publication,
for the discussions of the renormalization scheme and useful comments. We thank R. Schiel for
generating the plots of the nucleon distribution amplitude fig. \ref{fig:nucleonDA}.
This work was supported by the German Research Foundation (DFG), grant BR 2021/6-1 and
in part by the RFBR (grant 12-02-00613) and the Heisenberg-Landau Program.
%%%%%%%%%%%%%%%%%%%%%%%%%%%%%%%%%%%%%%%%%%%%%%%%%%%%%%%%%%%%%%%%%%%%%%%%%%%%%
%%%%%%%%%%%%%%%%%%%%%%%%%%%%%   Appendix   %%%%%%%%%%%%%%%%%%%%%%%%%%%%%%%%%%
%%%%%%%%%%%%%%%%%%%%%%%%%%%%%%%%%%%%%%%%%%%%%%%%%%%%%%%%%%%%%%%%%%%%%%%%%%%%%
\appendix
\renewcommand{\theequation}{\Alph{section}.\arabic{equation}}
\section*{Appendices}
%%%%%%%%%%%%%%%%%%%%%%%%%%%%%%%%%%%%%%%%%%%%%%%%%%%%%%%%%%%%%%%%%%%%%%%%%%%%%%%%%%%%%%%%%%
\section{Renormalization scheme for three-quark operators}\label{App:A}
%%%%%%%%%%%%%%%%%%%%%%%%%%%%%%%%%%%%%%%%%%%%%%%%%%%%%%%%%%%%%%%%%%%%%%%%%%%%%%%%%%%%%%%%%%
For simplicity we consider local three-quark operators without derivatives
\begin{align}
\mathcal{Q}_{\alpha\beta\gamma}=\epsilon^{ijk} q^{i,a}_\alpha q^{j,b}_\beta q^{k,c}_\gamma\,,
\end{align}
where $i,j,k$ are color and  $a,b,c$ flavor indices, respectively.
We will assume that $a\neq b\neq c$, i.e. the quarks have different flavor.
We imply using dimensional regularization with the space-time dimension
$d = 4-2\epsilon$ and adopt the notation
$$
     a(\mu) = \frac{\alpha_s(\mu)}{\pi}\,.
$$
The divergent part of the sum of Feynman diagrams for the Green function
\begin{align}
\langle 0|\mathcal{Q}_{\alpha\beta\gamma} \bar q_{\alpha'}(p_1)\bar q_{\beta'}(p_2)\bar q_{\beta'}(p_3)| 0 \rangle
\label{green}
\end{align}
after subtraction of the subdivergences can be cast into the form
\begin{align}
  \sum_{lmn} g_{lmn}(\epsilon)(\Gamma_{lmn})^{\alpha'\beta'\gamma'}_{\alpha\beta\gamma}
\label{diver}
\end{align}
where $g_{lmn}(\epsilon)$ are given by a series in $1/\epsilon$
\begin{align}
 g_{lmn}(\epsilon) = \sum_{p=1}^\infty \epsilon^{-p} a_{lmn}^{(p)}(a)
\end{align}
and the gamma-matrix structures $(\Gamma_{lmn})^{\alpha'\beta'\gamma'}_{\alpha\beta\gamma}$ are
defined as
\begin{align}
 (\Gamma_{lmn})^{\alpha'\beta'\gamma'}_{\alpha\beta\gamma} = \gamma^{(l)}_{\alpha\alpha'} \otimes \gamma^{(m)}_{\beta\beta'}
\otimes \gamma^{(n)}_{\gamma\gamma'}\,.
\label{Gamma}
\end{align}
Here
\begin{align}
  \gamma^{(n)}_{\mu_1,\mu_2,\ldots,\mu_n} = \gamma_{[\mu_1}\gamma_{\mu_2}\ldots\gamma_{\mu_n]}
\end{align}
are the antisymmetrized (over Lorentz indices) products of gamma-matrices, cf.~\cite{Dugan:1990df}.
In (\ref{Gamma}) it is assumed that all Lorentz indices of gamma-matrices are contracted between
themselves; one can show that there exists only one nontrivial way to contract all indices.
Following Ref.~\cite{Kraenkl:2011qb} we define the subtraction scheme by removing the singular
terms (\ref{diver}) from the correlation function (\ref{green}). Thus, the renormalized
operator, $[\mathcal{Q}]_{\alpha\beta\gamma}$, takes the form
\begin{eqnarray}
{}[\mathcal{Q}]_{\alpha\beta\gamma}(q)&=&\mathcal{Z}_{\alpha\beta\gamma}^{\alpha'\beta'\gamma'}
\mathcal{Q}_{\alpha'\beta'\gamma'}(q)
\nonumber\\&=&
\mathcal{Z}_{\alpha\beta\gamma}^{\alpha'\beta'\gamma'} Z_q^{-3}
\mathcal{Q}^B_{\alpha'\beta'\gamma'}(q_B)\,,
\end{eqnarray}
where
\begin{align}
 \mathcal{Z}_{\alpha\beta\gamma}^{\alpha'\beta'\gamma'} = 1 +
  \sum_{lmn} g_{lmn}(\epsilon)(\Gamma_{lmn})^{\alpha'\beta'\gamma'}_{\alpha\beta\gamma}
\end{align}
and $Z_q$ is the quark renormalization constant.
%To the two-loop accuracy in Feynman gauge~\cite{Egorian:1978zx}
%
%\begin{align}
%Z_q=1-a\frac{1}{6\epsilon}+a^2
%\frac1{32}\left[
%\frac{40}{9}\frac{1}{\epsilon^2}
%+\frac1\epsilon\left(
%-\frac{47}{3}+\frac23 n_F
%\right)
%\right]
%\end{align}
%
The renormalization group equation reads
\begin{align}\label{RGE}
\Big(\mu{\partial_\mu}+\beta(a){\partial_a}\Big)[\mathcal{Q}_{\alpha\beta\gamma}]
= - \gamma_{\alpha\beta\gamma}^{\alpha'\beta'\gamma'}[\mathcal{Q}_{\alpha'\beta'\gamma'}]
\end{align}
where $\beta(a)$ is the QCD beta-function
\begin{equation}
\beta(a) = \mu \partial_\mu a(\mu) = -2\epsilon a  - 2\beta_0 a^2+O(a^3)\,
\end{equation}
with $\beta_0 = 11/3 N_c-2/3 n_f$, and
the anomalous dimension matrix $\gamma_{\alpha\beta\gamma}^{\alpha'\beta'\gamma'}$
is defined as
\begin{equation}
{\mathcal{H}}=-\left(\mu\frac{d}{d\mu} {\mathbb{Z}}\right)\mathbb{Z}^{-1} =
-\beta(a)(\partial_a \mathbb{Z}) \mathbb{Z}^{-1}
\end{equation}
Here
\begin{align}
\mathbb{Z}=\mathcal{Z}Z_q^{-3} = 1 + a \mathbb{Z}^{(1)} + a^2 \mathbb{Z}^{(2)} + \mathcal{O}(a^3)\,.
\end{align}
Note that calculating the inverse matrix
\begin{align}
\mathbb{Z}^{-1} = 1 - a \mathbb{Z}^{(1)} - a^2 [\mathbb{Z}^{(1)}\mathbb{Z}^{(1)}+ \mathbb{Z}^{(2)}] + \mathcal{O}(a^3)\,
\end{align}
one must carry out all gamma-matrix
algebra in $d$-dimensions, which gives rise to finite (regular) contributions $\sim \epsilon^p$,
$p=0,1,\ldots$.
Such terms arise, in particular, because the product $\mathbb{Z}^{(1)}\mathbb{Z}^{(1)}$ has to be brought to the
standard form as an expansion in the basis of antisymmetrized gamma matrices (see above).
The resulting terms $\sim\epsilon$ must be taken into account.
This is different from the standard situation where $\mathbb{Z}^{-1}$
only contains poles $\sim 1/\epsilon^p$, $p=1,2,\ldots$, and there are no finite terms.
Thus the relation between the $Z$-factor and the anomalous dimension becomes somewhat
more complicated. To the two-loop accuracy one obtains~\cite{Kraenkl:2011qb}
\begin{align}
 \gamma \,=&\,\, 2a\epsilon\mathbb{Z}^{(1)} + 2a^2 [2\mathbb{Z}^{(2)}-\mathbb{Z}^{(1)}\mathbb{Z}^{(1)}]+\beta_0\mathbb{Z}^{(1)}
%\notag\\&
+ \mathcal{O}(a^3)\,.
\end{align}
Both terms in the square bracket  $[2\mathbb{Z}^{(2)}- \mathbb{Z}^{(1)}\mathbb{Z}^{(1)}]$
contain $1/\epsilon^2$ poles which have to cancel so that their difference only contains single poles.
The anomalous dimension matrix can be expanded in the contributions of different gamma-matrix
structures similar to Eq.~(\ref{diver}):
\begin{align}
 \gamma_{\alpha\beta\gamma}^{\alpha'\beta'\gamma'} = \sum_{lmn} \gamma_{lmn}(a)(\Gamma_{lmn})^{\alpha'\beta'\gamma'}_{\alpha\beta\gamma}\,.
\end{align}
Since $\gamma$ is finite, one can drop in this (final) expression all terms in $\gamma^{(n)}$ with $n>4$.
All structures (i.e. including those with $n\ge 5$) must be kept, however, in
the $\mathbb{Z}$-factors, at least in principle. In practice this complication appears starting at three loops.
Finally, the renormalization of three-quark operators in the usual Rarita-Schwinger representation
is obtained by applying the corresponding projection operators (in $d=4$ dimensions).
For example for the Ioffe current
\begin{eqnarray}
&& [\eta_I]_\gamma = (C\gamma^\mu)_{\alpha\beta} (\gamma_\mu)_{\gamma\gamma'} [\epsilon^{ijk} u^{i}_{\alpha} u^{j}_{\beta} d^{k}_{\gamma'}]\,.
\end{eqnarray}
For the two-loop anomalous dimensions of the couplings $f_N$ and $\lambda_1$ (see text)
one obtains in this scheme~\cite{Kraenkl:2011qb}
\begin{eqnarray}
 \gamma_{f_N}\, &=&\,\frac13 a + \left(\frac{23}{36}+\frac{7}{18}\beta_0\right) a^2 \,,
\nonumber\\
 \gamma_{\lambda_1}\, &=&\, - a  -  \left(\frac{19}{12}-\frac{1}{3}\beta_0\right) a^2 \,.
\end{eqnarray}
Generalization of thie KM renormalization scheme to nonlocal light-ray operators that define baryon DAs is
in principle straightforward. The calculations become, of course, much more involved as the $\mathcal{Z}$-factors
and anomalous dimensions become integral operators acting on quark coordinates. For higher-twist operators
one has also to take into account the mixing with light-ray operators including transverse derivatives and/or the
gluon field in addition to the three quarks, see Ref.~\cite{Braun:2008ia}.
%%%%%%%%%%%%%%%%%%%%%%%%%%%%%%%%%%%%%%%%%%%%%%%%%%%%%%%%%%%%%%%%%%%%%%%%%%%%%%%%%%%%%%%%%%
\section{Summary of nucleon distribution amplitudes}\label{App:NDA}
%%%%%%%%%%%%%%%%%%%%%%%%%%%%%%%%%%%%%%%%%%%%%%%%%%%%%%%%%%%%%%%%%%%%%%%%%%%%%%%%%%%%%%%%%%
In practical calculations it is convenient to work with the expression
for the renormalized three-quark light-ray operator with open Dirac indices.
The general expression for the nucleon matrix element contains 24 scalar functions~\cite{Braun:2000kw}
of which only 12, however, contribute to the LCSRs considered in this paper:
\begin{widetext}
\begin{eqnarray}
\label{defdisamp}
\lefteqn{ \hspace*{-1cm}
4 \langle 0| \epsilon^{ijk} u_\alpha^i(a_1 n) u_\beta^j(a_2 n) d_\gamma^k(a_3 n)
|P\rangle =}
\nonumber \\[1mm]
&=&
%m_N S_1 C_{\alpha \beta} \left(\gamma_5 N^+\right)_\gamma +
%m_N S_2 C_{\alpha \beta} \left(\gamma_5 N^-\right)_\gamma +
%m_N P_1 \left(\gamma_5 C\right)_{\alpha \beta} N^+_\gamma +
%m_N P_2 \left(\gamma_5 C \right)_{\alpha \beta} N^-_\gamma
%\nonumber \\ &&
V_1  \left(\slashed{p}C \right)_{\alpha \beta} \left(\gamma_5 N^+\right)_\gamma +
V_2  \left(\slashed{p}C \right)_{\alpha \beta} \left(\gamma_5 N^-\right)_\gamma +
\frac12 m_N V_3  \left(\gamma_\perp C \right)_{\alpha \beta}\left( \gamma^{\perp}
\gamma_5 N^+\right)_\gamma
\nonumber \\
&& +
\frac12 m_N V_4 \left(\gamma_\perp C \right)_{\alpha \beta}\left( \gamma^{\perp}
\gamma_5 N^-\right)_\gamma +
\frac{m_N^2}{2 p n} V_5 \left(\slashed{n}C \right)_{\alpha \beta} \left(\gamma_5 N^+\right)_\gamma +
\frac{m_N^2}{2 pn} V_6  \left(\slashed{n}C \right)_{\alpha \beta} \left(\gamma_5 N^-\right)_\gamma
\nonumber \\
&& +
A_1  \left(\slashed{p}\gamma_5 C \right)_{\alpha \beta} N^+_\gamma +
A_2  \left(\slashed{p}\gamma_5 C \right)_{\alpha \beta} N^-_\gamma +
\frac12 m_N A_3  \left(\gamma_\perp \gamma_5 C \right)_{\alpha \beta}\left( \gamma^{\perp} N^+\right)_\gamma
\nonumber \\
&& +
\frac12 m_N  A_4  \left(\gamma_\perp \gamma_5 C \right)_{\alpha \beta}\left( \gamma^{\perp} N^-\right)_\gamma +
\frac{m_N^2}{2 p n} A_5 \left(\slashed{n}\gamma_5 C \right)_{\alpha \beta} N^+_\gamma +
\frac{m_N^2}{2 pn}  A_6  \left(\slashed{n}\gamma_5 C \right)_{\alpha \beta} N^-_\gamma +\ldots
%\nonumber \\
%&& +
%T_1 \left(i \sigma_{\perp p} C\right)_{\alpha \beta} \left(\gamma^\perp\gamma_5 N^+\right)_\gamma +
%T_2 \left(i \sigma_{\perp\, p} C\right)_{\alpha \beta} \left(\gamma^\perp\gamma_5 N^-\right)_\gamma +
%\frac{m_N}{p n} T_3 \left(i \sigma_{p\, n} C\right)_{\alpha \beta} \left(\gamma_5 N^+\right)_\gamma
%\nonumber \\
%&&
%+ \frac{m_N}{p n} T_4 \left(i \sigma_{n\, p} C\right)_{\alpha \beta}
%\left(\gamma_5 N^-\right)_\gamma +
%\frac{m_N^2}{2 p n} T_5 \left(i \sigma_{\perp\, n} C\right)_{\alpha \beta} \left(\gamma^\perp\gamma_5 N^+\right)_\gamma +
%\frac{m_N^2}{2 pn}  T_6 \left(i \sigma_{\perp\, n} C\right)_{\alpha \beta} \left(\gamma^\perp\gamma_5 N^-\right)_\gamma
%\nonumber \\
%&& +
%\frac12 m_N T_7  \left(\sigma_{\perp\, \perp'} C\right)_{\alpha \beta} \left(\sigma^{\perp\, \perp'} \gamma_5 N^+\right)_\gamma +
%\frac12 m_N T_8  \left(\sigma_{\perp\, \perp'} C\right)_{\alpha \beta} \left(\sigma^{\perp\, \perp'} \gamma_5 N^-\right)_\gamma\,,
\end{eqnarray}
\end{widetext}
where for brevity we do not show the Wilson lines that make this operator gauge-invariant;
$\alpha,\beta,\gamma$ are Dirac indices and
we use a shorthand notation $ \sigma_{\perp\, n}\otimes \gamma^\perp = \sigma_{\mu\nu} n^\nu g_\perp^{\mu\alpha}\otimes \gamma_\alpha$ etc.
Each invariant function $F = V_i,A_i$ can be written as a Fourier integral
\begin{eqnarray}
\label{fourier}
F(a_j, Pn) = \int \! [dx]\, e^{-i Pn \sum_i x_i a_i} F(x_i)\,,
\end{eqnarray}
where $F(x_i)$ depend on the longitudinal momentum fractions $x_i$
carried by the quarks inside the nucleon.
The integration measure is defined in Eq.~(\ref{measure}).
The invariant functions $V_1,A_1$ correspond to the leading contribution of collinear twist three.
They are related to the nucleon DA $\varphi_N(x_i)$ defined in Eqs.~(\ref{varphi-N}),~(\ref{expand-varphi}) as
follows~\cite{Chernyak:1983ej}:
\begin{eqnarray}
V_1(1,2,3) &=& \frac12f_N\big[\varphi_N(1,2,3)+\varphi_N(2,1,3)\big]\,,
\nonumber\\
A_1(1,2,3) &=& \frac12f_N\big[\varphi_N(2,1,3)-\varphi_N(1,2,3)\big]\,.
%\nonumber\\
%T_1(1,2,3) &=& \frac12f_N\big[\varphi_N(1,3,2)+\varphi_N(2,3,1)\big]\,.
\label{VA1}
\end{eqnarray}
Here and below $F(1,2,3) \equiv F(x_1,x_2,x_3)$.
The functions $V_2,A_2,V_3,A_3$ correspond to the contributions of
collinear twist-4. They include contributions of ``genuine'' geometric twist-four
operators and Wandzura-Wilczek-type terms of geometric twist-three that are related
to the leading-twist DA, cf. Eqs.~(\ref{twist-4}),~(\ref{WW}).
One obtains~\cite{Braun:2000kw}
\begin{eqnarray}
 V_2(1,2,3) &=& \frac14f_N\big[\Phi^{WW}_4(1,2,3)+\Phi^{WW}_4(2,1,3)\big]
\nonumber\\&&{}
            +  \frac14\lambda_1^N\big[\Phi_4(1,2,3)+\Phi_4(2,1,3)\big] \,,
\nonumber\\
 A_2(1,2,3) &=& \frac14f_N\big[\Phi^{WW}_4(2,1,3)-\Phi^{WW}_4(1,2,3)\big]
\nonumber\\&&{}
            +  \frac14\lambda_1^N\big[\Phi_4(2,1,3)-\Phi_4(1,2,3)\big] \,,
\nonumber\\
 V_3(1,2,3) &=& \frac14f_N\big[\Psi^{WW}_4(1,2,3)+\Psi^{WW}_4(2,1,3)\big]
\nonumber\\&&{}
            -  \frac14\lambda_1^N\big[\Psi_4(1,2,3)+\Psi_4(2,1,3)\big] \,,
\nonumber\\
 A_3(1,2,3) &=& \frac14f_N\big[\Psi^{WW}_4(2,1,3)-\Psi^{WW}_4(1,2,3)\big]
\nonumber\\&&{} -  \frac14\lambda_1^N\big[\Psi_4(2,1,3)-\Psi_4(1,2,3)\big] \,.
%\nonumber\\
% T_2(1,2,3) &=& \frac{1}{12}\lambda_2^N\big[\Xi_4(3,1,2)+\Xi_4(3,2,1)\big] \,,
%\nonumber\\
%(T_3+T_7)(1,2,3) &=& \frac14f_N\big[\Psi^{WW}_4(3,1,2)+\Psi^{WW}_4(1,3,2) + \Phi^{WW}_4(2,3,1)+\Phi^{WW}_4(3,2,1)\big]
%\nonumber\\&&{}
%               - \frac14\lambda_1^N\big[\Psi_4(3,1,2)+\Psi_4(1,3,2)- \Phi_4(2,3,1)-\Phi_4(3,2,1)\big]
%\nonumber\\
%(T_3-T_7)(1,2,3) &=& \frac{1}{12}\lambda_2^N\big[\Xi_4(1,2,3)+\Xi_4(2,1,3)\big]
%\nonumber\\
%(S_1-P_1)(1,2,3) &=& \frac14f_N\big[\Psi^{WW}_4(3,1,2)-\Psi^{WW}_4(1,3,2) + \Phi^{WW}_4(2,3,1)-\Phi^{WW}_4(3,2,1)\big]
%\nonumber\\&&{}
%               - \frac14\lambda_1^N\big[\Psi_4(3,1,2)-\Psi_4(1,3,2)- \Phi_4(2,3,1)+\Phi_4(3,2,1)\big]
%\nonumber\\
%(S_1+P_1)(1,2,3) &=&
%               \frac{1}{12}\lambda_2^N\big[\Xi_4(1,2,3)-\Xi_4(2,1,3)\big]
\end{eqnarray}
In turn, collinear twist-5 DAs contain Wandzura-Wilczek-type contributions of
twist-three (WWW) and twist-four (WW) operators, but to our accuracy
no ``genuine'' geometric twist-five terms:
\begin{eqnarray}
 V_4(1,2,3) &=& \frac14f_N\big[\Psi^{WWW}_5(1,2,3)+\Psi^{WWW}_5(2,1,3)\big]
\nonumber\\&&{}
              -  \frac14\lambda_1^N\big[\Psi^{WW}_5(1,2,3)+\Psi^{WW}_5(2,1,3)\big] \,,
\nonumber\\
 A_4(1,2,3) &=& \frac14f_N\big[\Psi^{WWW}_5(2,1,3)-\Psi^{WWW}_5(1,2,3)\big]
\nonumber\\&&{}
              -  \frac14\lambda_1^N\big[\Psi^{WW}_5(2,1,3)-\Psi^{WW}_5(1,2,3)\big] \,,
\nonumber\\
 V_5(1,2,3) &=& \frac14f_N\big[\Phi^{WWW}_5(1,2,3)+\Phi^{WWW}_4(2,1,3)\big]
\nonumber\\&&{}
              +  \frac14\lambda_1^N\big[\Phi^{WW}_5(1,2,3)+\Phi^{WW}_5(2,1,3)\big] \,,
\nonumber\\
 A_5(1,2,3) &=& \frac14f_N\big[\Phi^{WWW}_4(2,1,3)-\Phi^{WWW}_4(1,2,3)\big]
\nonumber\\&&{}
              +  \frac14\lambda_1^N\big[\Phi^{WW}_4(2,1,3)-\Phi^{WW}_4(1,2,3)\big] \,.
\nonumber\\
\end{eqnarray}
The expressions presented here are more general as compared to the parametrization suggested in Ref~\cite{Braun:2000kw}
in that we employ exact expressions for the Wandzura-Wilczek-type contributions, cf. Eqs.~(\ref{WW}),  (\ref{WW5a}),  (\ref{WW5b}):
In the earlier work only the first two terms in their conformal expansion were taken into account.
As the result, the normalization integrals and the first moments of our DAs and those
given in~\cite{Braun:2000kw} coincide
$$ \int [dx]\, x_k F^{\rm this~work}_{2,3,4,5}(x_i) = \int [dx]\, x_k F^{\rm Ref.[29]}_{2,3,4,5}(x_i)$$
where $F=V,A$ and $k=1,2,3$, but our DAs also contain contributions of higher conformal partial waves
that are necessitated by the algebra of spin rotation and QCD equations of motion.
Apart from theoretical consistency, this important update allows us to use arbitrary models of the
leading-twist DAs, e.g. include second-order polynomials in the momentum fractions.
Taking into account the contributions of collinear twist-6 DAs $V_6,A_6$ is, strictly speaking,
beyond our accuracy. As an estimate we use the model of Ref.~\cite{Braun:2000kw}:
\begin{eqnarray}
V_6(x_i) &=& 2 \left[\phi_6^0 +  \phi_6^+(1- 3 x_3)\right],
\nonumber \\
A_6(x_i) &=& 2 (x_2 - x_1) \phi_6^- \,,
\end{eqnarray}
where
\begin{align}
\phi_6^0 =& f_N\,,
\nonumber\\
\phi_6^+ =& f_N \Big( 2 \varphi_{10} -\frac23 \varphi_{11}-\frac13\Big)
           +\lambda_1\Big(\frac15\eta_{10} -\frac13 \eta_{11}-\frac15\Big),
\nonumber\\
\phi_6^- =& f_N \Big( 2 \varphi_{10} + 2\phi_{11} +1 \Big)
           +\lambda_1 \Big(\frac15 \eta_{10} + \eta_{11}- \frac15 \Big).
\end{align}
The corresponding contributions to the LCSRs prove to be very small.
For completeness we give the relations between the shape parameters of first order ---
$\varphi_{10},\varphi_{11}$ for twist-three and $\eta_{10},\eta_{11}$  for twist-four --- used in this work, to
the parameters $V_1^d, A_1^u$ and $f_1^d,f_1^u$ used in Ref.~\cite{Braun:2000kw} and also the LCSR
calculations in \cite{Braun:2001tj,Braun:2006hz,PassekKumericki:2008sj}:
\begin{eqnarray}
   A_1^u & = & \varphi_{10}+\varphi_{11}\,,
\nonumber\\
   V_1^d &=& \frac13 - \varphi_{10} + \frac{1}{3}\phi_{11}\,,
\end{eqnarray}
\begin{eqnarray}
 f_1^d &=& \frac{3}{10} - \frac16 \frac{f_N}{\lambda_1} +\frac15 \eta_{10} -\frac13 \eta_{11}\,,
\nonumber\\
 f_1^u &=& \frac{1}{10} - \frac16 \frac{f_N}{\lambda_1} -\frac35 \eta_{10} -\frac13 \eta_{11}\,.
\nonumber\\
f_2^d &=&\frac{4}{15}+\frac25 \xi_{10}\,.
\end{eqnarray}
Numerical values of these parameters are discussed in the main text.
%%%%%%%%%%%%%%%%%%%%%%%%%%%%%%%%%%%%%%%%%%%%%%%%%%%%%%%%%%%%%%%%%%%%%%%%%%%%%%%%%%%%%%%%%%
\section{Operator Product Expansion of three-quark currents}\label{App:OPE}
%%%%%%%%%%%%%%%%%%%%%%%%%%%%%%%%%%%%%%%%%%%%%%%%%%%%%%%%%%%%%%%%%%%%%%%%%%%%%%%%%%%%%%%%%%
Matrix elements of three-quark operators at small non-light-like separations can be
reduced to the DAs. In the leading-order LCSRs there is a major simplification that two
of the quark coordinates always coincide. This case was considered in detail in
Refs.~\cite{Braun:2001tj,Braun:2006hz}. The relevant matrix elements can be written as
\begin{widetext}
\begin{eqnarray}
\label{dquark}
 - \langle{0}| \epsilon^{ijk} \left[u^i
C\gamma_\alpha u^j\right](0) d_\gamma^{k}(y)
          |P\rangle &=&
\left({\mathcal V}_1  + \frac{y^2m_N^2}{4} \mathcal{V}_1^{M(d)}\right) P_\alpha \left(\gamma_5 N\right)_\gamma +
\frac{\mathcal{V}_2 m_N}{2(Py)} P_\alpha
\left(\slashed{y} \gamma_5 N\right)_\gamma
+ \frac12 \mathcal{V}_3 m_N  \left(\gamma_\alpha \gamma_5 N\right)_\gamma
\nonumber
\\ &&{}\hspace*{-0.0cm}
+ \frac{\mathcal{V}_4 m_N^2}{4(Py)}   y_\alpha \left(\gamma_5 N\right)_\gamma
+ \frac{\mathcal{V}_5 m_N^2}{4(Py)} \left(i\sigma_{\alpha\lambda}y^\lambda \gamma_5 N\right)_\gamma
+ \frac{\mathcal{V}_6 m_N^3}{4(Py)^2}  y_\alpha \left(\slashed{y} \gamma_5 N\right)_\gamma \,,
\nonumber\\
- \langle{0}| \epsilon^{ijk} \left[u^i
C\gamma_\alpha \gamma_5 u^j\right](0) d_\gamma^{k}(y)
          |{P}\rangle &=&
\left({\mathcal A}_1  + \frac{y^2m_N^2}{4} \mathcal{A}_1^{M(d)}\right) P_\alpha \left( N\right)_\gamma +
\frac{\mathcal{A}_2 m_N}{2(Py)} P_\alpha \left(\slashed{y}N\right)_\gamma
+ \frac12 \mathcal{A}_3 m_N  \left(\gamma_\alpha N\right)_\gamma
\nonumber \\
&&{}\hspace*{-0.0cm}
+ \frac{\mathcal{A}_4 m_N^2}{4(Py)}   y_\alpha \left( N\right)_\gamma
+ \frac{\mathcal{A}_5 m_N^2}{4(Py)} \left(i\sigma_{\alpha\lambda}x^\lambda N\right)_\gamma
+ \frac{\mathcal{A}_6 m_N^3}{4(Py)^2}  y_\alpha \left(\slashed{y} N\right)_\gamma
\end{eqnarray}
\end{widetext}
and similar expressions for
$$ \langle{0}| \epsilon^{ijk} \left[u^i(0)  C\gamma_\alpha (\gamma_5) u^j(y)\right] d_\gamma^{k}(0)|P\rangle $$
with the replacement $\mathcal{V}_1^{M(d)}, \mathcal{A}_1^{M(d)}  \to  \mathcal{V}_1^{M(u)}, \mathcal{A}_1^{M(u)}$.
The invariant functions $\mathcal{V}_i,\mathcal{A}_i$ depend on the
quark coordinates $a_iy$ and can be written as
\begin{equation}
\mathcal{F}(a_i;Py) = \int [dx]\, e^{-iPy(x_1 a_1+ x_2 a_2 + x_3 a_3)} \mathcal{F}(x_i)\,.
\end{equation}
The ``calligraphic'' functions in the momentum fraction representation, $\mathcal{F}(x_i)$, can be expressed in terms of the nucleon DAs
introduced in App.~\ref{App:NDA} (at the scale $\mu^2 \sim 1/|y^2|$). One obtains~\cite{Braun:2001tj}
\begin{align}
&\mathcal{V}_1 = V_1\,,
\qquad
\mathcal{V}_2 = V_1-V_2-V_3\,,
\qquad \mathcal{V}_3 = V_3\,,
\nonumber\\
&\mathcal{V}_4 = -2 V_1 + V_3 + V_4 + 2 V_5\,,
\qquad \mathcal{V}_5 = V_4-V_3\,,
\nonumber\\
&\mathcal{V}_6 = -V_1+V_2+V_3+V_4+V_5-V_6
\end{align}
and, similarly,
\begin{align}
&\mathcal{A}_1 = A_1\,,
\qquad \mathcal{A}_2 = A_2-A_1-A_3\,,
\qquad \mathcal{A}_3 = A_3\,,
\nonumber\\
&\mathcal{A}_4 = -2 A_1 - A_3 - A_4 + 2 A_5\,,
\qquad \mathcal{A}_5 = A_3-A_4\,,
\nonumber\\
& \mathcal{A}_6 = A_1-A_2+A_3+A_4-A_5+A_6\,.
\end{align}
 We also use the following notations~\cite{Braun:2001tj}:
\begin{eqnarray}
\widetilde{F}(x_3)
 &=& \int\limits_1^{x_3}\!\!dx'_3\int\limits_0^{1- x^{'}_{3}}\!\!\!dx_1\, F(x_1,1-x_1-x'_3,x'_3)\,,
\nonumber\\
\widetilde{\!\widetilde{F}}(x_3) &=& \int\limits_1^{x_3}\!\!dx'_3 \int\limits_1^{x'_3}\!\!dx^{''}_3
\int\limits_0^{1- x^{''}_{3}}\!\!\!\!dx_1\, F(x_1,1-x_1-x^{''}_3,x^{''}_3)
\nonumber\\
\label{DAtilde}
\end{eqnarray}
and
\begin{eqnarray}
\widehat{F}(x_2)
&=& \int\limits_1^{x_2}\!\!dx'_2\int\limits_0^{1-x^{'}_{2}}\!\!\!dx_1 F(x_1,x'_2,1-x_1-x'_2)\,,
\nonumber\\
\widehat{\!\widehat{F}}(x_2)&=& \int\limits_1^{x_2}\!\!dx'_2 \int\limits_1^{x'_2}\!\!dx^{''}_2
\int\limits_0^{1-x^{''}_{2}}\!\!\!dx_1 F(x_1,x^{''}_2,1-x_1-x^{''})\,,
\nonumber\\
\label{DAhat}
\end{eqnarray}
where $F=A_k,V_k$ is a generic nucleon DA that depends on the three valence quark momentum fractions,
The calculation of $\mathcal{O}(y^2)$ corrections to the leading-twist contributions is explained in
detail in Ref.~\cite{Braun:2006hz}: The moments of $\mathcal{V}_1^{M(u,d)}(x_2)$, $\mathcal{A}_1^{M(u,d)}(x_2)$ can be
expressed in terms of the moments of twist-3 and twist-4 DAs. We have rederived these relations using a
somewhat different approach and confirmed the results. Using our modified expressions for the DA we obtain
\begin{align}
 \mathcal{V}_1^{M(u)}(x_2) \equiv &\int\limits_{0}^{1-x_2}\!dx_1 \, V_1^M(x_1,x_2,1-x_1-x_2)
\notag\\ =& ~x_2^2 (1-x_2)^3 \Big(\frac53 f_N  C_f^u + \frac{1}{12} \lambda_1 C_\lambda^u\Big),
\notag\\
 \mathcal{A}_1^{M(u)}(x_2) \equiv &\int\limits_{0}^{1-x_2}\!dx_1 \, A_1^M(x_1,x_2,1-x_1-x_2)
\notag\\ =& ~x_2^2(1-x_2)^3 \Big(\frac53 f_N  D_f^u + \frac{1}{12}\lambda_1 D_\lambda^u\Big),
\end{align}
\begin{align}
  \mathcal{V}_1^{M(d)}(x_3) \equiv &\int\limits_{0}^{1-x_3}\!dx_1 \, V_1^M(x_1,1-x_1-x_3,x_3)
\notag\\ =& ~x_3^2(1-x_3)^2 \Big(\frac53 {f_N} C_f^d + \frac{1}{12} \lambda_1  C_\lambda^d\Big),
\notag\\
  \mathcal{A}_1^{M(d)}(x_3) \equiv &\int\limits_{0}^{1-x_3}\!dx_1 \, A_1^M(x_1,1-x_1-x_3,x_3) ~=~ 0\,,
\label{eq:y20}
\end{align}
where
\begin{align}
C_\lambda^u =& -4 -3\eta_{10}(5x_2-3)-5\eta_{11}(x_2+1)\,,
\notag\\
C_f^u =&  - (4x_2-5)- \frac{21}{4}\varphi_{10}(9x_2^2-14x_2+3)
\notag\\ &+ \frac{7}{4} \varphi_{11}(9x_2^2-8x_2+1)\,,
\notag\\
D_\lambda^u =& -4 -3\eta_{10}(5x_2-3)-5\eta_{11}(9x_2-7)\,,
\notag\\
D_f^u    =&   \phantom{-} 1 - \frac{21}{4}\varphi_{10}(9x_2^2-14x_2+3)
\notag\\ &-\frac{7}{4}\varphi_{11}(27x_2^2-36x_2+7)\,
\end{align}
and
\begin{align}
C_\lambda^d =&  \phantom{-} 8 + 2 (5x_3-3)(3\eta_{10}-5 \eta_{11})\,,
\notag\\
C_f^d =  &    -2 (2x_3-3)+ \frac{7}{2}(9x_3^2-14x_3+3)(3\varphi_{10}-\varphi_{11})\,,
\label{eq:y21}
\end{align}
These expressions are somewhat simpler as compared to the results of Ref.~\cite{Braun:2006hz,remark1}
which have been obtained using truncated Wandzura-Wilczek contributions, although
the numerical difference is small.
For the calculation of correlation functions to the NLO accuracy, which is the subject of this work, we need to find a
generalization of Eqs.~(\ref{dquark}) to twist-four accuracy for arbitrary quark positions
\begin{align}
 y_1 =& a_1 n + \vec{b}_1\,,\notag\\   y_2 =& a_2 n + \vec{b}_2\,,\notag\\  y_3 =& a_3 n + \vec{b}_3\,,
\end{align}
where $\vec{b}_i$ are transverse vectors w.r.t. $n_\mu$ and $P_\mu$.
Let
\begin{widetext}
\begin{eqnarray}
 - \langle{0}| \epsilon^{ijk} \left[u^i(y_1) C\gamma_\alpha u^j(y_2)\right] d^{k}(y_3) |P\rangle
&=& \biggl\{  P_\alpha \mathbb{V}_1  + m_N \gamma_\alpha  \mathbb{V}_3
+ i m_N P_\alpha \Big[\mathbb{V}^{(1)}_2\slashed{y}_1 + \mathbb{V}^{(2)}_2\slashed{y}_2
+ \mathbb{V}^{(3)}_2\slashed{y}_3\Big]
+\ldots \biggr\}\gamma_5 N\,,
\nonumber\\
 - \langle{0}| \epsilon^{ijk} \left[u^i(y_1) C\gamma_\alpha\gamma_5 u^j(y_2)\right] d^{k}(y_3) |P\rangle
&=& \biggl\{  P_\alpha \mathbb{A}_1  + m_N \gamma_\alpha  \mathbb{A}_3
+ i m_N P_\alpha \Big[\mathbb{A}^{(1)}_2\slashed{y}_1 + \mathbb{A}^{(2)}_2\slashed{y}_2
+ \mathbb{A}^{(3)}_2\slashed{y}_3\Big]
+\ldots \biggr\} N\,,
\nonumber\\
\label{mathbbVA}
\end{eqnarray}
%\end{widetext}
where the ellipses stand for terms of twist higher than four. Note that the invariant functions $\mathbb{V}_i$
and $\mathbb{A}_i$ do not depend on transverse coordinates, e.g.
\begin{eqnarray}
 \mathbb{V}_i(y_iP) &=& \int [dx_i] e^{-iP\sum x_i y_i} \mathbb{V}_i(x_i)
%\nonumber\\ &=&
 = \int [dx_i] e^{-iPn\sum x_i a_i} \mathbb{V}_i(x_i)\,.
\end{eqnarray}
\end{widetext}
Translation invariance requires that (suppressing color indices)
\begin{eqnarray}
  &&\hspace*{-1.2cm}\langle 0|[u(y_1+z)C\gamma_\alpha(\gamma_5) u(y_2+z)]d(y_3+z)|P\rangle =
\nonumber\\ &&{} =  e^{-iPz}\langle 0|[u(y_1)C\gamma_\alpha(\gamma_5)u(y_2)) d(y_3)|P\rangle\,.
\end{eqnarray}
This condition is satisfied identically for $\mathbb{V}_1$, $\mathbb{V}_3$, $\mathbb{A}_1$, $\mathbb{A}_3$  and implies the relations
\begin{align}
 \mathbb{V}^{(1)}_2 + \mathbb{V}^{(2)}_2 + \mathbb{V}^{(3)}_2 = 0\,, &&  \mathbb{A}^{(1)}_2 + \mathbb{A}^{(2)}_2 + \mathbb{A}^{(3)}_2 = 0\,.
\end{align}
The parametrization of the matrix element in Eq.~(\ref{mathbbVA}) must reproduce
the known expression in (\ref{defdisamp}) in the light-cone limit $b_1=b_2=b_3=0$.
From this requirement it follows immediately that
\begin{align}
\mathbb{V}_1(x_i) = \mathcal{V}_1(x_i) = V_1(x_i)\,,
&&
\mathbb{V}_3(x_i) = \mathcal{V}_3(x_i) = V_3(x_i)\,.
\notag\\
\mathbb{A}_1(x_i) = \mathcal{A}_1(x_i) = A_1(x_i)\,,
&&
\mathbb{A}_3(x_i) = \mathcal{A}_3(x_i) = A_3(x_i)\,.
\end{align}
The derivation for $\mathbb{V}^{(k)}_2(x_i)$,  $\mathbb{A}^{(k)}_2(x_i)$ is somewhat more involved. We obtain
\begin{eqnarray}
 \mathbb{V}^{(1)}_2(x_i)&=& \frac{1}{4}\Big[x_3 V_2(x_i)
+ (x_2-x_1) {V}_3(x_i) - A_3(x_i)
\nonumber\\&&{}\hspace*{0.5cm}+ x_3 A_3(x_i)+ x_3 A_2(x_i) \Big],
\nonumber\\
 \mathbb{V}^{(2)}_2(x_i)&=& \frac{1}{4}\Big[
 x_3 V_2(x_i) + (x_1-x_2) {V}_3(x_i)+A_3(x_i)
\nonumber\\&&{}\hspace*{0.5cm}
- x_3 A_3(x_i)- x_3 A_2(x_i) \Big],
\nonumber\\
 \mathbb{V}^{(3)}_2 (x_i) & = & -\frac{1}{2} x_3 V_2(x_i)\,,
\label{mathbbV2}
\end{eqnarray}
and, similarly,
\begin{eqnarray}
\mathbb{A}^{(1)}_2(x_i)&=& \frac{1}{4}\Big[-x_3 A_2(x_i)
+ (x_2-x_1) {A}_3(x_i) - V_3(x_i)
\nonumber\\&&{}\hspace*{0.5cm}+ x_3 V_3(x_i)- x_3 V_2(x_i)
            \Big],
\nonumber\\
 \mathbb{A}^{(2)}_2(x_i)&=& \frac{1}{4}\Big[
 -x_3 A_2(x_i) + (x_1-x_2) {A}_3(x_i)
+V_3(x_i)
\nonumber\\&&{}\hspace*{0.5cm}- x_3 V_3(x_i)+ x_3 V_2(x_i) \Big],
\nonumber\\
 \mathbb{A}^{(3)}_2 (x_i) & = & \frac{1}{2} x_3 A_2(x_i)\,.
\label{mathbbA2}
\end{eqnarray}
One can show that
\begin{eqnarray}
 \frac{i}{2} \widetilde {\mathcal{V}}_{2}(x_3) &=& \int\limits_0^{1-x_3} dx_1 \, \mathbb{V}^{(3)}_2(x_1,1-x_1-x_3,x_3)\,,
\nonumber\\
 \frac{i}{2} \widehat {\mathcal{V}}_{2}(x_2) &=& \int\limits_0^{1-x_2} dx_1 \, \mathbb{V}^{(2)}_2(x_1,x_2,1-x_1-x_2)
\end{eqnarray}
(cf. (\ref{DAtilde}), (\ref{DAhat}))
and similar for $A$-functions. These relations are satisfied identically for the models of nucleon
DAs used in this work, but are violated for the DAs in~\cite{Braun:2006hz} because of the truncation
in Wandzura-Wilczek-type contributions.
%%%%%%%%%%%%%%%%%%%%%%%%%%%%%%%%%%%%%%%%%%%%%%%%%%%%%%%%%%%%%%%%%%%%%%%%%%%%%%%%%%%%%%%%%%
\section{Auxiliary functions}\label{App:D}
%%%%%%%%%%%%%%%%%%%%%%%%%%%%%%%%%%%%%%%%%%%%%%%%%%%%%%%%%%%%%%%%%%%%%%%%%%%%%%%%%%%%%%%%%%
The momentum dependence of the NLO corrections to the correlation function (\ref{correlator})
can conveniently be written in terms of the following functions:
\begin{align}
  g_{nk}(y,x;W) = \frac{\ln^n[1-yW-i\eta]}{(-1+ x W +i\eta)^k},
\notag\\
  h_{nk}(x;W) = \frac{\ln^n[1-x W-i\eta]}{(W +i\eta)^k}
\end{align}
with $n=0,1,2$ and $k=1,2,3$. For the particular case $n=0$ the first argument
becomes dummy; for simplicity of notation we write the corresponding entries as
\begin{align}
g_k(x;W) \equiv g_{0k}(\ast,x;W)\,,
\end{align}
cf. Eq.~(\ref{gk-functions}). Going over to the Borel parameter space and subtracting the continuum
corresponds to the substitutions
\begin{align}
  g_{nk} \to & G_{nk}(y,x;M^2)
%\notag\\ &
= \frac{1}{\pi}\!\int_0^{s_0}\! \frac{ds}{Q^2}\,e^{-s/M^2} \text{Im}\,g_{nk}(y,x,W)\,,
\notag\\
  h_{nk} \to & H_{nk}(x;M^2)
%\notag\\ &
= \frac{1}{\pi}\!\int_0^{s_0} \! \frac{ds}{Q^2}\,e^{-s/M^2} \text{Im}\,h_{nk}(x,W)\,,
\end{align}
where $s={P'}^2$ is the invariant mass of the quark-antiquark (+gluon) state,
$W = 1 + s/Q^2$, $M^2$ is the Borel parameter and $s_0$ the continuum threshold.
LCSRs involve integrals of the type
\begin{align}
   \mathbf{G}_{nk} =& \int [dx]\, \mathcal{F}(\underline{x})  G_{nk}(x_i+x_j,x_i;M^2)\,,
\notag\\
   \widetilde{\mathbf{G}}_{nk} =& \int [dx]\, \mathcal{F}(\underline{x})  G_{nk}(x_i,x_i;M^2)\,,
\notag\\
  \mathbf{H}_{nk} =& \int [dx]\, \mathcal{F}(\underline{x})  H_{nk}(x_i+x_j;M^2)\,,
\end{align}
where $\mathcal{F}(\underline{x}) = \mathcal{F}(x_i,x_j,1-x_i-x_j)$ is a function of quark momentum
fractions and $x_i, x_j \in \{x_1,x_2,x_3\}$.
In addition one needs
\begin{align}
   \widehat{\mathbf{G}}_{01} =& \int [dx]\, \mathcal{F}(\underline{x})  G_{01}(\ast,x_i+x_j;M^2)
\end{align}
(only this special case).
The corresponding expressions are collected below. We use the following notations:
\begin{align}
   &x_{ij}= x_i+x_j\,, \qquad \bar x = 1-x\,, \qquad x_0 = \frac{Q^2}{s_0+Q^2}\,,
\notag\\
   &E(x) = \exp\left[-\frac{\bar x Q^2}{x M^2}\right],
\notag\\
  &\Big[{\mathcal F}(x_i,x_j)\Big]_+ = \mathcal{F}(x_i,x_j) - \mathcal{F}(x_0,x_j)\,
\end{align}
and
\begin{align}
 \mathcal{F}\otimes \mathcal{G} =&  \int_{x_0}^1 dx_i \int_{0}^{1-x_i}dx_j \,\mathcal{F}(\underline{x}) \, \mathcal{G}(\underline{x})\,,
\notag\\
  \mathcal{F}\circledast \mathcal{G} =& \int_0^{x_0} dx_i \int_{x_0-x_i}^{1- x_i}dx_j \,\mathcal{F}(\underline{x}) \, \mathcal{G}(\underline{x})\,.
\end{align}
We obtain:
\begin{eqnarray}
  {\mathbf{G}}_{01} &=&
- {\mathcal F} \otimes \frac{E(x_{i})}{x_{i}}\,,
\nonumber\\
 \widehat{\mathbf{G}}_{01} &=&
- {\mathcal F} \Big[\otimes + \circledast \Big]\frac{E(x_{ij})}{x_{ij}}\,,
\end{eqnarray}
\begin{eqnarray}
 \mathbf{G}_{11} &=&
    \mathcal{F}\otimes \int\limits_{x_0}^{x_{ij}}\! dy \frac{E(y)-E(x_i)}{y(y-x_i)}
+   \mathcal{F}\circledast \int\limits_{x_0}^{x_{ij}}\frac{dy\, E(y)}{y(y-x_i)}
\nonumber\\&&{}
-  \mathcal{F}\otimes \frac{E(x_i)}{x_i}\Big[\ln\Big(\frac{x_i}{x_0}-1\Big) + \ln\frac{x_{ij}}{x_i}\Big],
%\end{eqnarray}
\\
%\begin{eqnarray}
\widetilde{ \mathbf{G}}_{11}&=&
{\mathcal F} \Big[\otimes + \circledast \Big] \int\limits_{x_0}^{x_{ij}} dy \frac{E(y)-E(x_{ij})}{y(y-x_{ij})}
\nonumber\\&&{}
- {\mathcal F} \Big[\otimes + \circledast \Big]\frac{E(x_{ij})}{x_{ij}} \ln\Big(\frac{x_{ij}}{x_0}-1\Big),
\end{eqnarray}
\begin{eqnarray}
 \mathbf{G}_{21} &=&
2  {\mathcal F} \otimes
 \int\limits_{x_0}^{x_{ij}} dy \frac{E(y)-E(x_i)}{y(y-x_i)} \ln\Big(\frac{x_{ij}}{y}-1\Big) -
\nonumber\\&&
- 2  {\mathcal F} \circledast
 \int\limits_{x_0}^{x_{ij}} dy \frac{E(y)}{y(y-x_i)} \ln\Big(\frac{x_{ij}}{y}-1\Big)
\nonumber\\&&
+ {\mathcal F} \otimes \frac{E(x_i)}{x_i} \biggl\{ \frac{\pi^2}{3}
- 2\text{Li}_2\Big(\frac{x_j x_0}{x_{ij}(x_0-x_i)}\Big)
\nonumber\\&&{} \hspace*{0.5cm}
- \Big[ \ln\Big(\frac{x_i}{x_0}-1\Big) + \ln\Big(\frac{x_{ij}}{x_i}\Big) \Big]^2 \biggr\},
\end{eqnarray}
\begin{eqnarray}
\mathbf{G}_{02}&=&
\frac{Q^2}{M^2}
{\mathcal F}\otimes \frac{E(x_i)}{x_i^2} +
E(x_0)\!\! \int\limits_{0}^{1-x_0} dx_j\, {\mathcal F}(x_0,x_j)\,.
\nonumber\\[-2mm]
\end{eqnarray}
\begin{widetext}
\begin{eqnarray}
\mathbf{G}_{12} &=&
 - \Big[\frac{x_0}{x_i} {\mathcal F}(x_i,x_j)\Big]_+ \Big[\otimes + \circledast \Big]
\frac{E(x_0)}{x_0-x_i}
 +{\mathcal F} \otimes \frac{x_{ij} \big(E(x_{ij}) - E(x_{i})\big)}{x_i x_j} +
 {\mathcal F} \circledast \frac{x_{ij} E(x_{ij})}{x_i x_j}
\nonumber\\&&{}
 -  E(x_0) \biggl[
\int\limits_{0}^{x_0} dx_j \ln\Big(\frac{x_j}{x_0}\Big)
- \int\limits_{0}^{1} dx_j \ln\Big|1-\frac{\bar x_j}{x_0}\Big|
- \int\limits_{0}^{\bar x_0} dx_j \ln\Big(\frac{x_j}{x_0}\Big)
\biggr]{\mathcal F}(x_0,x_j)
\nonumber\\&&{}
-\frac{Q^2}{M^2} {\mathcal F} \otimes \frac{1}{x_i} \int\limits_{x_0}^{x_{ij}} dy \frac{E(y)-E(x_{i})}{y(y-x_{i})} -
\frac{Q^2}{M^2} {\mathcal F} \circledast \frac{1}{x_i} \int\limits_{x_0}^{x_{ij}} dy \frac{E(y)}{y(y-x_{i})}\,,
\nonumber\\
\widetilde{\mathbf{G}}_{12}&=&
- \Big[\frac{x_0}{x_i} {\mathcal F}(x_i,x_j)\Big]_+ \otimes \frac{E(x_0)}{x_0-x_i}
+ E(x_0) \int\limits_{0}^{\bar x_0} dx_j\,{\mathcal F}(x_0,x_j) \ln\Big(\frac{\bar x_j}{x_0}-1\Big)
\nonumber\\&&
-\frac{Q^2}{M^2} {\mathcal F} \otimes \frac{1}{x_i} \int\limits_{x_0}^{x_i} dy \frac{E(y)-E(x_i)}{y(y-x_i)} +
\frac{Q^2}{M^2} {\mathcal F} \otimes \frac{E(x_i)}{x^2_i} \ln\Big(\frac{x_i}{x_0}-1\Big)
+ \mathbf{G}_{02}\,,
\end{eqnarray}
\begin{eqnarray}
\mathbf{G}_{22}&=&
- 2 \Big[ \frac{x_0}{x_i} \ln\Big(\frac{x_{ij}}{x_0}-1\Big) {\cal F}(x_i,x_j)\Big]_{+} \Big[\otimes + \circledast \Big] \frac{E(x_0)}{x_0-x_i}
+ E(x_0)\int\limits_{0}^{\bar x_0} dx_j
\Big[ \ln^2\Big(\frac{x_j}{x_0}\Big) - \pi^2 \Big]  {\cal F}(x_0,x_j)
\nonumber\\[-2mm]&&{}
- 2 E(x_0) \biggl[
\int\limits_{0}^{x_0} dx_j \ln\Big(\frac{x_j}{x_0}\Big) -
\int\limits_{0}^{1} dx_j \ln\Big|1-\frac{\bar x_j}{x_0}\Big|
\biggr]  \ln\Big(\frac{x_j}{x_0}\Big)
{\cal F}(x_0,x_j)
\nonumber\\[-2mm]&&{}
-{\cal F} \Big[\otimes + \circledast \Big] \frac{2 x^2_{ij}}{x_i x_j} \int\limits_{x_0}^{x_{ij}} dy \frac{E(y)-E(x_{ij})}{y(y-x_{ij})}
+{\cal F} \Big[\otimes + \circledast \Big] \frac{2 x_{ij} E(x_{ij})}{x_i x_j} \ln\Big( \frac{x_{ij}}{x_0}-1\Big)
\nonumber\\[-2mm]&&{}
+ {\cal F} \otimes \frac{2 x_{ij}}{x_j} \int\limits_{x_0}^{x_{ij}} dy \frac{E(y)-E(x_{i})}{y(y-x_{i})} +
{\cal F} \circledast \frac{2 x_{ij}}{x_j} \int\limits_{x_0}^{x_{ij}} dy \frac{E(y)}{y(y-x_{i})}
- {\cal F} \otimes \frac{2 x_{ij} E(x_i)}{x_i x_j}
\Big[ \ln\Big(\frac{x_i}{x_0}-1\Big) + \ln\Big(\frac{x_{ij}}{x_i} \Big)\Big] \, ,
\nonumber\\[-2mm]&&{}
-2 \frac{Q^2}{M^2} {\cal F} \otimes
\frac{1}{x_i} \int\limits_{x_0}^{x_{ij}} dy \frac{E(y)-E(x_i)}{y(y-x_i)} \ln\Big(\frac{x_{ij}}{y}-1\Big) -
2 \frac{Q^2}{M^2} {\cal F} \circledast
\frac{1}{x_i} \int\limits_{x_0}^{x_{ij}} dy \frac{E(y)}{y(y-x_i)} \ln\Big(\frac{x_{ij}}{y}-1\Big)
\nonumber\\&&{}
+\frac{Q^2}{M^2} {\cal F} \otimes \frac{E(x_i)}{x^2_i} \biggl\{ \frac{\pi^2}{3} -
\Big[ \ln\Big(\frac{x_i}{x_0}-1\Big) + \ln\Big(\frac{x_{ij}}{x_i}\Big) \Big]^2 -
2\text{Li}_2\Big(\frac{x_j x_0}{x_{ij}(x_0-x_i)}\Big) \biggr\}\,,
\end{eqnarray}
\begin{eqnarray}
\widetilde{\mathbf{G}}_{22}&=&
-2 \Big[ \frac{x_0}{x_i} {\cal F}(x_i,x_j)\Big]_{+} \otimes  \ln\Big(\frac{x_{i}}{x_0}-1\Big)\frac{E(x_0)}{x_0-x_i}
+E(x_0) \int\limits_{0}^{\bar x_0} dx_j
{\cal F}(x_0,x_j) \Big[\ln^2\Big(\frac{\bar x_j}{x_0}-1\Big) - \frac{\pi^2}{3}\Big]
\nonumber\\
&&
-2 \Big[\frac{x_0}{x_i}{\cal F}(x_i,x_j)\Big]_{+} \otimes \frac{E(x_0)}{x_0-x_i} +
2 E(x_0) \int\limits_{0}^{\bar x_0} dx_j
{\cal F}(x_0,x_j) \ln\Big(\frac{\bar x_j}{\bar x_j}-1\Big)
\nonumber\\
&&
- 2 \frac{Q^2}{M^2} {\cal F} \otimes
\frac{1}{x_i} \int\limits_{x_0}^{x_i} dy \frac{E(y)-E(x_i)}{y(y-x_i)} \ln\Big( \frac{x_i}{y}-1\Big) +
\frac{Q^2}{M^2} {\cal F} \otimes \frac{E(x_i)}{x^2_i} \Big[ \frac{\pi^2}{3} - \ln^2\Big(\frac{x_i}{x_0}-1 \Big)\Big]
\nonumber\\
&&
-2 \frac{Q^2}{M^2} {\cal F} \otimes \frac{1}{x_i} \int\limits_{x_0}^{x_i} dy \frac{E(y)-E(x_i)}{y(y-x_i)} +
2 \frac{Q^2}{M^2} {\cal F} \otimes \frac{E(x_i)}{x^2_i} \ln\Big(\frac{x_i}{x_0}-1\Big)
 + 2 \mathbf{G}_{02}\,,
\end{eqnarray}
\end{widetext}
and finally
\begin{eqnarray}
\mathbf{H}_{11} &=& - {\cal F} \Big[ \otimes + \circledast\Big] \int_{x_0}^{x_{ij}} dy\, \frac{E(y)}{y} \,,
\\
\mathbf{H}_{21} &=&  - 2 {\cal F} \Big[ \otimes + \circledast\Big]
\int_{x_0}^{x_{ij}} dy\, \ln\Big( \frac{x_{ij}}{y}-1\Big) \frac{E(y)}{y} \,,
\nonumber
\end{eqnarray}
\begin{eqnarray}
\mathbf{H}_{12} &=& - {\cal F} \Big[ \otimes + \circledast\Big] \int_{x_0}^{x_{ij}} dy\, E(y)\,,
\\
\mathbf{H}_{22} &=&  - {\cal F} \Big[ \otimes + \circledast\Big]
\int_{x_0}^{x_{ij}} dy\, \ln\Big( \frac{x_{ij}}{y}-1\Big) E(y)\,,
\nonumber
\end{eqnarray}
\begin{eqnarray}
\mathbf{H}_{13} &=&   - {\cal F} \Big[ \otimes + \circledast\Big]\int_{x_0}^{x_{ij}} dy \, y E(y)\, ,
\\
\mathbf{H}_{23} &=&  - {\cal F} \Big[ \otimes + \circledast\Big]
\int_{x_0}^{x_{ij}} dy \, y \ln\Big( \frac{x_{ij}}{y}-1\Big) E(y)\,.
\nonumber
\end{eqnarray}
Our results for  ${\mathbf G}_{11},{\mathbf G}_{21}, {\mathbf H}_{11}, {\mathbf H}_{12}, {\mathbf H}_{21}, {\mathbf H}_{22}$ differ from the
corresponding expressions $g_7-g_{12}$ in Ref.~\cite{PassekKumericki:2008sj} by extra terms from the $\circledast$ integration region;
in addition our expression for  ${\mathbf G}_{21}$ does not contain a contribution  $\sim \pi^2(1-\delta(x_j))$.
\begin{widetext}
%%%%%%%%%%%%%%%%%%%%%%%%%%%%%%%%%%%%%%%%%%%%%%%%%%%%%%%%%%%%%%%%%%%%%%%%%%%%%%%%%%%%%%%%%%
\section{Summary of NLO coefficient functions}\label{App:NLO-functions}
%%%%%%%%%%%%%%%%%%%%%%%%%%%%%%%%%%%%%%%%%%%%%%%%%%%%%%%%%%%%%%%%%%%%%%%%%%%%%%%%%%%%%%%%%%
The NLO corrections (\ref{AB-NLO}) to the correlation functions  $\mathcal{A}(Q^2,P'^2)$ and  $\mathcal{B}(Q^2,P'^2)$ can be
written as a sum of contributions of a given quark flavor $q = u,d$ and expanded in contributions of nucleon DAs as
shown in Eqs.~(\ref{NLO-A}), (\ref{NLO-B}.
Our results for the coefficient functions $C_q^{F}(x_i,W)$ are collected below. We use a shorthand notation
$L= \ln Q^2/\mu^2$ where $\mu^2$ is the factorization scale. The dependence on $W = 1+ P'^2/Q^2$ is not shown for brevity.
\begin{eqnarray}
\lefteqn{x_2 C^{\mathbb{V}_1}_d(x_i)=}
\nonumber\\ &=&
    2x_2 x_3 \Big[ 3 (L-2) g_1(x_3) + 2 (L-1) g_{11}(x_3,x_3) + g_{21}(x_3,x_3)\Big]
    + \Big[2x_2 + (4L-3)x_3\Big] h_{11}(x_3) + (3-4L) \bar x_1  h_{11}(\bar x_1)
\nonumber\\&&{}
+ 2 x_3 h_{21}(x_3) -2 \bar x_1 h_{21}(\bar x_1)
    - 2\Big[3 (x_2/x_3) (2L\!-\!3) + 5L-7 \Big] h_{12}(x_3) + 2 (5L\!-\!7) h_{12}(\bar x_1)
    - \Big[6 (x_2/x_3)+ 5 \Big] h_{22}(x_3)
\nonumber\\[1mm]&&{}
+ 5 h_{22}(\bar x_1) + (6/x_3)(L-2) h_{13}(x_3)  - (6/\bar x_1)(L-2) h_{13}(\bar x_1)  + (3/x_3) h_{23}(x_3) - (3/\bar x_1) h_{23}(\bar x_1)\,,
\end{eqnarray}
\begin{eqnarray}
\lefteqn{ x_1 x_3 C^{\mathbb{V}_1}_u(x_i) = }
\nonumber\\&=&
x_1 x_2 x_3 \Big[\big( 17-7 L \big) g_1(x_2) +
(1+2 L) g_{11}(\bar x_1,x_2)+2 (2 L-3) g_{11}(\bar x_3,x_2) + 2\big( 5-7 L \big) g_{11}(x_2,x_2) +g_{21}(\bar x_1,x_2)
\nonumber\\&&{}
+2 g_{21}(\bar x_3,x_2)-7 g_{21}(x_2,x_2)\Big]
- x_1 x_3\Big[ (1+2 L) h_{11}(\bar x_1)+2 (2 L-3) h_{11}(\bar x_3)+2 \big(5-7 L\big) h_{11}(x_2)\Big]
\nonumber\\&&
+(1+2 L) x_1 h_{12}(\bar x_1)+4 x_3 h_{12}(\bar x_3)
- \Big[4 x_3 + (x_1/x_2) \big[x_2(1+2 L) + 4 x_3(4\!-\!L)\big] \Big] h_{12}(x_2)
-2 (L\!-\!2 ) (x_1/\bar x_1) h_{13}(\bar x_1)
\nonumber\\&&
+ 2 (2 L-7) (x_3/\bar x_3) h_{13}(\bar x_3) +
(1/x_2)\Big[2 (L-2) x_1 + 2 (7-2 L) x_3 \Big] h_{13}(x_2)
- x_1 x_3 \Big[ h_{21}(\bar x_1)+2 h_{21}(\bar x_3)- 7 h_{21}(x_2)\Big]
\nonumber\\&&
+ x_1 h_{22}(\bar x_1)+x_1 \Big[2 (x_3/x_2)-1\Big] h_{22}(x_2)
- (x_1/\bar x_1) h_{23}(\bar x_1) +
2 (x_3/\bar x_3) h_{23}(\bar x_3) +
\Big[ (x_1 - 2 x_3)/x_2 \Big] h_{23}(x_2)\,,
\end{eqnarray}
\begin{eqnarray}
\lefteqn{x_2 C^{\mathbb{V}_3}_d(x_i)=}
\nonumber\\&=&
2 x_2 x_3 \Big[
 (5-3 L)  g_1(x_3)+ (3-4 L)  g_{11}(\bar x_1,x_3)+2 (2 L-1)  g_{11}(x_3,x_3)
-2  g_{21}(\bar x_1,x_3)+2  g_{21}(x_3,x_3)
\Big]
\nonumber\\&&{}
+
2 (4 L-3) (2 x_2+x_3) h_{11}(\bar x_1)+
\Big[8(1-2 L) x_2+2 (3-4 L) x_3\Big] h_{11}(x_3)
+4 (2 x_2+x_3) \Big[ h_{21}(\bar x_1)-  h_{21}(x_3)\Big]
\nonumber\\
&&
+
6 (3-4 L) h_{12}(\bar x_1)+
6 \Big[4 L-3+4 (x_2/x_3) (L-1)\Big] h_{12}(x_3)-12 h_{22}(\bar x_1)+
12(\bar x_1/x_3) h_{22}(x_3)
\nonumber\\
&&
+
(4/\bar x_1)(2 L -1) h_{13}(\bar x_1)-
(4/x_3)(2L-1) h_{13}(x_3)+(4/\bar x_1) h_{23}(\bar x_1)
-(4/x_3) h_{23}(x_3)\,,
\end{eqnarray}
\begin{eqnarray}
\lefteqn{x_1 x_3 C^{\mathbb{V}_3}_u(x_i)=}
\nonumber\\
&=&
2 x_1 x_2 x_3 \Big[5 (L-3) g_1(x_2)+2(1-2 L) g_{11}(\bar x_1,x_2)+(5-4 L) g_{11}(\bar x_3,x_2)
+2 (8 L - 5) g_{11}(x_2,x_2)
\nonumber\\&&{}
- 2g_{21}(\bar x_1,x_2)-2g_{21}(\bar x_3,x_2)+8 g_{21}(x_2,x_2)\Big]
+2 x_3 \Big[(6 L-8) x_1+(2 L-3) x_2\Big] h_{11}(\bar x_3)
\nonumber\\&&{}
+4 x_1 \Big[L x_2+(3 L-1) x_3\Big] h_{11}(\bar x_1)
-2\Big[4x_1x_3(5L-3)+ x_2 x_3(2L-3)+ 2 x_1 x_2 L\Big]  h_{11}(x_2)
%-2 x_3 \Big[6 (L-1) x_1+(2 L -3) x_2\Big] h_{11}(x_2)
%-4 x_1 \Big[L x_2+(7 L -3) x_3\Big] h_{11}(x_2)
\nonumber\\&&{}
+2 x_1 (x_2+3 x_3) h_{21}(\bar x_1)
+2 x_3 (3 x_1+x_2) h_{21}(\bar x_3)
-2\Big[10 x_1 x_3 + x_2 \bar x_2\Big]h_{21}(x_2)
%-2 x_3 (3 x_1+x_2) h_{21}(x_2)
%- 2 x_1 (x_2+7 x_3) h_{21}(x_2)
-4 (3+2 L) x_1 h_{12}(\bar x_1)
\nonumber\\&&{}
+2x_3 (15-8 L)h_{12}(\bar x_3)
+2(x_3/x_2) \Big[4 (L-1) x_1+(8 L - 15) x_2\Big] h_{12}(x_2)
+4(x_1/x_2) \Big[(3+2 L)x_2+6 x_3\Big] h_{12}(x_2)
\nonumber\\&&{}
-4 x_1 h_{22}(\bar x_1)-8 x_3 h_{22}(\bar x_3)
+(4/x_2) \Big[2 x_2 x_3 + x_1\bar x_1\Big] h_{22}(x_2)
+12 (x_1/\bar x_1) h_{13}(\bar x_1)
+8 (x_3/\bar x_3) (L-2)  h_{13}(\bar x_3)
\nonumber\\&&{}
- (4/x_2)\big[3 x_1 + 2 x_3 (L-2)\Big] h_{13}(x_2)
+4 (x_3/\bar x_3) h_{23}(\bar x_3)-4 (x_3/x_2) h_{23}(x_2)\,,
\end{eqnarray}
\begin{eqnarray}
\lefteqn{x_1 x_2 C^{\mathbb{V}^{(1)}_2}_u(x_i)=}
\nonumber\\&=&
-8 \bar x_3 x_2 g_1(\bar x_3)+2 x_2^2 \Big[4 (L-2)+(2 L-5) x_1\Big]g_1(x_2)-4 x_2^2\Big[(L-3) + (2 L-3) x_1\Big] g_{11}(\bar x_3,x_2)
\nonumber\\&&{}
+ 2 x_2^2 (1+2 x_1) \Big[ 2 (L-1) g_{11}(x_2,x_2)- g_{21}(\bar x_3,x_2)+ g_{21}(x_2,x_2)\Big]
+4 x_2 \Big[(L-3)+(2 L-3) x_1\Big] h_{11}(\bar x_3)
\nonumber\\&&{}
- 2 x_2  (1+2 x_1)\Big[
2 (L-1) h_{11}(x_2)- h_{21}(\bar x_3) + 2  h_{21}(x_2)\Big]
+2 x_2 \Big[(4L-12)/\bar x_3 + (4 L-13)/x_1  - 4\Big] h_{12}(\bar x_3)
\nonumber\\&&{}
+2 \Big[4(1+  x_2 - L) + (x_2/x_1)(13-4 L)\Big] h_{12}(x_2)
+4 \Big[1-(x_1/\bar x_3)+(x_2/x_1)\Big] h_{22}(\bar x_3)-4 \Big[1+(x_2/x_1)\Big] h_{22}(x_2)
\nonumber\\&&
+2\Big[ (x_1/\bar x_3^2) (8 L-25) -  2 (x_2 /\bar x_3)(2L - 7) - (1/x_1)(8L-25)\Big]h_{13}(\bar x_3)
+2 \Big[2(2L-7) +(1/x_1) (8 L-25)\Big] h_{13}(x_2)
\nonumber\\&&{}
+ 4 \Big[2 (x_1/\bar x_3^2)  - (x_2 /\bar x_3) - (2/x_1)\Big] h_{23}(\bar x_3)
+4\Big[1+(2/x_1)\Big] h_{23}(x_2)\,,
\end{eqnarray}
\begin{eqnarray}
\lefteqn{x_2 C^{\mathbb{V}^{(2)}_2}_d(x_i)=}
\nonumber\\&=&
4 \bar x_1 g_1(\bar x_1)+2 x_3 (3-4 L) g_1(x_3)-8 x_3 g_{11}(\bar x_1,x_3)
+ 2\Big[4+(4 L-3 ) \bar x_1\Big] h_{11}(\bar x_1)
-2\Big[4 (L-1)\bar x_1 + x_3\Big] h_{11}(x_3)
\nonumber\\
&&
+4 \bar x_1 \Big[h_{21}(\bar x_1)- h_{21}(x_3)\Big]
+2\Big[2 (7-5 L)  +(11-2 L)/x_2 +2(5-2 L)/\bar x_1 \Big] h_{12}(\bar x_1)
\nonumber\\&&{}
-2 \Big[2(7-5 L)+(11-2 L)/x_2 + 4 (1-L)(1+2 x_2)/x_3\Big] h_{12}(x_3)
-\Big[10+(4/\bar x_1)+(2/x_2)\Big] h_{22}(\bar x_1)
\nonumber\\&&{}
+2 \Big[5+ (1/x_2)+ 2(1+2x_2)/x_3\Big] h_{22}(x_3)
+(4/\bar x_1) \Big[(3 L-8)(1/\bar x_1+ 1/x_2) +3 (L-2) \Big] h_{13}(\bar x_1)
\nonumber\\&&{}
-(4/x_3)\Big[(3 L\!-\!8)/x_2  +3 (L\!-\!2)\Big] h_{13}(x_3)
+(6/\bar x_1)\Big[ 1/\bar x_1 + 1/x_2 +1\Big] h_{23}(\bar x_1)
- (6/x_3)\Big[1/x_2 +1\Big] h_{23}(x_3)\,,
\end{eqnarray}
\begin{eqnarray}
\lefteqn{x_1 x_3 C^{\mathbb{V}^{(2)}_2}_u(x_i)=}
\nonumber\\&=&
4 \bar x_1 x_1 g_1(\bar x_1)-8 \bar x_3 x_3 g_1(\bar x_3)
+ 2 \Big[4 x_2 x_3- 2 x_1 x_2 + x_1x_3\big[13 L-25 +(7 L-17) x_2\big]\big)\Big] g_1(x_2)
\nonumber\\&&{}
-2 x_1 \Big[2 x_3 L + x_2 x_3 (1+2 L) -2x_2\big]\Big] g_{11}(\bar x_1,x_2)
-2 x_3 \Big[x_1 (1+2 x_2)(2 L-3)-2 x_2\Big]  g_{11}(\bar x_3,x_2)
\nonumber\\&&{}
+4 \Big[3x_1x_3 (2L-1) +x_1x_2x_3 (7 L-5 ) -x_2 \bar x_2 \Big] g_{11}(x_2,x_2)
- 2 x_1  x_3 \Big[
(1+x_2) g_{21}(\bar x_1,x_2) +  (1+2 x_2) g_{21}(\bar x_3,x_2)
\nonumber\\[1mm]&&{}
 - (6+7 x_2) g_{21}(x_2,x_2) + (1+5 L) g_2(x_2) + 2 L  g_{12}(\bar x_1,x_2) -  (3-2 L) g_{12}(\bar x_3,x_2)
-2 (4 L - 7)  g_{12}(x_2,x_2)
\nonumber\\[1mm]&&{}
+  g_{22}(\bar x_1,x_2) + g_{22}(\bar x_3,x_2) - 4 g_{22}(x_2,x_2)
\Big]
+ 2 x_1 \Big[x_3(1+2 L) -2 (1+L)\Big] h_{11}(\bar x_1)
\nonumber\\&&{}
+ 2 x_3 \Big[1-2 L+ 2x_1 (2 L-3)\Big] h_{11}(\bar x_3)
-
(2/x_2) \Big[x_3(x_2-4x_1)(2 L -1) + 2 x_1 x_2 \big[1 + L +x_3(5\!-\!7 L)\big]\Big] h_{11}(x_2)
\nonumber\\&&{}
-2 x_1 \bar x_3  h_{21}(\bar x_1)
-2 (1-2 x_1) x_3 h_{21}(\bar x_3)
+(2/x_2) \Big[x_2 \bar x_2 -x_1 x_3 (4+7 x_2)\Big] h_{21}(x_2)
- 2 x_1 (1+2 L-2/\bar x_1) h_{12}(\bar x_1)
\nonumber\\&&{}
 - 8 x_3\Big[(3 - L)/\bar x_3 + 1\Big]  h_{12}(\bar x_3)
+ (2/x_2^2) \Big[4 x_2 x_3 \big[(3-L) + x_2 + x_1(4-L)\big]
             + x_1 x_2 \big[x_2(1+2L)-2 \big]
\nonumber\\&&{}
             + 2x_1x_3 (13-6 L)\Big] h_{12}(x_2)
-2 x_1 h_{22}(\bar x_1)+4 (x_3/\bar x_3) h_{22}(\bar x_3)
+ (2/x_2) \Big[x_1x_2  -2 x_3\big[1 + x_1 + 3(x_1/x_2)\big] \Big] h_{22}(x_2)
\nonumber\\&&{}
-2 (x_1/\bar x_1^2) \Big[9-2 L+2 \bar x_1 (2-L)\Big] h_{13}(\bar x_1)
+2 (x_3/\bar x_3^2) \Big[25-8 L+ 2 \bar x_3 (7-2 L)\Big]  h_{13}(\bar x_3)
\nonumber\\
&&
+
(2/x_2^2) \Big[ x_1 \big[9-2 L +2 x_2 (2-L)\big]- x_3 \big[25-8 L+2 x_2 (7-2 L)\big] \Big] h_{13}(x_2)
+2(x_1/\bar x_1^2) (1+\bar x_1) h_{23}(\bar x_1)
\nonumber\\
&&
-4(x_3/\bar x_3^2) (2+\bar x_3)  h_{23}(\bar x_3)
-(2/x_2^2) \Big[x_1 (1+x_2)-2 x_3 (2+x_2)\Big] h_{23}(x_2)\,,
\end{eqnarray}
\begin{eqnarray}
\lefteqn{x_2 C^{\mathbb{V}^{(3)}_2}_d(x_i)=}
\nonumber\\&=&
4 \bar x_1 g_1(\bar x_1)+2 x_2 \Big[ 25-13 L+ 6x_3 (2-L) - 2(x_3/x_2) \Big] g_1(x_3)
+\Big[2 x_2 (4 L-3) - 8 x_3\Big] g_{11}(\bar x_1,x_3)
\nonumber\\&&{}
+2 x_2
\Big[
\big[6(1-2 L)+ 4 x_3 (1-L) + 4(x_3/x_2)\big] g_{11}(x_3,x_3)
+ 2 g_{21}(\bar x_1,x_3)-2  (3+x_3) g_{21}(x_3,x_3)
\nonumber\\&&{}\hspace*{1.0cm}
+ (1+5 L) g_2(x_3) - (3-4 L) g_{12}(\bar x_1,x_3)+ 2 (7-4 L) g_{12}(x_3,x_3)+ 2 g_{22}(\bar x_1,x_3) -4 g_{22}(x_3,x_3)
\Big]
\nonumber\\&&{}
+2 \Big[ (1+4 L)-\bar x_1 (3-4 L) \bar x_1\Big] h_{11}(\bar x_1)
- 2\Big[4(x_2/x_3) (1-2 L)+ 2x_2 + 1+4 L-x_3 (3-4 L)\Big] h_{11}(x_3)
\nonumber\\&&{}
+4 (1+\bar x_1) h_{21}(\bar x_1)
+4 \Big[ 2 x_2/x_3 - (1+x_3)\Big] h_{21}(x_3)
+4 \Big[ 7-5 L + (5-2 L)/\bar x_1 \Big] h_{12}(\bar x_1)
\nonumber\\&&{}
+(4/x_3^2) \Big[x_3(2 L -5) +x^2_3 (5 L-7)+ x_2(6 L - 13) +3 x_2 x_3 (2 L - 3)\Big] h_{12}(x_3)
-2\Big[5+ 2/\bar x_1 \Big] h_{22}(\bar x_1)
\nonumber\\&&{}
+ (2/x_3^2) \Big[6 x_2 (1+x_3)+x_3 (2+5 x_3 ) \Big] h_{22}(x_3)
+(4/\bar x_1^2) \Big[3 L - 8 +3 \bar x_1 (L-2) \Big] h_{13}(\bar x_1)
\nonumber\\&&{}
- (4/x_3^2) \Big[ 3 L -8 +3 (L-2) x_3\Big] h_{13}(x_3)
+ (6/\bar x_1^2) (1+\bar x_1) h_{23}(\bar x_1)
- (6/x_3^2) (1+x_3) h_{23}(x_3)\,,
\end{eqnarray}
\begin{eqnarray}
\lefteqn{x_3 C^{\mathbb{V}^{(3)}_2}_u(x_i)=}
\nonumber\\&=&
4 \bar x_1 g_1(\bar x_1)
 +2x_2 \Big[ 5+x_3 (8-3 L)\Big] g_1(x_2)
 -2x_2 \Big[ 2(1-L)+x_3 (1+2 L)\Big] g_{11}(\bar x_1,x_2)
- 4 x_2 \bar x_3 (L-1) g_{11}(x_2,x_2)
\nonumber\\&&{}
+ 2 x_2 \bar x_3 \Big[g_{21}(\bar x_1,x_2) \!-\! g_{21}(x_2,x_2)\Big]
+ 2 \Big[2(1\!-\!L)+x_3 (1\!+\!2 L)\Big] h_{11}(\bar x_1)
- 2\bar x_3\Big[2 (1\!-\!L)h_{11}(x_2) + h_{21}(\bar x_1)\!-\! h_{21}(x_2)\Big]
\nonumber\\&&{}
-2 \Big[1+2 L - 2/\bar x_1 + 2(L\!-\!1)/x_3\Big] h_{12}(\bar x_1)
+2 \Big[ 1+2 L + 2 (L\!-\!1)(1/x_3 + 2 x_3/x_2)\Big] h_{12}(x_2)
-2 \Big[ 1 + 1/x_3 \Big] h_{22}(\bar x_1)
\nonumber\\&&{}
+ 2 \Big[ 1+ 1/x_3 + 2 x_3/x_2 \Big] h_{22}(x_2)
+ (2/\bar x_1) \Big[ \big(1/\bar x_1+ 1/x_3\big) (2L-9)+ 2 (L-2) \Big] h_{13}(\bar x_1)
\nonumber\\&&{}
+ (2/x_2)\Big[ (9-2 L)/x_3  - 2 (L-2)\Big] h_{13}(x_2)
+ (2/\bar x_1)\Big[1 + 1/\bar x_1 + 1/x_3 \Big] h_{23}(\bar x_1)
-(2/x_2) \Big[ 1+1/x_3\Big] h_{23}(x_2)\,,
\end{eqnarray}
%%%%%%%%%%%%%%%%%%%%%%%%%%%%%%%%%%%%%%%%%%%%%%%%%%%%%%%%%%%%%%%%%%%%%%%%%%%%%%%%%%%%%%%%%%%%%%%%%%%%
%%%%%%%%%%%%%%%%%%%%%%%%%%%%%%
%   Axial Projection
%%%%%%%%%%%%%%%%%%%%%%%%%%%%%%
%%%%%%%%%%%%%%%%%%%%%%%%%%%%%%%%%%%%%%%%%%%%%%%%%%%%%%%%%%%%%%%%%%%%%%%%%%%%%%%%%%%%%%%%%%%%%%%%%%%%
\begin{eqnarray}
\lefteqn{x_2 C^{\mathbb{A}_1}_d(x_i)=}
\nonumber\\&=&
3 \bar x_1 h_{11}(\bar x_1)-3 x_3 h_{11}(x_3)
+ 2(3 L-10) h_{12}(\bar x_1)-2(3L-10) h_{12}(x_3)+3 h_{22}(\bar x_1)-3 h_{22}(x_3) \hspace*{3.5cm}\phantom{.}
\nonumber\\[2mm]
&&
-(6/\bar x_1) (L-3) h_{13}(\bar x_1)
+(6/x_3) (L-3) h_{13}(x_3)
-(3/\bar x_1) h_{23}(\bar x_1)
+(3/x_3) h_{23}(x_3)\,,
\end{eqnarray}
\begin{eqnarray}
\lefteqn{x_2 x_3 C^{\mathbb{A}_1}_u(x_i)=}
\nonumber\\&=&
x_2^2 x_3 \Big[(3-L) g_1(x_2)+(1-2 L) g_{11}(\bar x_1,x_2)+2 (L-1) g_{11}(x_2,x_2)-g_{21}(\bar x_1,x_2)+g_{21}(x_2,x_2)\Big]
-(5+2 L) x_2 h_{12}(\bar x_1)
\nonumber\\&&{}
+ x_2 x_3 \Big[ (2 L\!-\!1) h_{11}(\bar x_1)-2 (L\!-\!1) h_{11}(x_2)+h_{21}(\bar x_1)-h_{21}(x_2)\Big]
+\Big[(5+2 L) x_2+4 (L\!-\!1) x_3\Big] h_{12}(x_2)
\nonumber\\[1mm]&&{}
-x_2 h_{22}(\bar x_1)+(\bar x_1+ x_3) h_{22}(x_2)
-2 (x_2/\bar x_1)(L\!-\!5)  h_{13}(\bar x_1)
+2 (L\!-\!5) h_{13}(x_2)-(x_2/\bar x_1) h_{23}(\bar x_1)+h_{23}(x_2)\,,
\end{eqnarray}
\begin{eqnarray}
\lefteqn{x_2 C^{\mathbb{A}_3}_d(x_i)=}
\nonumber\\&=&
2 x_2 x_3 \Big[(8-3 L) g_1(x_3)-3 g_{11}(\bar x_1,x_3)\Big]
+ 6 (\bar x_1+x_3) h_{11}(\bar x_1)-6 x_3 h_{11}(x_3)
+ 2(4 L-21 ) h_{12}(\bar x_1)    \hspace*{3cm}\phantom{.}
\nonumber\\&&{}
+2 \Big[21-4 L+4x_2(L-1)/x_3\Big] h_{12}(x_3)
+4 h_{22}(\bar x_1)-4\Big[1 - x_2/x_3\Big] h_{22}(x_3)
+(4/\bar x_1)(7-4 L) h_{13}(\bar x_1)
\nonumber\\&&{}
+(4/x_3) (2 L-7) h_{13}(x_3)
-(4/\bar x_1) h_{23}(\bar x_1)+(4/x_3) h_{23}(x_3)\,,
\end{eqnarray}
\begin{eqnarray}
\lefteqn{x_1 x_3 C^{\mathbb{A}_3}_u(x_i)=}
\nonumber\\&=&
2 x_1 x_2 x_3 \Big[(25-11 L) g_1(x_2)-2 g_{11}(\bar x_1,x_2)+g_{11}(\bar x_3,x_2)+2(3-4 L) g_{11}(x_2,x_2)-4 g_{21}(x_2,x_2)\Big]
           \hspace*{3cm}\phantom{.}
\nonumber\\&&{}
-2 \Big[x_2 x_3 (2 L-3) +2 x_1 \big[x_2 L -2 x_3 (L-1)\big]\Big] h_{11}(x_2)
+  4 x_1 \big[x_2 L +x_3 (1+L)\big] h_{11}(\bar x_1)
\nonumber\\&&{}
+ 2x_3 \Big[2 x_1 (L-2)+x_2 (2 L-3)\Big] h_{11}(\bar x_3)
+2\bar x_1 x_1 h_{21}(\bar x_1)+2 \bar x_3 x_3 h_{21}(\bar x_3)
-2 (x_1 x_2-3 x_1 x_3 +\bar x_3 x_3) h_{21}(x_2)
\nonumber\\&&{}
+ (2/x_2) \Big[ (8 L-15) x_2 x_3+2 x_1 \big[x_2 (3+2 L)+2 x_3  (5 L-8) \big]\Big] h_{12}(x_2)-
4 x_1 \Big[ (3+2 L) h_{12}(\bar x_1)+h_{22}(\bar x_1) \Big]
\nonumber\\&&{}
-2 x_3 \Big[ (8 L-15) h_{12}(\bar x_3)+4 h_{22}(\bar x_3)\Big]
+ 4\Big[2 x_3+x_1 +5 x_1 x_3/x_2 \Big] h_{22}(x_2)
+ 12 (x_1/\bar x_1) h_{13}(\bar x_1)
\nonumber\\&&{}
-(4/x_2) \Big[3 x_1+2 x_3 (L-2)\Big] h_{13}(x_2)
+ 4 (x_3/\bar x_3) \big[ 2 (L-2) h_{13}(\bar x_3)+h_{23}(\bar x_3)\big]- 4 (x_3/x_2) h_{23}(x_2)\,,
\end{eqnarray}
\begin{eqnarray}
\lefteqn{x_1 C^{\mathbb{A}^{(1)}_2}_u(x_i)=}
\nonumber\\&=&
2 x_2\Big[
2 g_1(x_2)+2 (L-1) g_{11}(\bar x_3,x_2)- 2 (L-1) g_{11}(x_2,x_2) +  g_{21}(\bar x_3,x_2)- g_{21}(x_2,x_2)
    \Big]
                   \hspace*{4.0cm}\phantom{.}
\nonumber\\&&{}
+
4 (1-L) h_{11}(\bar x_3)+4 (L-1) h_{11}(x_2)-2 h_{21}(\bar x_3)+2 h_{21}(x_2)
+\Big[ 8 (L-2)/\bar x_3+ 2(4 L -11)/x_1\Big] h_{12}(\bar x_3)
\nonumber\\&&{}
+\Big[ 2(11-4 L)/x_1+ 8(1-L)/x_2\Big] h_{12}(x_2)
+4 \Big[  1/\bar x_3+1/x_1\Big] h_{22}(\bar x_3) -4\Big[ 1/x_1+1 /x_2\Big] h_{22}(x_2)
\nonumber\\&&{}
-(2/ \bar x_3) \Big[ 1/x_1+1/ \bar x_3 \Big] \Big[ (4 L-9 ) h_{13}(\bar x_3) + 2 h_{23}(\bar x_3)\Big]
+ \big[2/(x_1 x_2)\big]\Big[(4 L-9) h_{13}(x_2) + 2 h_{23}(x_2)\Big],
\end{eqnarray}
\begin{eqnarray}
\lefteqn{x_2 C^{\mathbb{A}^{(2)}_2}_d(x_i)=}
\nonumber\\&=&
-4 \bar x_1 g_1(\bar x_1)-6 x_3 g_1(x_3)
+6 \bar x_1 h_{11}(\bar x_1)-6 x_3 h_{11}(x_3)
+2\Big[ 2(3 L-10) +2 (2 L-5)/\bar x_1+ (6 L-13)/x_2\Big] h_{12}(\bar x_1)
\nonumber\\&&{}
-2 \Big[2(3 L\!-\!10)  + 4(L\!-\!1)/x_3  + (6 L\!-\!13)/x_2 \Big] h_{12}(x_3)
+ 2 \Big[ 3+2/\bar x_1 + 3/x_2 \Big] h_{22}(\bar x_1) -\!2\Big[ 3+ 2/x_3 + 3/x_2\Big] h_{22}(x_3)
\nonumber\\&&{}
-(6/\bar x_1)\Big[1 + 1/ x_2 +1/ \bar x_1\Big]\Big[ 2 (L-3)h_{13}(\bar x_1) +  h_{23}(\bar x_1)\Big]
+(6/x_3) \Big[ 1+1/x_2\Big] \Big[ 2(L-3) h_{13}(x_3)+ h_{23}(x_3)\Big],
\end{eqnarray}
\begin{eqnarray}
\lefteqn{x_1 x_3 C^{\mathbb{A}^{(2)}_2}_u(x_i)=}
\nonumber\\&=&
2 x_1 \Big[x_3 (13 L-25)+ 2 x_2 -x_2 x_3 (L-3)\Big] g_1(x_2)
-4  \Big[ x_2 \bar x_2 + 3 x_1 x_3(1-2 L) + x_1 x_2 x_3 (1-L)\Big] g_{11}(x_2,x_2)
                    \hspace*{0.5cm}\phantom{.}
\nonumber\\&&{}
-4 x_1 \bar x_1  g_1(\bar x_1)
-2 x_1  \Big[2x_3  L - 2 x_2 - x_2 x_3 (1-2 L)\Big] g_{11}(\bar x_1,x_2)
+ 2 x_3 \Big[(3-2 L) x_1+2 x_2\Big] g_{11}(\bar x_3,x_2)
\nonumber\\&&{}
-2 x_1  \Big[ (1+x_2) x_3 g_{21}(\bar x_1,x_2) +  x_3 g_{21}(\bar x_3,x_2)-2 x_3 (6+x_2)g_{21}(x_2,x_2)\Big]
- 2 x_1 x_3 \Big[
(1+5 L) g_2(x_2)
\nonumber\\&&{}
+2 L g_{12}(\bar x_1,x_2)-(3-2 L) g_{12}(\bar x_3,x_2)
+2 (7-4 L) g_{12}(x_2,x_2)+g_{22}(\bar x_1,x_2)+g_{22}(\bar x_3,x_2)+4 g_{22}(x_2,x_2)
\Big]
\nonumber\\&&{}
+  2 \Big[ (4x_1-x_2) (x_3/x_2) (1-2 L) + 2 x_1 (1+L) + 2 x_1 x_3 (1-L) \Big] h_{11}(x_2)
+2 x_3 \Big[ (1-2 L) h_{11}(\bar x_3)-h_{21}(\bar x_3)\Big]
\nonumber\\&&{}
- 2x_1  \Big[
\big[ 2 (1+L)+x_3 (1-2 L) \big] h_{11}(\bar x_1)+\bar x_3 h_{21}(\bar x_1) \Big]
+(2/x_2) \Big[ x_2 \bar x_2 -4 x_1 x_3 - x_1 x_2 x_3\Big] h_{21}(x_2)
\nonumber\\&&{}
+ (2/x_2) \Big[ 4 (2-L) x_3 + 2 x_1  + x_1 x_2 (5+2 L)+ 2 x_1 x_3 (13-6 L)/x_2 + 4 x_1 x_3  (L-1)
\Big] h_{12}(x_2)
\nonumber\\&&{}
-2x_1 \Big[ 2/\bar x_1 + 5+2 L\Big] h_{12}(\bar x_1)
+4 (x_3/\bar x_3) \Big[ 2 (L-2) h_{12}(\bar x_3)+h_{22}(\bar x_3)\Big]
-2x_1   h_{22}(\bar x_1)
\nonumber\\&&{}
+ (2/x_2)\Big[ x_1 x_2 -2 x_3\bar x_1 -6 x_1 x_3/x_2\Big] h_{22}(x_2)
+(2/x_2^2) \Big[ x_1 (2 L-7)+2x_1 x_2 (L-5) +x_3 (4 L-9)\Big]  h_{13}(x_2)
\nonumber\\&&{}
- 2(x_1/\bar x_1^2)
\Big[
\big[ 2 L-7 +2 \bar x_1 (L-5) \big] h_{13}(\bar x_1)+
\big[ 1+\bar x_1\big] h_{23}(\bar x_1)
\Big]
-2(x_3/\bar x_3^2) \Big[ (4 L-9 ) h_{13}(\bar x_3)+2 h_{23}(\bar x_3)\Big]
\nonumber\\&&{}
+(2/x^2_2) \Big[ x_1 (1+x_2)+2 x_3\Big] h_{23}(x_2)\,,
\end{eqnarray}
\begin{eqnarray}
\lefteqn{x_2 C^{\mathbb{A}^{(3)}_2}_d(x_i)=}
\nonumber\\&=&
-4 \bar x_1 g_1(\bar x_1)+2 \Big[(3 L-8) x_2+ 2 x_3\Big] g_1(x_3)+6 x_2 g_{11}(\bar x_1,x_3)
+ 2 (3 L-8) x_2 g_2(x_3)+6 x_2 g_{12}(\bar x_1,x_3)
                    \hspace*{2.0cm}\phantom{.}
\nonumber\\&&{}
+6 (1+\bar x_1) h_{11}(\bar x_1)-6 (1+x_3) h_{11}(x_3)
+4 \Big[3 L-10+(2 L-5)/\bar x_1\Big] h_{12}(\bar x_1)
+ 4 \Big[(5-2 L)/x_3 + 10-3 L\Big] h_{12}(x_3)
\nonumber\\&&{}
+2\Big[3+2/\bar x_1\Big] h_{22}(\bar x_1)-2\Big[3+2/x_3\Big] h_{22}(x_3)
-(12/\bar x_1^2) (1+\bar x_1) (L\!-\!3) h_{13}(\bar x_1)
+(12/x_3^2) (1+x_3) (L\!-\!3)  h_{13}(x_3)
\nonumber\\[1mm]&&
-(6/\bar x_1^2)(1+\bar x_1) h_{23}(\bar x_1)
+(6/x_3^2) (1+x_3) h_{23}(x_3)\,,
\end{eqnarray}
\begin{eqnarray}
\lefteqn{x_3 C^{\mathbb{A}^{(3)}_2}_u(x_i)=}
\nonumber\\&=&
-4 \bar x_1 g_1(\bar x_1)-2 x_2 \Big[ 5-4 L+(L-3) x_3 \Big] g_1(x_2)
+2 x_2 \Big[ 2(3-L)+(1-2 L) x_3\Big] g_{11}(\bar x_1,x_2)
                    \hspace*{3.5cm}\phantom{.}
\nonumber\\&&{}
+2 x_2 (1+x_3) \Big[ 2(L-1)  g_{11}(x_2,x_2) - g_{21}(\bar x_1,x_2)+ g_{21}(x_2,x_2)\Big]
+ 2\Big[2 (L-3)+ x_3 (2L-1)\Big] h_{11}(\bar x_1)
\nonumber\\&&{}
-2(1+x_3)\Big[2 (L-1) h_{11}(x_2)- h_{21}(\bar x_1)+ h_{21}(x_2)\Big]
-2\Big[5+2 L+ 2/\bar x_1+ 2 L/x_3 \Big] h_{12}(\bar x_1)
\nonumber\\&&{}
+2\Big[5+2 L+ 2 L/x_3+ 4 x_3(L-1)/x_2\Big] h_{12}(x_2)
-2(1+1/x_3) h_{22}(\bar x_1)+2 (1+1/x_3 +2 x_3/x_2) h_{22}(x_2)
\nonumber\\&&{}
-(2/\bar x_1) \Big[(2 L-7)/x_3 +   2 (L-5) + (2L-7)/\bar x_1\Big] h_{13}(\bar x_1)
+(2/x_2)\Big[ (2 L-7)/x_3 +2 (L-5 )\Big] h_{13}(x_2)
\nonumber\\&&{}
-(2/\bar x_1) \Big[ (1+x_3)/x_3 + (1-x_3\bar x_3)/\bar x_1\Big] h_{23}(\bar x_1)
+(2/x_2) \big(1+1/x_3\big) h_{23}(x_2)\,,
\end{eqnarray}
%
%%%%%%%%%%%%%%%%%%%%%%%%%%%%%%%%%%%%%%%%%%%%%%%%%%%%%%%%%%%%%%%%%
%
%  D-coefficient functions for the B-function
%
%%%%%%%%%%%%%%%%%%%%%%%%%%%%%%%%%%%%%%%%%%%%%%%%%%%%%%%%%%%%%%%%%
%
\begin{eqnarray}
\lefteqn{x_2 x_3 D^{\mathbb{V}_1}_d(x_i)=}
\nonumber\\&=&
x_2 x_3 \Big[
(3 L-7) g_1(x_3)+(3-4 L) g_{11}(\bar x_1,x_3)+2 (4 L-3) g_{11}(x_3,x_3)
-2 g_{21}(\bar x_1,x_3)+4 g_{21}(x_3,x_3)\Big]
                    \hspace*{2.0cm}\phantom{.}
\nonumber\\&&{}
-2 x_3 L h_{11}(\bar x_1)
+ 2 \Big[ x_2 (3-2 L) + x_3 L \Big] h_{11}(x_3)-x_3 h_{21}(\bar x_1)
-(2 x_2-x_3) h_{21}(x_3)
+ (x_3/\bar x_1) (5-2 L) h_{12}(\bar x_1)
\nonumber\\&&{}
-(5-2 L) h_{12}(x_3)-(x_3/\bar x_1) h_{22}(\bar x_1) + h_{22}(x_3)\,,
\end{eqnarray}
\begin{eqnarray}
\lefteqn{x_1 x_2 D^{\mathbb{V}_1}_u(x_i)=}
\nonumber\\&=&
4 (3-4 L) x_1 x_2 g_{11}(x_2,x_2)+
x_1 x_2 \Big[
2 L g_{11}(\bar x_1,x_2)+(2 L-3) g_{11}(\bar x_3,x_2)
+g_{21}(\bar x_1,x_2)+g_{21}(\bar x_3,x_2)-4 g_{21}(x_2,x_2)
\Big]
\nonumber\\&&{}
+
(2x_1x_2/x_3) L h_{11}(\bar x_1)
+2 x_1 \Big[2 L-3 - (x_2/x_3) L \Big] h_{11}(x_2)
+(x_1x_2/x_3) h_{21}(\bar x_1)
+x_1 \big(2-x_2/x_3 \big) h_{21}(x_2)
\nonumber\\&&{}
+
\big[2x_1x_2/(\bar x_1 x_3)\big] h_{12}(\bar x_1)
+ (2L-7)\Big[(x_2/\bar x_3)h_{12}(\bar x_3)-h_{12}(x_2)\Big]
- 2 (x_1/x_3) h_{12}(x_2) +(x_2/\bar x_3) h_{22}(\bar x_3)-h_{22}(x_2)
\nonumber\\
\end{eqnarray}
\begin{eqnarray}
\lefteqn{x_2 D^{\mathbb{A}_1}_d(x_i)=}
\nonumber\\&=&
x_2 (3 L-8) g_1(x_3) + 3 x_2 g_{11}(\bar x_1,x_3)
+ 2 (L-1) \Big[ h_{11}(\bar x_1)- h_{11}(x_3)\Big] + h_{21}(\bar x_1) - h_{21}(x_3)
                    \hspace*{3.5cm}\phantom{.}
\nonumber\\
&&
+
(7-2 L)\Big[ (1/\bar x_1) h_{12}(\bar x_1) - (1/x_3) h_{12}(x_3)\Big]
-(1/\bar x_1) h_{22}(\bar x_1) + (1/x_3) h_{22}(x_3)\,,
\end{eqnarray}
\begin{eqnarray}
\lefteqn{x_1 x_2 D^{\mathbb{A}_1}_u(x_i)=}
\nonumber\\&=&
 x_1 x_2 \Big[
(3 L-7) g_1(x_2)+2 (4 L-3) g_{11}(x_2,x_2)
-2 L g_{11}(\bar x_1,x_2)+(3-2 L) g_{11}(\bar x_3,x_2)-g_{21}(\bar x_1,x_2)-g_{21}(\bar x_3,x_2)
\nonumber\\&&{}
+4 g_{21}(x_2,x_2) \Big]
- (2x_1x_2/x_3) (L-1) h_{11}(\bar x_1)-2 x_2 h_{11}(\bar x_3)+2 x_2 h_{11}(x_2)
+ 2x_1\Big[ (x_2/x_3)(L-1)+ 3-2 L \Big] h_{11}(x_2)
\nonumber\\&&{}
- (x_1x_2/x_3) h_{21}(\bar x_1) - x_1 \big( 2- x_2/x_3\big) h_{21}(x_2)
+ (5-2 L) \Big[(x_2/\bar x_3) h_{12}(\bar x_3) - h_{12}(x_2)\Big]
-(x_2/\bar x_3) h_{22}(\bar x_3) + h_{22}(x_2)\,,
\nonumber\\
\end{eqnarray}
%%%%%%%%%%%%%%%%%%%%%%%%%%%%%%%%%%%%%%%%%%%%%%%%%%%%%%%%%%%%%%%%%%%%%%%%%%%%%%%%%%%%%%%%%%%%%%%%%%%%
\end{widetext}
%\addcontentsline{toc}{section}{References}

%%%%%%%%%%%%
\end{document}